\documentclass[11pt, a4paper]{article}
\pdfoutput=1

\usepackage{jcappub}

\usepackage{comment}
\usepackage{amsmath}
\usepackage{amsfonts}
\usepackage{amssymb}
\usepackage{graphicx}
\usepackage{mathrsfs}
\usepackage{latexsym}
\usepackage{color}
\usepackage{slashed, cancel}
\usepackage{hyperref}
\usepackage{cleveref}
\usepackage{float}
\usepackage{tikz}
\usepackage{bbold}%For identity-matrix symbol
\usepackage{soul} %For strikethrough, \st{text}
\usepackage{mathtools}% http://ctan.org/pkg/mathtools
\usepackage{support-caption} 
\usepackage{subcaption}
\usepackage{enumitem}
\usepackage{journals}
\allowdisplaybreaks

\usepackage[utf8]{inputenc} 
\usepackage[version=3]{mhchem}
\hypersetup{urlcolor=blue, colorlinks=true} % Colors hyperlinks in blue - change to black if annoying

\usepackage{fancyhdr}
\renewcommand{\arraystretch}{1}
\numberwithin{equation}{section}

\definecolor{rossos}{rgb}{0.8,0.2,0.3}
\definecolor{bluscuro}{rgb}{0.15, 0.2, .85}
\definecolor{bluchiaro}{cmyk}{1,.3,0.,0.1}
\hypersetup{colorlinks, citecolor=bluscuro, linkcolor=bluscuro, urlcolor=bluscuro}

\makeatother   % Cancel the effect of \makeatletter
\newcommand{\GeV}{{\rm \,GeV}}
\newcommand{\TeV}{{\rm \,TeV}}
\newcommand{\MeV}{{\rm \,MeV}}
\newcommand{\keV}{{\rm \,keV}}

\newcommand{\cm}{{\rm \,cm}}
\newcommand{\fm}{{\rm \,fm}}
\newcommand{\km}{{\rm \,km}}

\newcommand{\g}{{\rm \,g}}
\newcommand{\K}{{\rm \,K}}

\newcommand{\Msun}{M_\odot}
\newcommand{\Mstar}{M_\star}
\newcommand{\Rstar}{R_\star}
\newcommand{\vstar}{v_\star}

\newcommand{\muFn}{\mu_{F,n}}
\newcommand{\muFp}{\mu_{F,p}}
\newcommand{\muFe}{\mu_{F,e}}
\newcommand{\muFmu}{\mu_{F,\mu}}
\newcommand{\muFi}{\mu_{F,i}}
\newcommand{\qomax}{q_0^{\rm MAX}}
\newcommand{\qonorm}{q_0^{\rm norm}}
\newcommand{\fMB}{f_{\rm MB}}
\newcommand{\fFD}{f_{\rm FD}}

\newcommand{\erf}{{\rm \,Erf}}
\newcommand{\optdepth}{\tau_{\chi}}

 \def\be   {\begin{equation}}   \def\ee   {\end{equation}}
 \def\ba   {\begin{array}}      \def\ea   {\end{array}}
 \def\bea  {\begin{eqnarray}}   \def\eea  {\end{eqnarray}}
 \def\bean {\begin{eqnarray*}}  \def\eean {\end{eqnarray*}}
\begin{document}

%\today

\title{Improved Treatment of Dark Matter Capture in Neutron Stars}

\author[a]{Nicole F.\ Bell,}
\author[b]{Giorgio Busoni,}
\author[a]{Sandra Robles}
\author[a]{and Michael Virgato}
\affiliation[a]{ARC Centre of Excellence for Dark Matter Particle Physics, \\
School of Physics, The University of Melbourne, Victoria 3010, Australia}
\affiliation[b]{Max-Planck-Institut fur Kernphysik, Saupfercheckweg 1, 69117 Heidelberg, Germany.}

\emailAdd{\tt n.bell@unimelb.edu.au}
\emailAdd{\tt giorgio.busoni@mpi-hd.mpg.de}
\emailAdd{\tt sandra.robles@unimelb.edu.au}
\emailAdd{\tt mvirgato@student.unimelb.edu.au}

\abstract{
Neutron stars provide a cosmic laboratory  to study the nature of dark matter particles and their interactions.  Dark matter can be captured by neutron stars via scattering, where kinetic energy is transferred to the star.  This can have a number of observational consequences, such as the heating of old neutron stars to infra-red temperatures. Previous treatments of the capture process have employed various approximation or simplifications. 
We present here an improved treatment of dark matter capture, valid for a wide dark matter mass range, that correctly incorporates all relevant physical effects.  These include  gravitational focusing, a fully relativistic scattering treatment, Pauli blocking, neutron star opacity and multi-scattering effects.  We provide general expressions that enable the exact capture rate to be calculated numerically, and derive simplified expressions that are valid for particular interaction types or mass regimes and that greatly increase the computational efficiency. Our formalism is applicable to the scattering of dark matter from any neutron star constituents, or to the capture of dark matter in other compact objects.
}

\maketitle

\section{Introduction}

There is a long history of using stars as cosmic laboratories for fundamental physics and, in particular, as a means of probing the nature of dark matter (DM).  If DM particles couple to visible matter, they will scatter with the constituents of stars. These collisions can result in sufficient energy loss that the DM particles become gravitationally bound to the star, and thus a population of DM is accumulated in the star over time~\citep{Gould:1987ju,Gould:1987ir,Jungman:1995df,Kumar:2012uh,Kappl:2011kz,Busoni:2013kaa,Bramante:2017xlb}.  Importantly, the rate of capture of DM particles is controlled by the size of the scattering cross section with either the nucleons or leptons in the star. This provides interesting complementary with terrestrial direct detection searches, which look for exactly the same scattering interactions in either nuclear or electron recoil experiments, albeit in a different kinematic regime.

The accumulation of DM in stars has a range of potentially observable consequences.  For the case of the Sun, accumulated DM could be detected via its annihilation to neutrinos~\citep{Tanaka:2011uf,Choi:2015ara,Adrian-Martinez:2016ujo,Adrian-Martinez:2016gti,Aartsen:2016zhm} or to other particles which escape the Sun~\citep{Batell:2009zp,Schuster:2009au,Bell:2011sn,Feng:2016ijc,Leane:2017vag}.  In  some cases, the presence of DM can also alter the energy transport in the Sun~\citep{Gould:1989hm, Gould:1989ez, Vincent:2013lua, Geytenbeek:2016nfg}.
The capture of DM in a neutron star (NS) may have various dramatic consequences, ranging from the collapse of neutron stars to black holes~\citep{ Goldman:1989nd,Kouvaris:2010jy,Kouvaris:2011fi,McDermott:2011jp,Guver:2012ba,Bell:2013xk, Bramante:2013nma,Bertoni:2013bsa,Garani:2018kkd}, 
or a modification of the rate of neutron star mergers~\citep{Bramante:2017ulk}. (Modification of the gravitational wave signatures from binary neutron star mergers is also possible~\citep{Ellis:2017jgp, Ellis:2018bkr,Nelson:2018xtr}, though this would require either a larger DM fraction than we consider here, or a DM halo which extends beyond the star, as might be obtained with light or self-interacting DM.)

Recent work has focused on the {\it kinetic heating} of NSs that results from DM capture~\citep{Baryakhtar:2017dbj,Raj:2017wrv,Bell:2018pkk,Camargo:2019wou,Bell:2019pyc,Garani:2019fpa,Acevedo:2019agu,Joglekar:2019vzy,Gonzalez}. The kinetic energy transferred from the DM particles to the star can cause heating of the star to temperatures of order 2000~K, which may be observable with forthcoming telescopes~\citep{Baryakhtar:2017dbj}.  This allows potential sensitivity to DM-nucleon cross sections of order $10^{-45}\cm^2$, which, depending on the type of interaction, is either comparable to current direct detection experiments or significantly more sensitive.
Furthermore, kinetic heating from DM scattering on the lepton constituents of a NS would greatly surpass the sensitivity of current and forthcoming electron-recoil type direct detection experiments, for all interaction types~\citep{Bell:2019pyc}. 

In previous work, the rate of capture of DM in NSs has been computed at various levels of approximation.  In this paper, we shall improve and extend the existing calculations in the literature, to properly incorporate the relevant physical effects. 

The scattering of DM particles in NSs occurs in an interesting kinematic regime. Because DM particles are non-relativistic in the Universe today, most DM scattering scenarios involve low velocities and very small momentum exchange, e.g., direct detection experiments or the capture of DM in the Sun. However, DM particles are accelerated to quasi-relativistic speeds upon infall to a NS. Moreover, the constituents of the neutron star may themselves be relativistic, particularly in the case of highly degenerate leptons. With the exception of the recent ref.~\citep{Joglekar:2019vzy},  most previous calculations assume the dark matter scatters from nonrelativistic targets.  The treatment of the scattering interaction that we provide here is fully relativistic, with Lorentz invariance correctly encoded.

There are other important kinematic effects that must be taken in account. At small mass, the DM scattering rate is suppressed by Pauli blocking in the highly degenerate NS medium, while at large  mass, a single scattering interaction is insufficient to lead to capture. We shall provide an  improved treatment of these two effects, which, in the past, have usually been handled in a schematic way. In addition, due to the NS gravitational field, we must include gravitational focusing of the dark matter trajectories \cite{Goldman:1989nd,Kouvaris:2007ay}. Finally, due to its high density, we cannot always treat the NS star as optically thin. Indeed, for the scattering cross sections for which DM capture in NSs become efficient, i.e., where the capture probability is of order 1, opacity effects are important. Interestingly, such cross sections are broadly comparable to those that may be probed in future terrestrial direct detection experiments.

The aim of this paper is to provide, for the first time, a realistic calculation that correctly includes gravitational focusing, a fully relativistic scattering treatment, Pauli  blocking, NS opacity and multi-scattering effects. In doing so we will provide exact expressions for the numerical evaluation of the capture rate, as well as a number of approximations that are valid in particular mass or cross section regimes. 
The examples we shall provide will assume scattering from the neutron component of the NS but, in fact, our formalism can be applied in a straightforward way to scattering from any NS constituents, including degenerate electrons or more exotic species, or to DM capture in other compact objects.

Our paper is organised as follows: We discuss relevant details of the NS composition and equation of state in Section~\ref{sec:NS}.  In Section~\ref{sec:capoptthin} we write down an exact relativistic expression for the capture rate, including Pauli blocking, in the optically thin limit, while in Section~\ref{sec:largemassandsigma} we then modify this capture rate to account for the NS opacity and multi-scattering effects. Our results are summarised in Section~\ref{sec:results} and our conclusions presented in Section~\ref{sec:conclusions}.

\section{Neutron Stars}
\label{sec:NS}

Neutron stars are the most compact stars known in the Universe. They are born in core-collapse supernova explosions of massive stars. Our knowledge of NSs has improved over recent decades but, as we shall detail below, their exact composition is still unknown. 

\subsection{Internal Structure}
\label{sec:intstruct}
NSs are primarily composed of degenerate matter. The standard picture for NS composition assumes that below a thin atmosphere, two concentric regions are found: a locally homogeneous core that accounts for $\sim 99\%$ of the mass of the star, and a thin crust $\sim 1\km$ thick~\cite{Haensel:2007yy,Chamel:2008ca}.  

The crust can be further characterised as two shells. The outermost shell, called the outer crust, is comprised of ionised heavy nuclei in a Coulomb lattice and non-relativistic degenerate electrons. Its surface, in the absence of accretion, is expected to be made of completely ionised $\ce{^{56}Fe}$, while the inner layers contain increasingly neutron-rich nuclei until the neutron drip density, $\rho_{ND}\sim 4.3 \times 10^{11} {\g \cm^{-3}}$, is reached~\cite{Ruester:2005fm,RocaMaza:2008ja,Pearson:2011,Kreim:2013rqa,Chamel:2015oqa}. 
This defines the transition to the inner crust, which is expected to be inhomogeneous and composed of dense nuclear structures or clusters, and a dilute gas of free neutrons and  relativistic electrons. 
Neutrons in the inner crust are expected to be superfluid~\cite{Page:2013hxa}.
Approaching the crust-core boundary, the so-called \emph{pasta phases}, nucleon cluster structures with different topologies, are expected to be found~\cite{Ravenhall:1983uh,Hashimoto:1984,Lassaut:1987,Oyamatsu:1993,Watanabe:2004tr,Avancini:2010ch,Grill:2014aea}. The density of this boundary, $\rho_{cc}$, is of order half the nuclear saturation density, $\rho_0=2.8 \times 10^{14} {\g \cm^{-3}}$. 

Even though the crust contributes only $\sim 1\%$ to the NS mass, it plays a crucial role in our understanding of NSs, since physics phenomena in the core are not observable unless some effect is transmitted through the crust. 
In particular, at low temperature, electrons, lattice phonons 
 and superfluid phonons of the neutron gas in the inner crust are the most relevant excitations in the determination of the NS thermodynamical properties  \cite{Page:2012zt}. In addition, while neutron superfluidity in the crust is suspected to be responsible for the glitches observed in pulsars, superfluidity in the core can explain the deceleration in the spin-down of pulsars~\cite{Page:2013hxa}. 

In the NS core, nucleon clusters dissolve into a superfluid liquid made of neutrons together with an admixture of protons and electrons in $\beta$ equilibrium. When the electron Fermi energy exceeds the muon mass, at number densities $n\gtrsim0.12\fm^{-3}$, muons start to appear.  At higher densities, an inner solid core containing meson condensates, hyperons or quark matter may or may not be present in massive neutron stars~\cite{Haensel:2007yy,Weber:2006ep}.

\subsection{Equation of State}
The NS equation of state (EoS) relates the pressure, $P$, to other fundamental parameters. With the sole exception of the outermost layers (a few meters thick) of a NS and newly-born NSs, the pressure in the strongly degenerate matter is independent of the temperature. Then, the microphysics governing particle interactions across different layers of a NS is encapsulated in one-parameter EoS, $P=P(\rho)$, where $P$ and $\rho$ are pressure and density, respectively. Calculations of the EoS are frequently reported in tabular form in terms of   the baryon number density, $n_b$, i.e. $P=P(n_b)$, $\rho=\rho(n_b)$.   
The EoS is the key ingredient for NS structure calculations; its precise determination, however, is an open problem in nuclear astrophysics and is limited by our understanding of the behaviour of nuclear forces in such extreme conditions. 
While the EoS of the outer crust is based on  experimental data and is rather well established, physics beyond the neutron drip point cannot be replicated in the laboratory, and theoretical models are used instead. Thus, the EoS of the inner crust and the core are calculated in a reliable way using methods of nuclear many-body theory. Nevertheless, even when considering the simplest NS core made of neutrons, protons, electrons and muons, the reliability of this EoS decreases at densities significantly higher than $\rho_0$, primarily due to our lack of knowledge of strong interactions in superdense matter. The only way to constrain these models is through observations~\cite{Watts:2016uzu}.  

Several EoSs can be found in the literature, see e.g. refs.~\citep{Akmal:1998cf,RikovskaStone:2006ta,Goriely:2010bm,Kojo:2014rca,Baym:2017whm,Annala:2019eax}. In these models, unified EoSs are valid in all regions of the stellar interior. They are obtained by performing many-body calculations based on a single effective nuclear Hamiltonian \cite{Haensel:2004nu}. In this paper, we have considered the unified equations of state for cold non-accreting matter developed by the Brussels-Montreal group \cite{Goriely:2010bm,Pearson:2011,Pearson:2012hz,Goriely:2013} using  the nuclear energy-density functional theory, whose analytical fits are given in refs.~\cite{Potekhin:2013qqa,Pearson:2018tkr}. 
These fits provide us with an excellent tool for evaluating NS microscopic properties without directly performing the underlying nuclear physics calculations.

\subsection{Neutron Star models}
\label{sec:NSmodels}
A given EoS is characterised by a single parameter,  the central density, $\rho_c$, and families of EoSs can be constructed by varying this parameter. 
In order to determine the NS structure, the EoS, $P=P(n_b)$, $\rho=\rho(n_b)$, is coupled to the general relativistic form of the hydrostatic equilibrium equation, known as the Tolman-Oppenheimer-Volkoff (TOV) equations \cite{Tolman:1939jz,Oppenheimer:1939ne}
\begin{eqnarray}
    \frac{dP}{dr}&=&- \rho(r)c^2\left[1+\frac{P(r)}{\rho(r)c^2}\right]\frac{d\Phi}{dr}, \label{eq:TOV} \\
    \frac{d\Phi}{dr}&=&\frac{G M(r)}{ c^2r^2}\left[1+\frac{4\pi P(r) r^3 }{M(r) c^2}\right]\left[1-\frac{2GM(r)}{ c^2 r}\right]^{-1},     
 \label{eq:gravpot}
\end{eqnarray}
and the mass equation
\begin{equation}
    \frac{dM}{dr}=4\pi r^2 \rho(r),
 \label{eq:NSmass}
\end{equation}
where $M(r)$ is the mass contained within a sphere of radius $r$ and $\Phi(r)$ is the gravitational potential. Note that we are assuming a non-rotating, non-magnetized, spherically symmetric NS, therefore we use the Schwarzschild metric 
\begin{equation}
ds^2= -d\tau^2 = -B(r) c^2 dt^2 + A(r)dr^2 +r^2 d\Omega^2,    \label{eq: Schwarzschild} 
\end{equation}
where
\begin{eqnarray}
A(r) &=& \left[1-\frac{2GM(r)}{c^2 r}\right]^{-1}, \label{eq:Arequation}\\
B(r) &=& e^{2\Phi},\\
\frac{d}{dr}B(r) &=& \frac{2G}{c^2 r^2}\left[M(r)+\frac{4\pi}{c^2}P(r)r^3\right] \left[1-\frac{2GM(r)} {c^2 r}\right]^{-1} B(r). \label{eq:Brequation}
\end{eqnarray}

The coupled  differential equation system is integrated from the centre, with $\rho(0)=\rho_c$ as a free parameter, out to the outermost layer of the outer crust, where $\rho=10^6\g\cm^{-3}$ (this layer is only few meters thick and accounts for only $\sim10^{-12}\Msun$). At that density the NS radius, $\Rstar$, and the gravitational mass of the star $\Mstar=M(r=\Rstar)$ are determined. 
The calculated mass and radius can then be compared with those inferred from astrophysical observations, such as low mass x-ray binaries  \cite{Steiner:2010fz,Steiner:2012xt,Lattimer:2013hma,Ozel:2015fia,Ozel:2016oaf,Miller:2016pom}, isolated NSs  \cite{Bogdanov:2019ixe,Riley:2019yda,Miller:2019cac,Raaijmakers:2019qny}, and more recently using  gravitational wave (GW) data and its respective electromagnetic (EM) counterpart from the binary NS merger event GW170817~\cite{TheLIGOScientific:2017qsa,Abbott:2018wiz,Monitor:2017mdv,Radice:2018ozg}. As argued in ref.~\cite{Bell:2019pyc}, from the set of EoS functionals in refs.~\cite{Potekhin:2013qqa,Pearson:2018tkr} we have chosen functional BSk24 as our benchmark EoS family since it gives slightly better NS mass fits to observational data than BSk25. BSk25 is also allowed by current observations. 
 Other functionals in refs.~\cite{Potekhin:2013qqa,Pearson:2018tkr} are ruled out by observational data. E.g., BSk19 cannot account for massive NSs, while BSk22 is excluded by the constraint on the tidal deformability parameter imposed by GW170817 data \cite{Koppel:2019pys}. BSk20 and BSk21 are very similar to BSk26 and BSk24, respectively~\cite{Perot:2019gwl}.

%%%%%%%%%%%%%%%%%%%%%%%%%%%%%%%%%%%%%%%%
\begin{table}[tb]
\centering
\begin{tabular}{|l|c|c|c|c|}
\hline
\bf EoS & \bf BSk24-1 & \bf BSk24-2 & \bf BSk24-3 & \bf BSk24-4 \\ \hline
$\rho_c$ $[\rm{g \, cm^{-3}}]$ & $5.94 \times 10^{14}$   & $7.76 \times 10^{14}$ & $1.04 \times 10^{15}$ & $1.42 \times 10^{15}$  \\
$\Mstar$ $[\Msun]$ & 1.000 & 1.500 & 1.900 & 2.160  \\
$\Rstar$ [km] & 12.215  & 12.593 & 12.419 & 11.965 \\
$B(\Rstar)$ & 0.763 & 0.648 & 0.548 & 0.467\\
$c_s(0)$ $[c]$ & 0.511 & 0.628 & 0.734 & 0.835 \\
\hline
\end{tabular} 
\caption{Benchmark NSs, for four different  configurations of the equations of state (EoS) for cold non-accreting neutron stars with Brussels–Montreal functionals BSk24 \cite{Pearson:2018tkr}. EoS configurations are determined by the central mass-energy density $\rho_c$. 
}
\label{tab:eos}
\end{table} 
%%%%%%%%%%%%%%%%%%%%%%%%%%%%%%%%%%%%%%%%

Particle number fractions and chemical potentials for the different species ($n$, $p$, $e$ and $\mu$)  are calculated as functions of the baryon number density $n_b$. 
The relevant analytic functions for $Y_i$  and $\muFi$ in the core were derived in Appendix~C of ref.~\cite{Pearson:2018tkr} for BSk24, BSk25 and other functionals,  under the conditions of beta equilibrium 
\begin{eqnarray}
\muFn(n_b,Y_p) &=& \muFp(n_b,Y_p) +  \muFe(n_b,Y_e), \\
\muFn(n_b,Y_p) &=& \muFp(n_b,Y_p) +  \muFmu(n_b,Y_\mu), 
\end{eqnarray}
and charge neutrality 
\begin{equation}
 Y_p(n_b)=Y_e(n_b)+Y_\mu(n_b),   
\end{equation}
where $Y_i$ is the number of species $i$ per nucleon and $\muFi$ its corresponding chemical potential. (See ref.~\cite{Pearson:2018tkr} for further details on the calculation of these quantities.) 
Note that $Y_n(n_b)=1-Y_p(n_b)$. The chemical potentials in the inner and outer crust were also calculated  assuming beta equilibrium; see ref.~\cite{Pearson:2018tkr} for further details. 

%%%%%%%%%%%%%%%%%%%%%%%%%%%%%%%%%%%%%%%%%%%%%%%%%%
\begin{figure}[t] 
\centering
\includegraphics[width=0.485\textwidth]{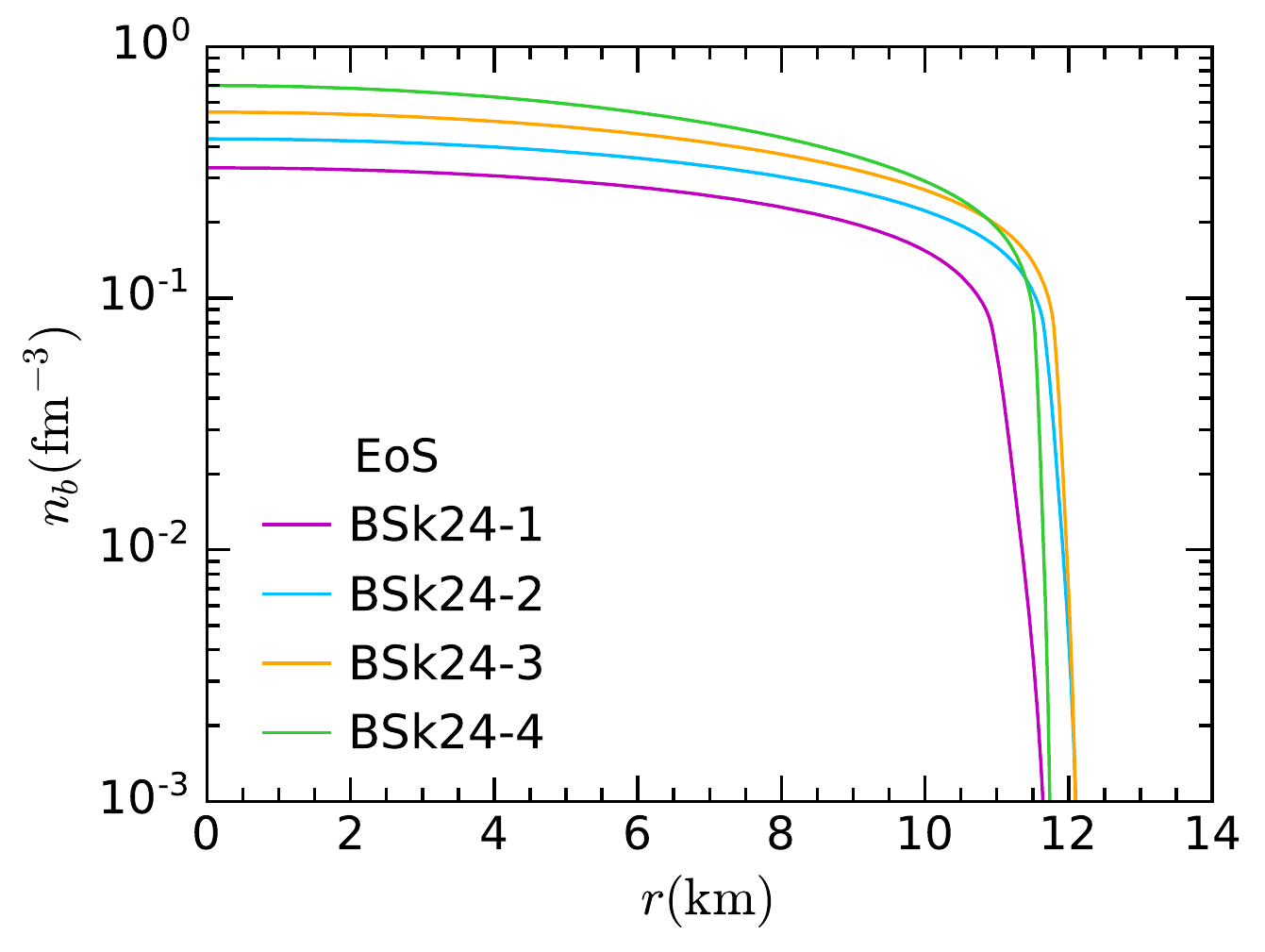}
\includegraphics[width=0.475\textwidth]{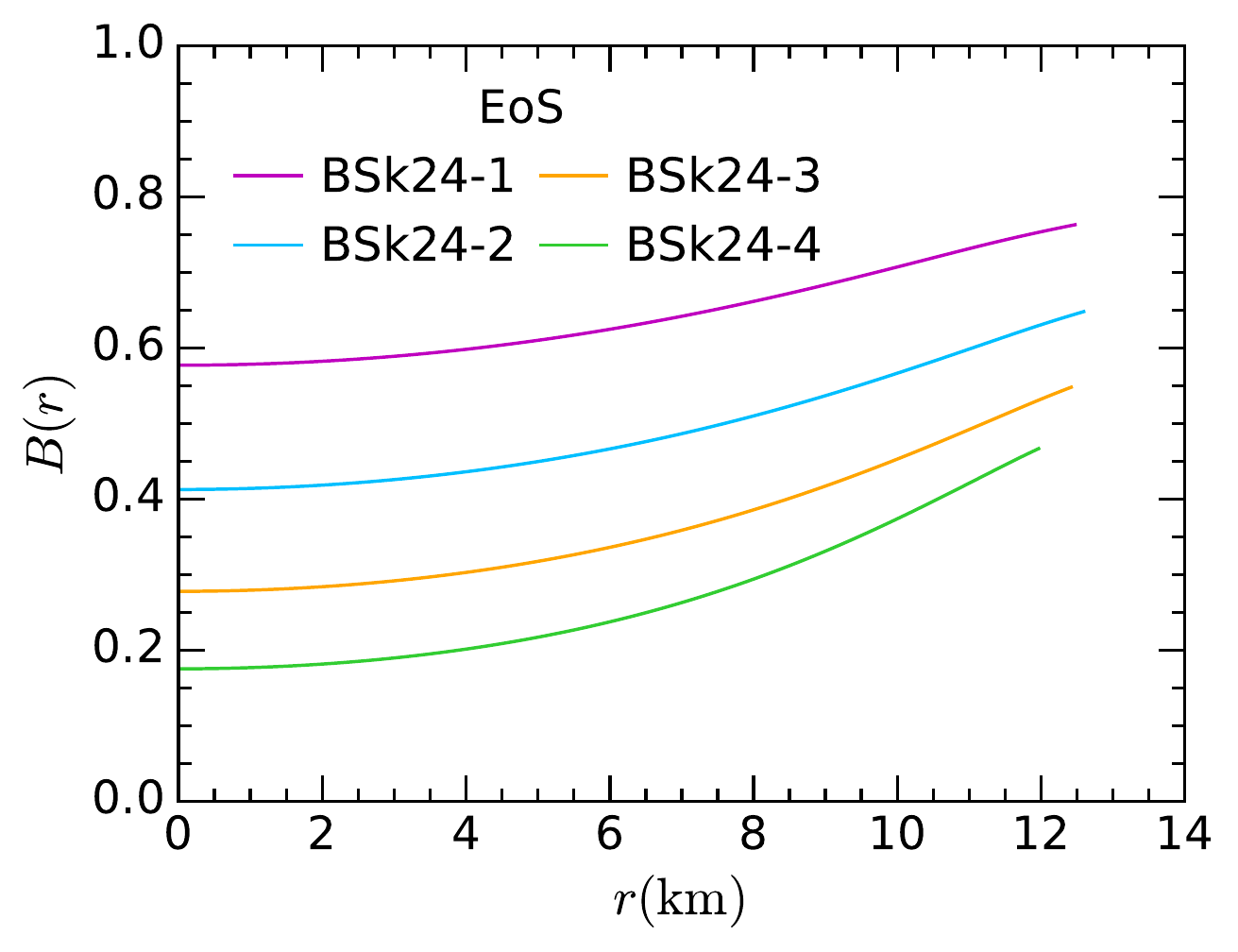}
\includegraphics[width=0.485\textwidth]{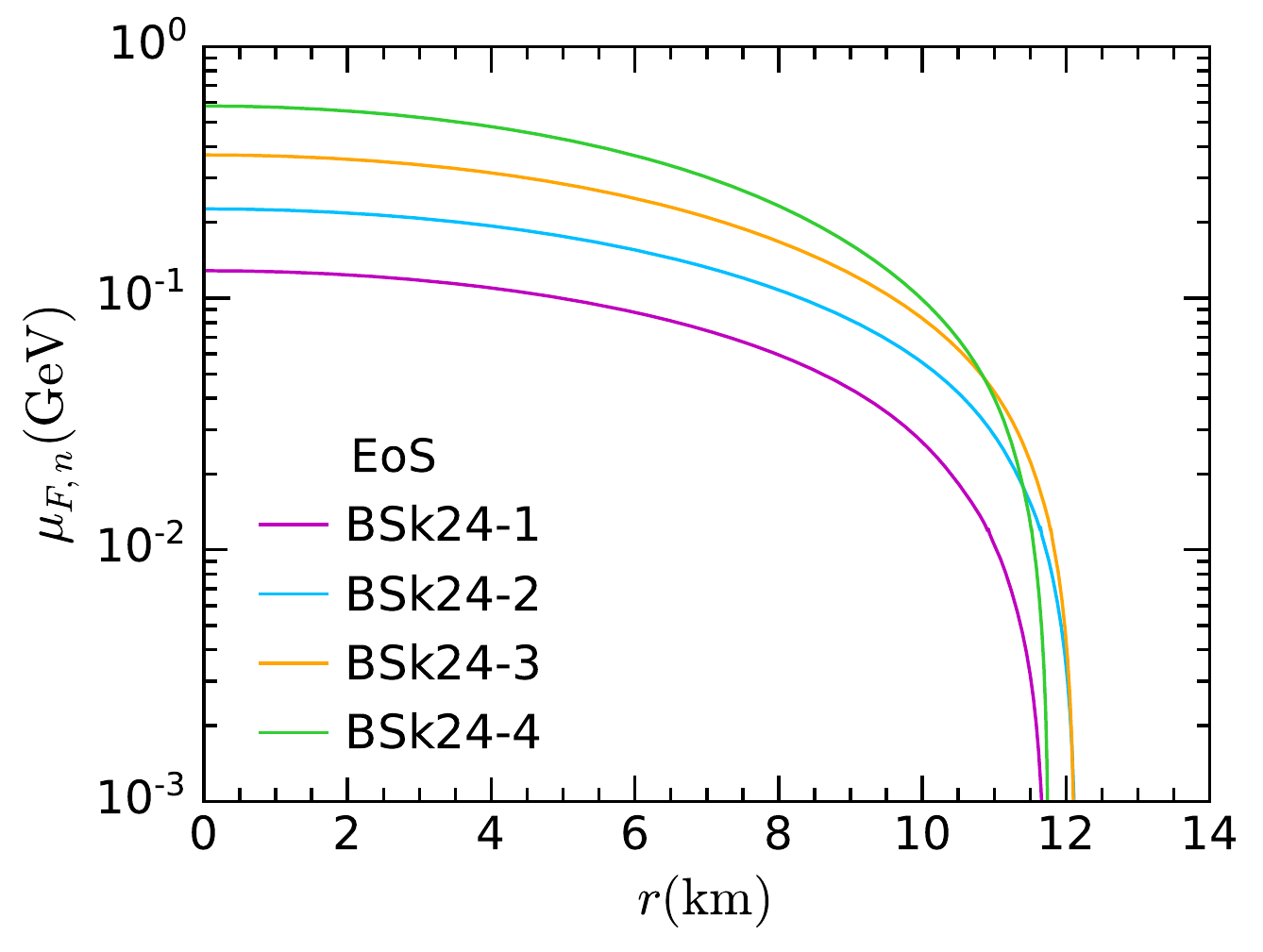}
\includegraphics[width=0.485\textwidth]{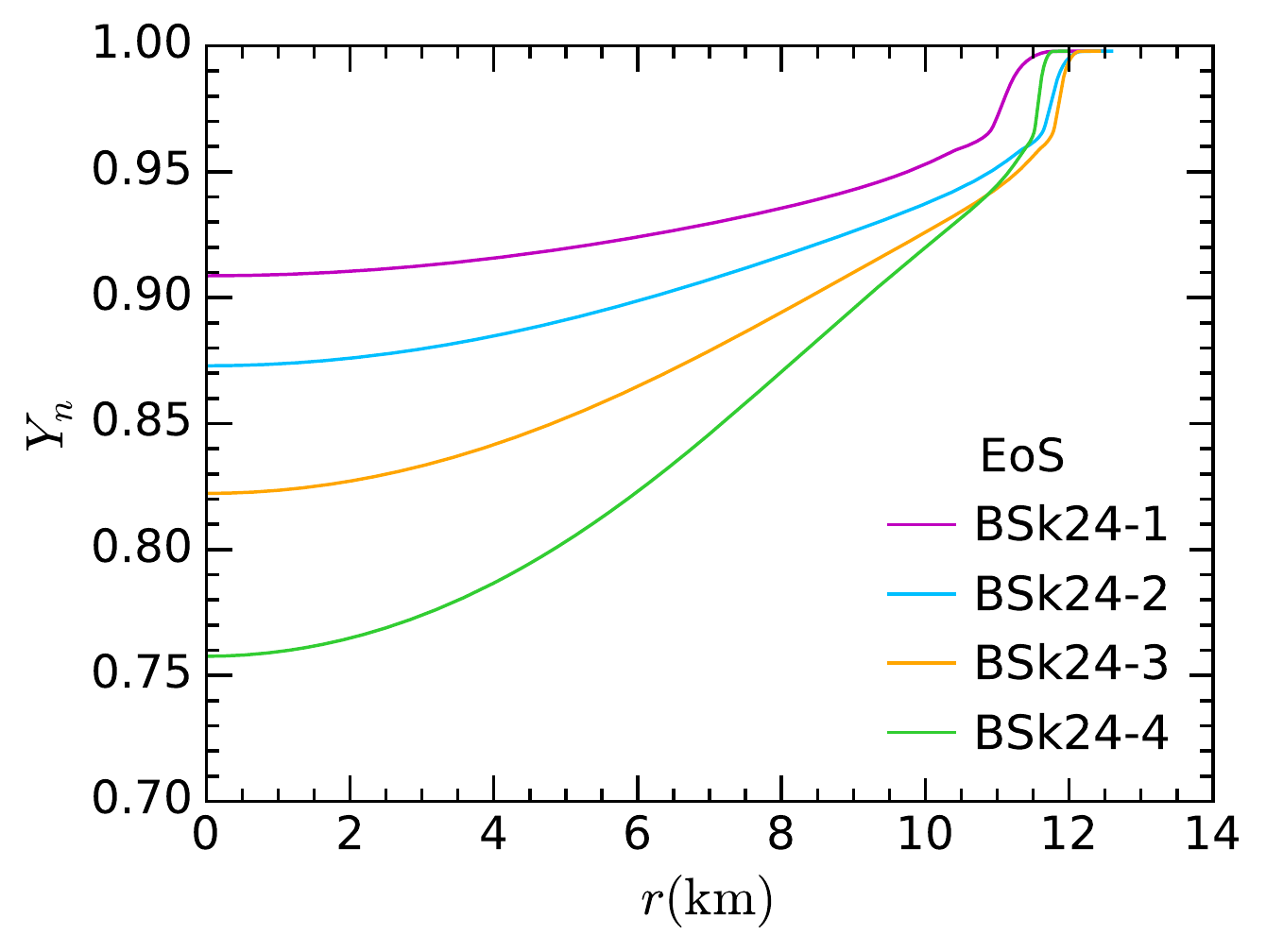}
\caption{Top left: Baryon number density profile for the different configurations of the  BSk24 functional in Table~\ref{tab:eos}. Top right: $B$ radial profile. Bottom left: Neutron chemical potential as a function of the NS radius.
Bottom right: $Y_n$ abundance as a function of the NS radius, where the neutron  fraction is computed with respect to the baryon number ($N_n+N_p$).  }
\label{fig:NSradprofs}
\end{figure}  
%%%%%%%%%%%%%%%%%%%%%%%%%%%%%%%%%%%%%%%%%%%%%%%%%%

Using the analytical fits for BSk24, implemented as {\tt FORTRAN} subroutines  by the  authors of ref.~\cite{Pearson:2018tkr}\footnote{These public available subroutines can be found at \url{http://www.ioffe.ru/astro/NSG/BSk/}.},  we solve the TOV equations \ref{eq:TOV}-\ref{eq:NSmass}. At every step of the adaptive fourth order Runge-Kutta integration of the differential equation system previously described, we calculate $n_b$,  $\muFi$ and $Y_i$. It is worth remarking that different fits apply to the core, inner and outer crust. In this way, we have obtained  
 radial profiles  from the NS centre out to the outermost layers of the crust, relevant for the calculations in the following sections.   These profiles vary with the EoS choice, determined by $\rho_c$. The $B(r)$ profile is obtained by solving Eq.~\ref{eq:Brequation} subject to the boundary condition
 \begin{equation}
B(\Rstar) = 1-\frac{2GM_\star}{c^2 \Rstar}.    \label{eq:BRstar}
\end{equation} 
To exemplify those calculations, we have chosen four configurations of the functional BSk24, given in Table~\ref{tab:eos}. Note that the maximum NS mass is restricted to  $\Mstar\lesssim2.16\Msun$ by the GW170817 event~\cite{Margalit:2017dij,Shibata:2017xdx,Ruiz:2017due,Rezzolla:2017aly,Shibata:2019ctb}. For completeness, we also provide in Table~\ref{tab:eos} the central value for the speed of sound, $c_s^2(r)=\dfrac{\partial P}{\partial \rho}(r)$. Note that all the chosen NS configurations are below the causality limit $c_s\leq c$~\cite{Haensel:2007yy}. Aside from nucleons, leptons and exotic matter, DM can scatter off the different phonon modes present in a NS (in the crust and the core) as mentioned in section~\ref{sec:intstruct}. The response function of these interactions depends on $c_s$~\cite{Bertoni:2013bsa}. In the following sections, we do not consider this particular case since a more realistic treatment of the pairing in the EoS, essential for superfluidity studies  (such as that of the BSk functionals in ref.~\cite{Goriely:2016sdz} for which no analytical fits are available) would be desirable.

In Fig.~\ref{fig:NSradprofs}, we show the resultant profiles for the baryon number density, the neutron abundance, $Y_n$, and the neutron chemical potential, $\muFn$ and $B(r)$. 
 In fact, the analytical parametrizations in ref.~\cite{Pearson:2018tkr} were obtained using precision fits not only for the core but also for the inner and outer crust. The radial profiles shown in  Fig.~\ref{fig:NSradprofs} include the three regions and will be used in the following sections to compute the DM capture rate.
The influence of the choice of EoS on our final capture rates will be illustrated later in Section~\ref{sec:results}.

\section{Capture Rate in the Optically Thin Limit}
\label{sec:capoptthin}

The capture  and interaction rates, $C$ and $\Omega^-$ respectively, for DM scattering on nucleons were first calculated by Gould  \citep{Gould:1987ju,Gould:1987ir} for the Sun and the Earth.  The early calculations considered a constant DM-nucleon cross section, and were later generalised for arbitrary cross sections~\citep{Garani:2017jcj,Busoni:2017mhe}. 
Recently, increasing interest in capture of DM in NSs has motivated several authors to modify the original derivation in order to obtain expressions valid for NSs. When dealing with capture in NSs, there are two main issues to take into account. Firstly, quantum degeneracy has an important effect. The nucleons in NSs are in a quasi-degenerate state. (Likewise, the electron component is highly degenerate.) As such, the lowest energy levels are nearly completely full, and are therefore not available as final nucleon states in DM-nucleon scattering interactions. For some parameters, this can severely suppress the interaction rate. Secondly, DM particles are accelerated to relativistic speeds as they approach the NS, hence a non-relativistic description is not adequate. Moreover, the neutron star constituents on which the DM scatters may also be relativistic, as is the case for the highly degenerate electron component of the star. 
Ref.~\citep{Garani:2018kkd}  addresses the first issue, by modifying Gould's original result for the interaction rate to correctly include the Fermi-Dirac (FD) distribution, $\fFD$, for the initial and final  nucleon states, such that the DM scattering rate is computed taking only the free nucleon final states into account.

We carefully address both of the above issues by
deriving an exact expression that allows us to calculate  the DM capture and interaction rates for any differential cross section and any relativistic/non-relativistic,  degenerate/non-degenerate NS constituent. In what follows, we will focus on DM scattering from neutrons. However, we note that our approach can be applied to scattering on any other NS constituents. 

\subsection{Capture Rate}
\label{sec:caprate}

We first use the TOV equations~\ref{eq:TOV} and \ref{eq:NSmass} within the star, and the Schwarzschild metric, Eq.~\ref{eq: Schwarzschild}, outside, to obtain an expression for the capture rate as a function of the interaction rate.
To that end, we recast Gould's formalism, replacing all variables at a finite distance $r$ with their general relativistic counterparts and using conservation laws for energy and angular momentum. This will result in a capture rate that incorporates gravitational focusing, i.e. the fact that the DM flux is focused due to the NS gravitational potential, with the trajectory of approaching DM bent toward the star.

The  proper time $d\tau$ spent by a  DM particle of mass $m_\chi$ that moves from the radial coordinate $r$ to $r+dr$,  assuming that the DM speed is $\sim0$ at infinity, is 
\begin{eqnarray}
d\tau = \sqrt{B(r)}\frac{dr}{\dot{r}} = \frac{dr}{\sqrt{\frac{1}{A(r)}\left[1-B(r)\left(1+\frac{J^2}{m_{\chi}^2r^2}\right)\right]}}.  
\label{eq:elapproptime}
\end{eqnarray}
The escape velocity, in terms of the proper time, can be defined as
\begin{eqnarray}
v_{esc}^{2}(r) = 
A(r)\left(\frac{dr}{d\tau}\right)^2 + r^2\left(\frac{d\phi}{d\tau}\right)^2 = 1-B(r).
\label{eq:vescsq}
\end{eqnarray}
Then, following Gould's approach, the number of DM particles, $N_\chi$, captured per unit of proper time by a thin shell of radius $r$ and thickness $dr$ is 
\begin{equation}
\dfrac{dC}{dr} = \frac{dN_\chi}{d\tau} = 2\pi\frac{\rho_\chi}{m_\chi} \frac{\fMB(u_\chi) du_\chi}{u_\chi} \frac{JdJ}{m_\chi^2} d\tau\Omega^{-}(r),  
 \label{eq:Crate_r}
\end{equation}
where $\rho_\chi$ is the local DM density, assumed to be $\rho_\chi=0.4\GeV\cm^{-3}$, $J$ is the DM angular momentum, and $\Omega^{-}(r)$ the DM interaction rate. The quantity  $\fMB(u_\chi)du_\chi$ is the relative velocity distribution between NS particle species and DM particles away from the NS gravitational field, which we assume to be Maxwell-Boltzmann (for further details see ref.~\cite{Busoni:2017mhe}), and reads, 
\begin{equation}
\fMB(u_\chi)du_\chi = \frac{u_\chi}{\vstar v_d}\sqrt{\frac{3}{2\pi}} \left[ e^{-\frac{3(u_\chi-\vstar)^2}{2 v_d^2}}-e^{-\frac{3(u_\chi+\vstar)^2}{2 v_d^2}}\right],   
\label{eq:DMveldist}
\end{equation}
where $\vstar$ is the NS velocity and $v_d$ is the DM velocity dispersion. 

Substituting Eq.~\ref{eq:elapproptime} and the maximum value of $J$ (also  obtained from Eq.~\ref{eq:elapproptime}) 
 \begin{equation}
    J_{max}^2 = \frac{1-B(r)}{B(r)} m_\chi^2 r^2,
\end{equation} 
into Eq.~\ref{eq:Crate_r},
and integrating over the DM relative velocity and the angular momentum, we find  
\begin{eqnarray}
\dfrac{dC}{dr}&=& 4\pi r^2 \sqrt{A(r)}  \frac{\rho_\chi}{m_\chi} \frac{1}{\vstar} \frac{\sqrt{1-B(r)}}{B(r)}  {\rm Erf}\left(\sqrt{\frac{3}{2}}\frac{\vstar}{v_d}\right)  \Omega^{-}(r) \, dr.
\end{eqnarray}
Note that the factor $1/B(r)$ in $J_{max}^2$ is due to gravitational focusing \cite{Kouvaris:2007ay} and that the total number of neutrons within a NS, $N_n$,  has to be calculated in the following way, 
\begin{eqnarray}
\int_0^{\Rstar}  r^2  n_n(r) \sqrt{A(r)}  dr =N_{n},
\label{eq:densitynorm}
\end{eqnarray}
We therefore reabsorb the factor $\sqrt{A(r)}$ within the neutron number density, $n_n(r)$, resulting in a total capture rate of
\begin{equation}
C = \frac{4\pi}{\vstar} \frac{\rho_\chi}{m_\chi} {\rm Erf }\left(\sqrt{\frac{3}{2}}\frac{\vstar}{v_d}\right)\int_0^{\Rstar}  r^2 \frac{\sqrt{1-B(r)}}{B(r)} \Omega^{-}(r)  \, dr. \label{eq:captureclsimplrel}
\end{equation}
The overall $1/B(r)$ correction factor in the final expression for $C$ is in agreement with previous derivations of the capture rate~\cite{Goldman:1989nd,Kouvaris:2007ay}.

\subsection{Interaction Rate}
\label{sec:intratenumeric}

The next step is to derive an expression for the interaction rate as a function of the differential cross section in terms of the Mandelstam variables $s$ and $t$. 
We start from the non-relativistic expression for the interaction rate,  $\Omega^{-}(r)$, and modify it to obtain a relativistic treatment which is correctly Lorentz invariant.
The rate at which a DM particle with velocity $w$  scatters off a neutron target with velocity $u_n$  to a final velocity $v$ is
\begin{equation}
    \Omega^{-}(r) = \int dv \frac{d\sigma}{dv} |\vec{w}-\vec{u}_n| n_n(r) \fMB(u_n) d^3u_n,
\end{equation}
where $\fMB(u_n)d^3u_n$ is the Maxwell-Boltzmann (MB) velocity distribution of the target, in this case neutrons. Switching to the FD distribution, $\fFD$, requires we use the properly normalised neutron number density instead of $n_n(r)f_{MB}(u_n)d^3u_n$.  We therefore make the replacement 
\begin{equation}
    n_n(r)\fMB(u_n)d^3u_n\rightarrow d^3p \frac{g_s}{(2\pi)^3}\fFD(E_n,r),
\label{eq:MBtoFD}    
\end{equation}
where $g_s=2$ is the number of neutron spin states, $p$ is the momentum of the incoming target and $E_n$ is its corresponding  energy. 
Note that the dependence of $\fFD$ on $r$ stems from the radial dependence of the target chemical potential (see Fig.~\ref{fig:NSradprofs}). 
We then obtain
\begin{equation}
    \frac{dN_n}{dV} = d^3p \frac{g_s}{(2\pi)^3}\fFD(E_n,r) = \frac{p E_n dE_n d\cos\theta_{uw}}{2\pi^2}\fFD(E_n,r),
\label{eq:numdensfree}    
\end{equation}
where $\cos\theta_{uw}$ is the cosine of the relative angle between the incoming DM particle and the neutron target. This angle can be traded for the centre of mass energy, $s$, in the following way, 
\begin{equation}
 \frac{d\cos\theta_{uw}}{ds} = \frac{1}{2p \sqrt{E_\chi^2-m_\chi^2}} = \frac{1}{2p m_\chi}\sqrt{\frac{B(r)}{1-B(r)}}. \label{eq:dcosuw}
\end{equation}
In addition, we have to calculate $|\vec{w}-\vec{u}_n|$ using relativistic kinematics, 
\begin{equation}
 |\vec{w}-\vec{u}_n|_{rel} = \frac{\sqrt{s^2-2s(1+\mu^2)m_n^2+(1-\mu^2)^2m_n^4}}{s-(1+\mu^2)m_n^2},  
 \label{eq:wurel}
\end{equation}
where  $\mu=\frac{m_\chi}{m_n}$ and $m_n$ is the neutron mass. 

Next, we can rewrite the differential DM-target cross section  as
\begin{equation}
    dv \frac{d\sigma}{dv} = d\cos\theta_{cm} \frac{d\sigma}{d\cos\theta_{cm}} = dt \frac{d\sigma}{d\cos\theta_{cm}} \frac{d\cos\theta_{cm}}{dt},
\end{equation}
where $t$ is the four-momentum exchanged in the collision, $\theta_{cm}$ is the scattering angle in the centre of mass frame and 
\begin{equation}
 \frac{d\cos\theta_{cm}}{dt} = \frac{2s}{s^2-2s(1+\mu^2)m_n^2+(1-\mu^2)^2m_n^4}. \label{eq:dcoscm}
\end{equation}
Then, $\Omega^-(r)$ reads
\begin{equation}
    \Omega^{-}(r) = \int dt dE_n ds \frac{d\sigma}{d\cos\theta_{cm}} \frac{d\cos\theta_{cm}}{dt} |\vec{w}-\vec{u}_n|_{rel}\frac{d\cos \theta_{uw}}{ds} \frac{p E_n }{2\pi^2}\fFD(E_n,r)(1-\fFD(E^{'}_n,r)). \label{eq:omegaminusfinal}
\end{equation}
Note that we have included the Pauli suppression factor, $1-\fFD(E^{'}_n,r)$, for the neutron final distribution. Here,  
$E_n^{'}$ is the target  energy after the collision, and can be obtained as a function of $E_n,t,s$ and $r$ from kinematics. We do not report the complete expression for $E^{'}_n$ due to its length.

Substituting Eqs.~\ref{eq:dcosuw}, \ref{eq:wurel} and \ref{eq:dcoscm} into \ref{eq:omegaminusfinal}, we obtain
\begin{eqnarray}
\Omega^{-}(r) &=& \int dt dE_n ds \frac{d\sigma}{d\cos\theta_{cm}} \frac{E_n}{2\pi^2m_\chi}\sqrt{\frac{B(r)}{1-B(r)}}\frac{s}{\beta(s)\gamma(s)}\fFD(E_n,r)(1-\fFD(E^{'}_n,r)),\label{eq:omegampauli}
\end{eqnarray}
where 
\begin{eqnarray}
    \beta(s) &=& s-\left(m_n^2+m_\chi^2\right),\\
    \gamma(s) &=& \sqrt{\beta^2(s)-4m_n^2m_\chi^2}.
\end{eqnarray}
The integration intervals  are
\begin{eqnarray}
t_{max} &=& 0,\\
t_{min} &=& -\frac{\beta^2(s)-4m_n^2m_\chi^2}{s},\\
s_{min} &=& m_n^2+m_\chi^2 + 2\frac{E_nm_\chi}{\sqrt{B(r)}}-2\sqrt{\frac{1-B(r)}{B(r)}}m_\chi\sqrt{E_n^2-m_n^2},\\
s_{max} &=& m_n^2+m_\chi^2 + 2\frac{E_nm_\chi}{\sqrt{B(r)}}+2\sqrt{\frac{1-B(r)}{B(r)}}m_\chi\sqrt{E_n^2-m_n^2},
\end{eqnarray}
and $E_n>0$.  In general, the integration range for the neutron energy $E_n$ is  $\left[m_n,\frac{m_n}{\sqrt{B(r)}}\right]$. 

Note that when dealing with NSs at low temperatures, we can take the $T\rightarrow0$ limit by replacing the FD distributions with $\Theta$ functions. In this case, $\fFD(E_n)$ restricts the target initial kinetic energy range to $[0,\mu_{F,n}]$, the factor $1-\fFD(E^{'}_n)$ is approximated by $\Theta(E^{'}_n-\mu_{F,n})$, and the integration range for $E_n$ is $[m_n,m_n+\mu_{F,n}]$.  In fact, the zero temperature  approximation  holds for temperatures up to $T\sim 10^6\K$ for the DM mass range considered here.

Finally,  since we are going to use a realistic neutron number density profile, as described in section~\ref{sec:NSmodels}, $n_{n}(r)$, 
we correct the target number density with the factor $\zeta(r)=\frac{n_{n}(r)}{n_{free}(r)}$ as in ref.~\cite{Garani:2018kkd}, where $n_{free}(r)$ is obtained by
integrating Eq.~\ref{eq:numdensfree} over $E_n$, in the limit $T\rightarrow0$, 
\begin{eqnarray}
    n_{free}(r) &=& \frac{[\mu_{F,n}(r)(2m_n+\mu_{F,n}(r))]^{3/2}}{3\pi^2}.
\end{eqnarray}
Then, the final expression for the interaction rate is
\begin{equation}
\Omega^{-}(r) = \int dt dE_n ds \zeta(r) \frac{d\sigma}{d\cos\theta_{cm}} \frac{E_n}{2\pi^2m_\chi}\sqrt{\frac{B(r)}{1-B(r)}}\frac{s}{\beta(s)\gamma(s)}\fFD(E_n,r)(1-\fFD(E^{'}_n,r))\label{eq:omegampaulitext}.
\end{equation} 
This expression resembles that  of ref.~\citep{Garani:2018kkd}, but uses a relativistic formalism instead. In Appendix~\ref{sec:weakfieldlimit}, we show that Eq.~\ref{eq:omegampauli} reduces to the classical expression for the interaction rate in the non-relativistic limit.

It is important to note that these results all assume that a DM particle will be captured after a single scattering (which, for scattering on nucleons is true for $m_\chi\lesssim 10^6\GeV$), and that the probability of multiple scattering is negligible (which holds for $\sigma \ll \sigma_{th}$). The value of the threshold cross section, $\sigma_{th}$, is defined as the cross section for which the resulting (optically thin) capture rate is equal to the geometric limit \cite{Bell:2018pkk},  
\begin{eqnarray}
C_{geom} &=&  \frac{\pi R_\star^2(1-B(R_\star))}{v_\star B(R_\star)} \frac{\rho_\chi}{m_\chi} \erf\left(\sqrt{\frac{3}{2}}\frac{v_\star}{v_d}\right).
\label{eq:capturegeom}
\end{eqnarray}
Note the $1/B(\Rstar)$ factor in the equation above. In stars and planets where classical Newtonian mechanics can be applied, gravitational focusing would result in a factor  $v_{esc}^2/\vstar =  (1-B(\Rstar))/\vstar$ in Eq.~\ref{eq:capturegeom}, where we have used Eqs.~\ref{eq:vescsq} and \ref{eq:BRstar}. 
%, in the geometric limit. 
In neutron stars, on the other hand, general relativity introduces an additional factor of $1/B(\Rstar)$, which can be obtained from the derivation of the flux of DM particles accreted to a NS with a Schwarzschild metric (Eq.~\ref{eq:captureclsimplrel})~\cite{Goldman:1989nd,Kouvaris:2007ay}.

For scattering on neutrons, the threshold cross section is approximately
\begin{eqnarray}
\sigma_{th} &=&  \begin{cases}
\, \sigma_{ref} \frac{\GeV}{m_\chi} \quad &m_\chi \lesssim 1\GeV \quad \ \ \text{ Pauli blocking  regime}, \\
\, \sigma_{ref} \quad &1\GeV \lesssim m_\chi \lesssim 10^6\GeV,  \\
\, \sigma_{ref} \frac{m_\chi}{10^6\GeV} \quad &m_\chi\gtrsim 10^6\GeV \quad  \text{Multiscattering regime},
\end{cases}
\label{eq:sigmath}
\end{eqnarray}
where
\begin{equation}
    \sigma_{ref}\sim 1.7 \times 10^{-45}\cm^2.
\end{equation}
For scattering on other targets, Pauli blocking is relevant for $\qomax \lesssim  \mu_{\rm target}$ 
while multi-scattering is relevant for $m_\chi \gtrsim \qomax/v_{\star}^2$, where $\qomax$ quantifies the energy transfer in a collision, as will be discussed later.  In addition, because the other target species have a lower abundance than neutrons, the reference cross section, $\sigma_{ref}$, will be higher. The values of $\sigma_{th}$ in Eq.~\ref{eq:sigmath}, and their regions of applicability, can thus be altered appropriately for other target species of interest.

\subsection{Differential Interaction Rate}

In the previous section, we have calculated the interaction rate, $\Omega^-$, assuming the initial DM energy $E_\chi$ takes its pre-capture value. However, we may also be interested in a generalised expression for the interaction rate, valid for arbitrary DM energy. This will be required when we consider capture via multiple scattering. It would also be necessary for studying the subsequent scattering interactions that follow capture, and which allow the dark matter to thermalise in the NS.
In principle, it is possible to calculate this rate numerically by binning $\Omega^-$, Eq.~\ref{eq:omegampaulitext}, in the energy loss, i.e.   multiplying $\Omega^-$ by $\frac{1}{E_i-E_j}\Theta(E_i+E_n-E_n^{'})\Theta(E_n^{'}-E_n-E_j)$ and integrating over the bin $[E_j,E_i]$. 
However, for matrix elements that are independent of $s$, it is possible to derive analytic expressions for the differential rate. In order to do so, we use the definition of the scattering rate in ref.~\cite{Reddy:1997yr,Bertoni:2013bsa} 
\begin{eqnarray}
\Gamma^- = 2\int \frac{d^3k^{'}}{(2\pi)^3}\int \frac{d^3p}{(2\pi)^3}\int \frac{d^3p^{'}}{(2\pi)^3} \frac{|\overline{M}|^2}{(2E_\chi)(2E^{'}_\chi)(2E_n)(2E^{'}_n)}(2\pi)^4\delta^4\left(k_\mu+p_\mu-k_\mu^{'}-p_\mu^{'}\right) \nonumber \\* \times\fFD(E_n)(1-\fFD(E^{'}_n)),
\label{eq:scattrate}
\end{eqnarray}
where $|\overline{M}|^2$ is the squared matrix element,
$k^\mu=(E_\chi,\vec{k})$ and $k^{'\mu}=(E^{'}_\chi,\vec{k^{'}})$ are the DM initial and final momenta, and $p^\mu=(E_n,\vec{p})$ and $p^{'\mu}=(E^{'}_n,\vec{p^{'}})$ are the target particle initial and final momenta, respectively.
Rearranging  terms, and multiplying and dividing by
$v_{rel}=|\vec{w}-\vec{u}_n|$,  it can be easily demonstrated that $\Gamma^-$ is equivalent to $\Omega^-$,  Eq.~\ref{eq:omegaminusfinal}.

The advantage of Eq.~\ref{eq:omegaminusfinal} is that it can be used to calculate the capture rate for any operator. The disadvantage is that this computation has to be evaluated numerically, which can be computationally intensive. For this reason, shall now use Eq.~\ref{eq:scattrate} to derive analytic expressions that will allow us to speed up computations and, in addition, to calculate the shape of the interaction rate as a function of the energy loss. The limitation of this approach is that our analytic expressions are applicable only when the squared matrix element is independent of the center of mass energy $s$, i.e., when $|\overline{M}|^2$ is either constant or depends on the transferred momentum $t$.

The interaction rate $\Gamma^{-}$, for $d\sigma \propto t^n$, is
\begin{eqnarray}
\Gamma^{-}(E_\chi) &\propto& \frac{1}{2^7\pi^3E_\chi k }\int_0^{E_\chi-m_\chi}q_0 dq_0 \int \frac{t_E^n dt_E }{\sqrt{q_0^2+t_E}}  \left[1-g\left(\frac{E_n^{\,t^{-}}-\mu_{F,n}}{q_0}\right)\right],
\label{eq:gammafinaltext}
\end{eqnarray}
for elastic scattering with $t_E=-t=q^2-q_0^2$, where $q_0=E_n^{'}-E_n$ is the DM energy loss, 
\begin{equation}
E_n^{\, t^{-}} = -\left(m_n+\frac{q_0}{2}\right) + \sqrt{\left(m_n+\frac{q_0}{2}\right)^2+\left(\frac{\sqrt{q^2-q_0^2}}{2}-\frac{m_n q_0}{\sqrt{q^2-q_0^2}}\right)^2}, 
\end{equation}
is the minimum energy of the neutron before the collision, obtained from kinematics,  and $g(x)$ is a step function with a smooth transition, 
\begin{align}
g(x) =  \begin{cases}
\, 1 \quad &x>0, \\
\, 1+x \quad &-1<x<0,\\
\, 0 \quad &x<-1.
\end{cases}
\end{align}
The integral over $t_E$ can be solved analytically; the integration intervals and the exact expressions can be found in Appendix~\ref{sec:intrate}. 
Our result for $\Gamma^-$ is an extension of that presented in ref.~\citep{Bertoni:2013bsa}, where the interaction rate was calculated only in the case of low energy and a constant matrix element. It is valid at all energy ranges  and for all matrix elements that are independent of $s$.
The differential interaction rate $\frac{d\Gamma}{d q_0}(E_\chi,q_0)$ is then just the integrand of Eq.~\ref{eq:gammafinaltext}.
We will use $\frac{d\Gamma}{d q_0}$ to obtain normalised shapes for the differential interaction spectrum, while we will use $\Omega^-$ when we  need the total interaction rate.

%%%%%%%%%%%%%%%%%%%%%%%%%%%%%%%%%%%%%%%%%%%
\begin{table}[t]
\centering
{\renewcommand{\arraystretch}{1.3}
\begin{tabular}{ | c | c | c | c |}
  \hline                        
  Name & Operator & Coupling & $|\overline{M}|^2(s,t)$   \\   \hline
  D1 & $\bar\chi  \chi\;\bar q  q $ & ${y_q}/{\Lambda^2}$ & $\frac{c_N^S}{\Lambda^4} \frac{\left(4 m_{\chi }^2-t\right) \left(4 m_{\chi }^2-\mu ^2
   t\right)}{\mu ^2}$ \\  \hline
  D2 & $\bar\chi \gamma^5 \chi\;\bar q q $ & $i{y_q}/{\Lambda^2}$ & $\frac{c_N^S}{\Lambda^4} \frac{t \left(\mu ^2 t-4 m_{\chi }^2\right)}{\mu ^2}$ \\  \hline
  D3 & $\bar\chi \chi\;\bar q \gamma^5  q $&  $i{y_q}/{\Lambda^2}$ &  $\frac{c_N^P }{\Lambda^4} t \left(t-4 m_{\chi }^2\right)$ \\  \hline
  D4 & $\bar\chi \gamma^5 \chi\; \bar q \gamma^5 q $ & ${y_q}/{\Lambda^2}$  & $\frac{c_N^P}{\Lambda^4} t^2$ \\  \hline
  D5 & $\bar \chi \gamma_\mu \chi\; \bar q \gamma^\mu q$ & ${1}/{\Lambda^2}$ &  $2\frac{c_N^V}{\Lambda^4} \frac{2 \left(\mu ^2+1\right)^2 m_{\chi }^4-4 \left(\mu ^2+1\right) \mu ^2 s m_{\chi }^2+\mu ^4 \left(2 s^2+2 s t+t^2\right)}{\mu^4}$ \\  \hline
  D6 & $\bar\chi \gamma_\mu \gamma^5 \chi\; \bar  q \gamma^\mu q $ & ${1}/{\Lambda^2}$ &  $2\frac{c_N^V}{\Lambda^4} \frac{2 \left(\mu ^2-1\right)^2 m_{\chi }^4-4 \mu ^2 m_{\chi }^2 \left(\mu ^2 s+s+\mu ^2 t\right)+\mu ^4 \left(2 s^2+2 s
   t+t^2\right)}{\mu^4}$  \\  \hline
  D7 & $\bar \chi \gamma_\mu  \chi\; \bar q \gamma^\mu\gamma^5  q$ & ${1}/{\Lambda^2}$ &  $2\frac{c_N^A}{\Lambda^4} \frac{2 \left(\mu ^2-1\right)^2 m_{\chi }^4-4 \mu ^2 m_{\chi }^2 \left(\mu ^2 s+s+t\right)+\mu ^4 \left(2 s^2+2 s t+t^2\right)}{\mu^4}$ \\  \hline
  D8 & $\bar \chi \gamma_\mu \gamma^5 \chi\; \bar q \gamma^\mu \gamma^5 q $ & ${1}/{\Lambda^2}$ &  $2\frac{c_N^A}{\Lambda^4} \frac{2 \left(\mu ^4+10 \mu ^2+1\right) m_{\chi }^4-4 \left(\mu ^2+1\right) \mu ^2
   m_{\chi }^2 (s+t)+\mu ^4 \left(2 s^2+2 s t+t^2\right)}{\mu ^4}$ \\  \hline
  D9 & $\bar \chi \sigma_{\mu\nu} \chi\; \bar q \sigma^{\mu\nu} q $ & ${1}/{\Lambda^2}$ & $8\frac{c_N^T }{\Lambda^4} \frac{4 \left(\mu ^4+4 \mu ^2+1\right) m_{\chi }^4-2 \left(\mu ^2+1\right) \mu ^2 m_{\chi
   }^2 (4 s+t)+\mu ^4 (2 s+t)^2}{\mu ^4}$  \\  \hline
 D10 & $\bar \chi \sigma_{\mu\nu} \gamma^5\chi\; \bar q \sigma^{\mu\nu} q \;$ & ${i}/{\Lambda^2}$ &  $8\frac{c_N^T }{\Lambda^4} \frac{4 \left(\mu ^2-1\right)^2 m_{\chi }^4-2 \left(\mu ^2+1\right) \mu ^2 m_{\chi }^2 (4 s+t)+\mu ^4 (2 s+t)^2}{\mu^4}$ \\  \hline
\end{tabular}}
\caption{EFT dimension 6 operators and squared matrix elements for the scattering of Dirac DM from nuclei \cite{Goodman:2010ku}. The effective couplings for each operator are given as a function of the quark Yukawa coupling, $y_q$, and the cutoff scale, $\Lambda$. The fourth column shows the squared matrix elements at high energy as a function of the Mandelstam variables $s$ and $t$. The coefficients $c_N^S$, $c_N^P$, $c_N^V$ $c_N^A$ and $c_N^T$ are given in ref.~\cite{DelNobile:2013sia}.  
\label{tab:operatorshe}}
\end{table}
%%%%%%%%%%%%%%%%%%%%%%%%%%%%%%%%%%%%%%%%%%%

%%%%%%%%%%%%%%%%%%%%%%%%%%%%%%%%%%%%%%%
\begin{figure}[t]
    \centering
    \includegraphics[width=.46\textwidth]{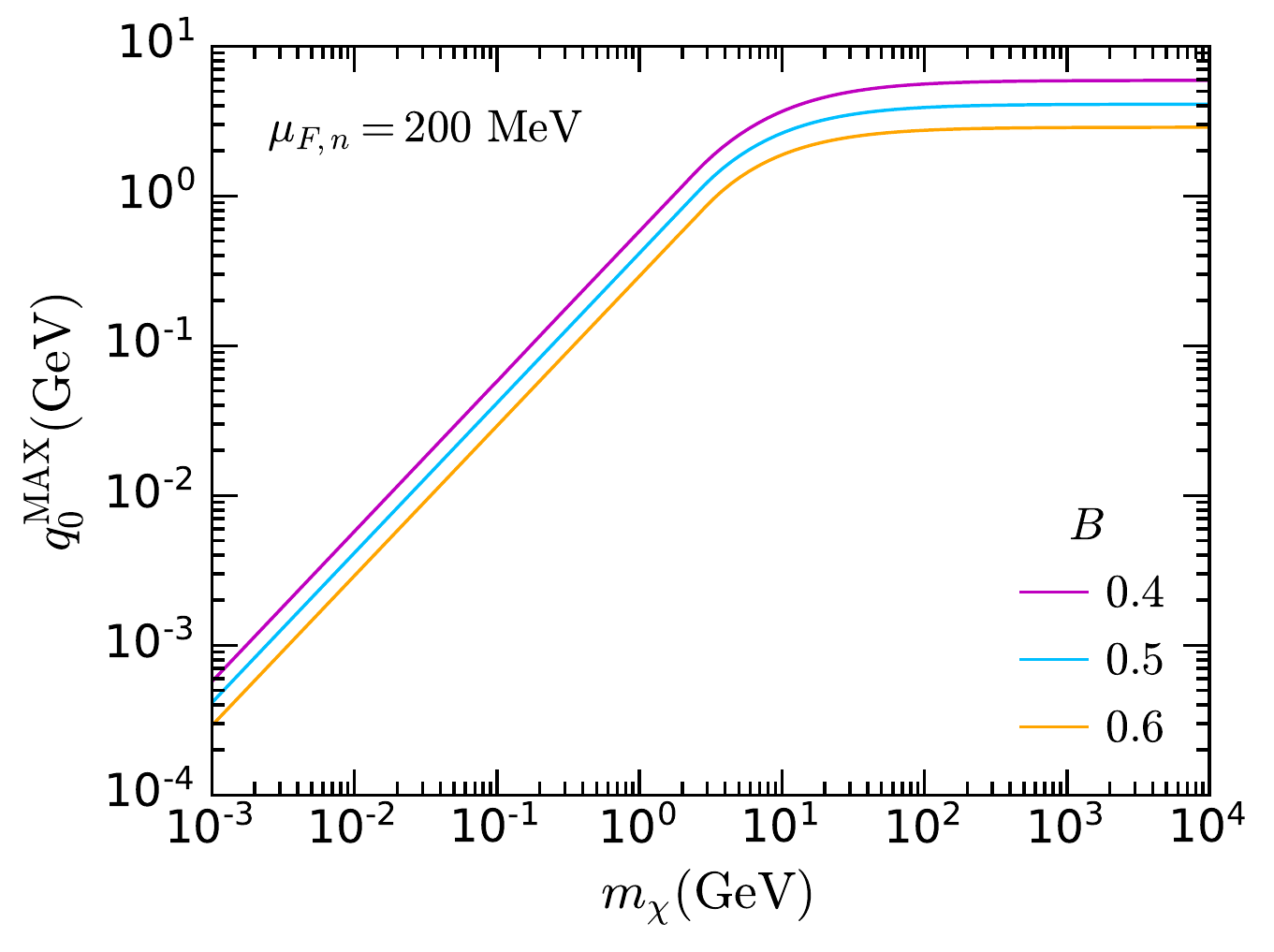}
\includegraphics[width=.45\textwidth]{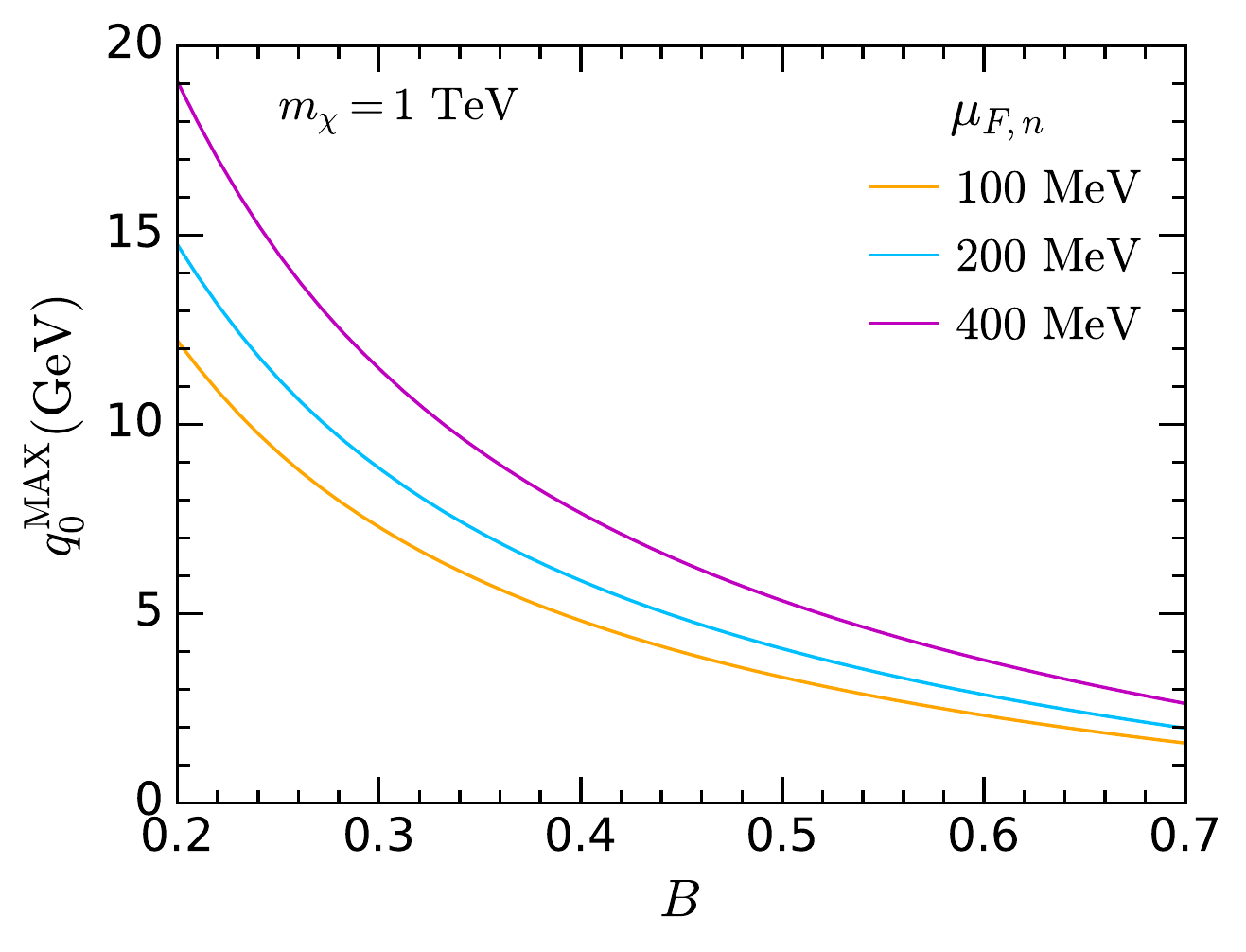}

    \caption{Left: $\qomax$ vs. $m_\chi$ for $\muFn=200\MeV$ and different values of B. 
    Right: $\qomax$ as a function of $B$ for different values of $\mu_{F,n}$ and $m_\chi=1\TeV$.}
    \label{fig:q0max}
\end{figure}
%%%%%%%%%%%%%%%%%%%%%%%%%%%%%%%%%%%%%%

%%%%%%%%%%%%%%%%%%%%%%%%%%%%%%%%%%%%%%
\begin{figure}[t]
    \centering
    \includegraphics[width=.85\textwidth]{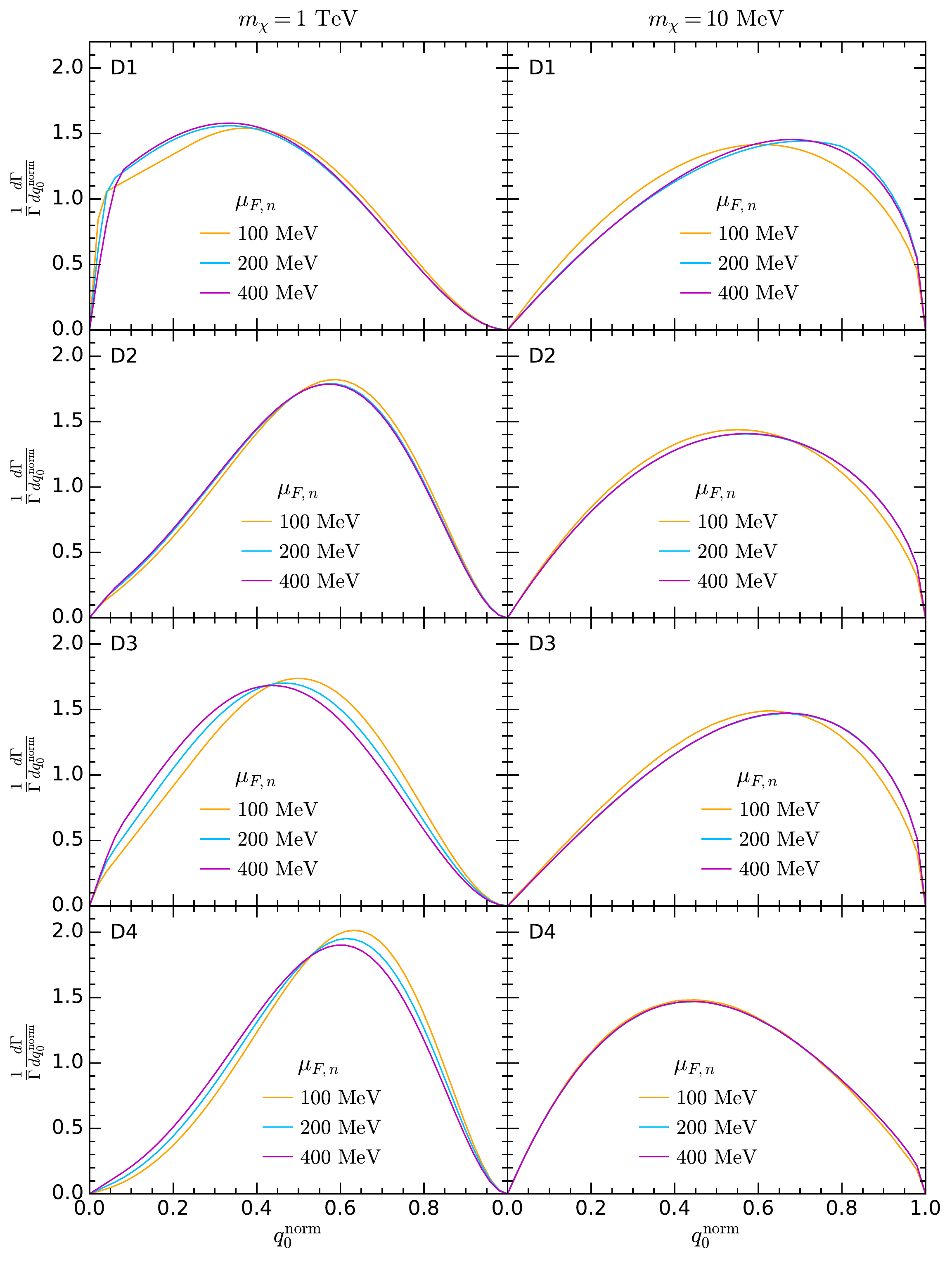}    
    \caption{Normalised differential interaction rates $\frac{1}{\Gamma}\frac{d\Gamma}{d\qonorm}$ as a function of $\qonorm$ for different values of $\muFn$, $m_\chi=1\TeV$ (left) and $m_\chi=10\MeV$ (right), $B=0.5$ and  operators D1 (first row), D2 (second row), D3 (third row) and D4 (fourth row). Profiles do not depend on $m_\chi$ in the limits $m_\chi\gg m_n$ (left) and $m_\chi\ll m_n$ (right).  }
    \label{fig:diffintratesd14}
\end{figure}
%%%%%%%%%%%%%%%%%%%%%%%%%%%%%%%%%%%%%%

Kinematics, and the phase space allowed by $g(x)$ in Eq.~\ref{eq:gammafinaltext}, determine the maximum energy that a DM particle can lose in one scattering interaction, $\qomax$ (see Appendix~\ref{sec:intrate}).
For DM capture, the value of $\qomax$ depends primarily on the DM mass, as is illustrated in the left panel of Fig.~\ref{fig:q0max}. We can see that for low $m_\chi$, $\qomax\propto m_\chi$, while, for $m_\chi\gg m_n$, $\qomax\sim 3 - 6 \GeV$. 
In the limits $m_\chi\gg m_n$ and $m_\chi\ll m_n$,  all normalised differential interaction rates $\frac{1}{\Gamma}\frac{d\Gamma}{d q_0}$ are independent of $m_\chi$. 
Both $\qomax$ and $\frac{d\Gamma}{dq_0}$ also depend on $\muFn$ and $B$. Changing $\muFn$ has a very mild effect on the value of $\qomax$ (see right panel of Fig.~\ref{fig:q0max}) and on the shape of the normalised spectrum (see Fig.~\ref{fig:diffintratesd14}). On the other hand, increasing $B$ has the main effect of reducing   $\qomax$ (see right panel of Fig.~\ref{fig:q0max}), but only a mild effect on the shape of the profile expressed as a function of the normalised energy loss 
\begin{equation}
    \qonorm = \frac{q_0}{\qomax}. 
\end{equation}

We can apply our results for $\frac{d\Gamma}{d q_0}$  to DM-neutron interactions  whose differential cross sections depend only on the transferred momentum $t=(k^\mu-k^{'\mu})^2$ and not on the centre of mass energy $s=(p^\mu+k^\mu)^2$. 
Of the lowest order EFT operators  listed in Table~\ref{tab:operatorshe}, only D1, D2, D3 and D4 satisfy this criterion.  
In the left hand panels of Fig.~\ref{fig:diffintratesd14} we show the normalised differential rates as a function of $\qonorm$ for the four operators D1-D4, in the limit $m_\chi\gg m_n$. We can observe that D1 has a softer spectrum, while  D2 and D4 spectra are peaked towards higher  values of $q_0$, around $\qonorm\sim0.6$. Varying the chemical potential $\muFn$ has a very mild effect, shifting the spectrum to lower values of $q_0$ with increasing values of $\muFn$.
Note that at small values of $\qonorm$ there is a sudden change in the slope of the normalised differential rate, which occurs for all operators but is more evident in D1 (top left panel). This is due to the zero temperature approximation, implicit in Eq.~\ref{eq:gammafinaltext}, where Heaviside functions were used to approximate Fermi Dirac distributions (see Appendix~\ref{sec:intrate}); using a finite temperature would produce a smoother spectrum at small $\qonorm$. 
In the right hand panels of Fig.~\ref{fig:diffintratesd14}, we explore the low DM mass region $m_\chi\ll m_n$. 
In this case, all operators give rise to similar profiles, the sole difference being that the peak of the profile is now shifted to lower  $\qonorm$ for D4 in contrast to D1, with intermediate values for D2 and D3. This is a consequence of Pauli blocking, as explicitly manifest in  Eq.~\ref{eq:Dn} for the appropriate integration interval. In fact, this effect depends on the specific power of $t$ that dominates the spectrum, with profiles with lower $n$ ($d\sigma\propto t^n$) peaked at higher $\qonorm$ (see Fig.~\ref{fig:diffgamma}, right panels). For D4 we have $|\overline{M}|\propto t^2$, while the matrix elements of D2 and D3 are linear combinations of $t$ and $t^2$ and that for D1 is a combination of $t^n$ with $n=0,1,2$. Comparing the right panels of Fig.~\ref{fig:diffintratesd14} with Fig.~\ref{fig:diffgamma}, we observe that the lowest power of $t$ determines the shape of the final differential interaction rate. Finally, varying $\muFn$ has a very mild effect, this time shifting the spectrum mostly to higher values of $q_0$ for higher $\muFn$.

%%%%%%%%%%%%%%%%%%%%%%%%%%%%%%%%%%%%%%%%%%%%%%%%%%%%%%%%%%%%%%%%%%%%%%%%%%%%%%%%%%%%
%%%%%%%%%%%%%%%%%%%%%%%%%%%%%%%%%%%%%%%%%%%%%%%%%%%%%%%%%%%%%%%%%%%%%%%%%%%%%%%%%%%%
\subsection{Pauli Blocking}
\label{sec:pauliblocking}
%%%%%%%%%%%%%%%%%%%%%%%%%%%%%%%%%%%%%%%%%%%%%%%%%%%%%%%%%%%%%%%%%%%%%%%%%%%%%%%%%%%%
%%%%%%%%%%%%%%%%%%%%%%%%%%%%%%%%%%%%%%%%%%%%%%%%%%%%%%%%%%%%%%%%%%%%%%%%%%%%%%%%%%%%

The DM interaction rate, Eq.~\ref{eq:scattrate}, is proportional to the number of target particles (nucleons/leptons) in the initial state with energy $E_n$, and to the number of free targets with final state energy $E_n+q_0$. 
In the $T\rightarrow 0$ approximation, all energy levels are either full or empty. Thus, in this limit, one necessarily has $\Gamma^-\rightarrow 0$ for $q_0\rightarrow 0$. 
This  is a consequence of Pauli blocking (PB), and is not exhibited in the classical rate calculation. 
It is worth noting that Pauli suppression only affects the differential rate $\frac{d\Gamma}{d q_0}$ when  $q_0\le \muFn$.

%%%%%%%%%%%%%%%%%%%%%%%%%%%%%%%%%%%%%%%%%
\begin{figure}[th]
    \centering
    \includegraphics[width=.4575\textwidth]{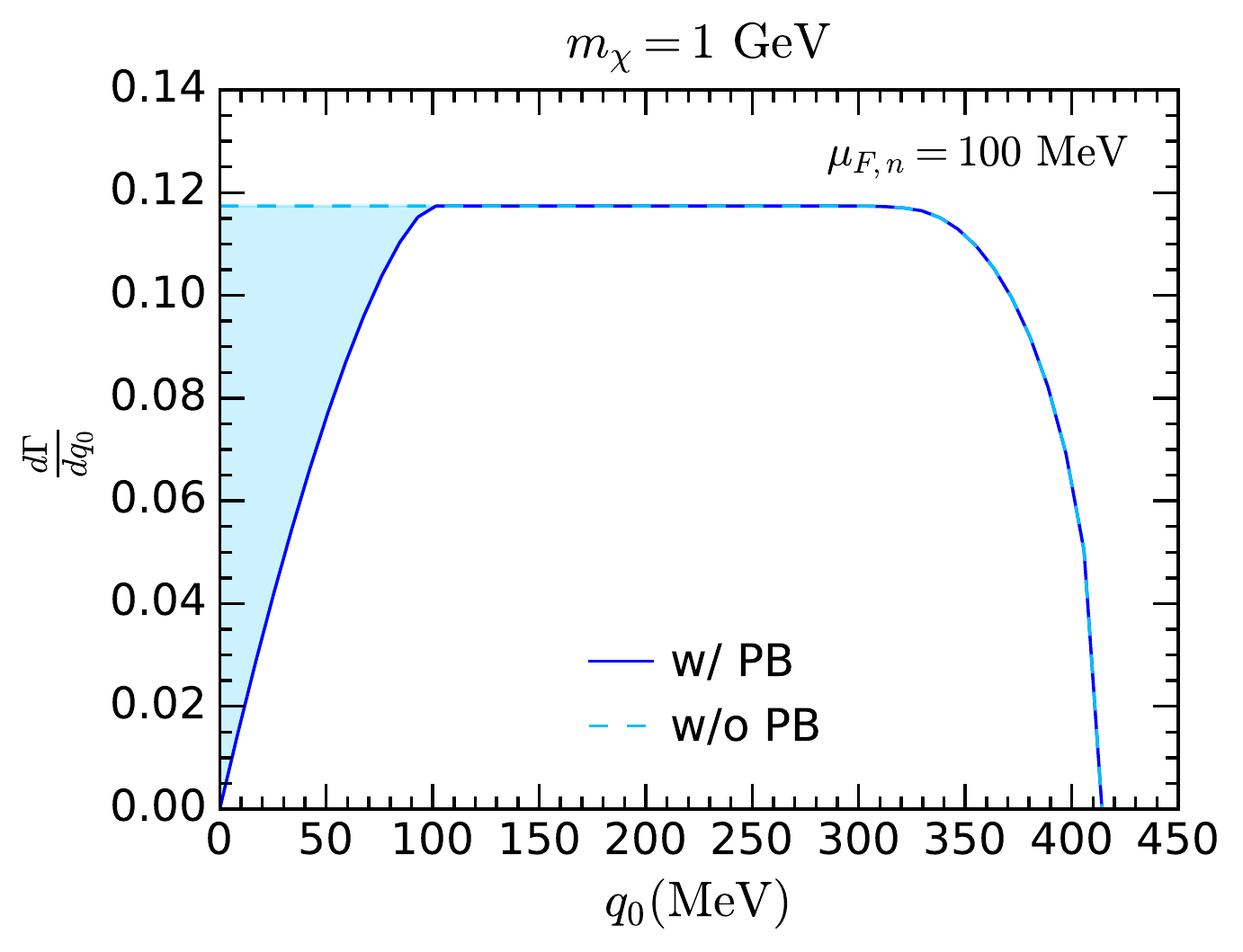}
    \includegraphics[width=.45\textwidth]{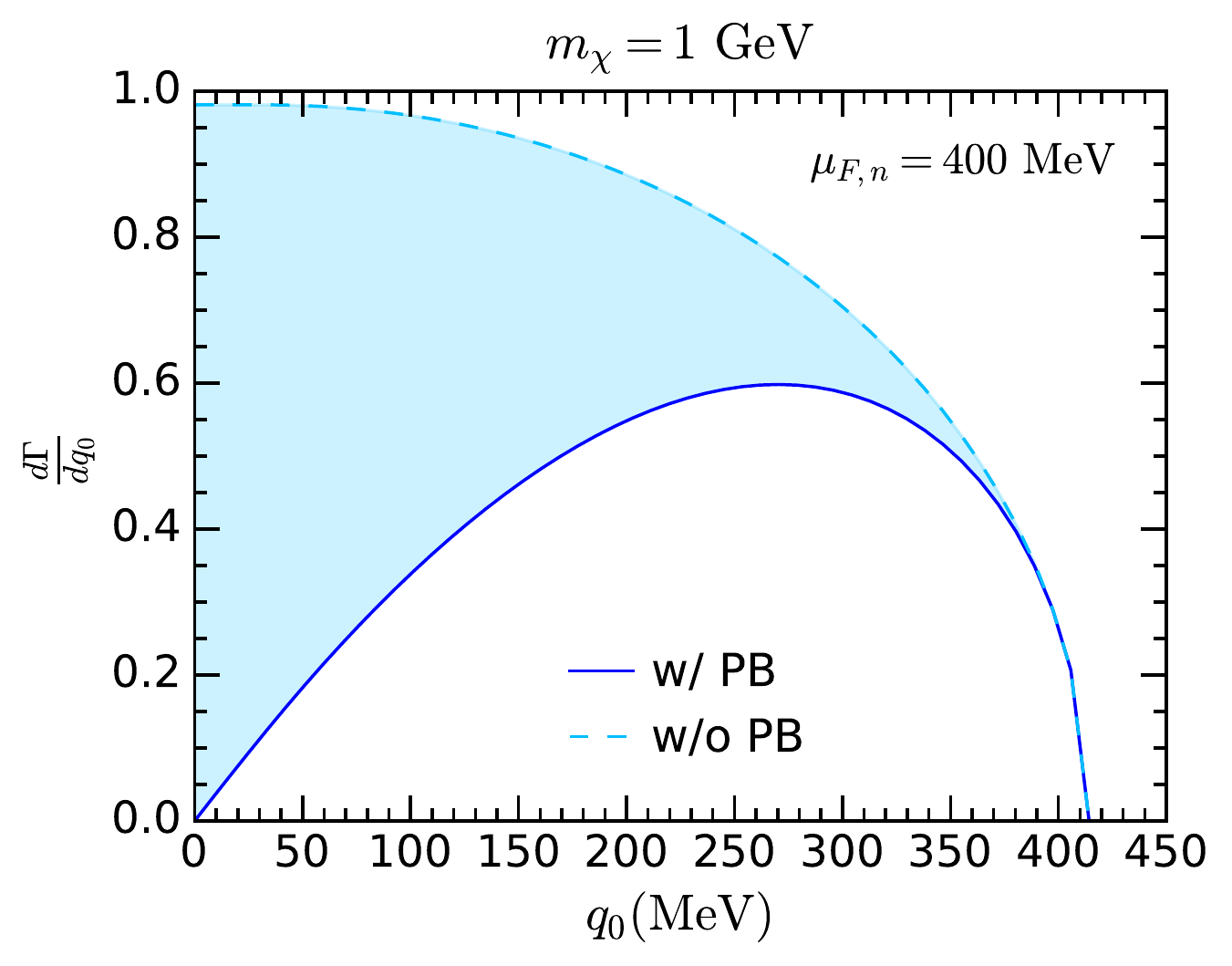}\\
    \includegraphics[width=.45\textwidth]{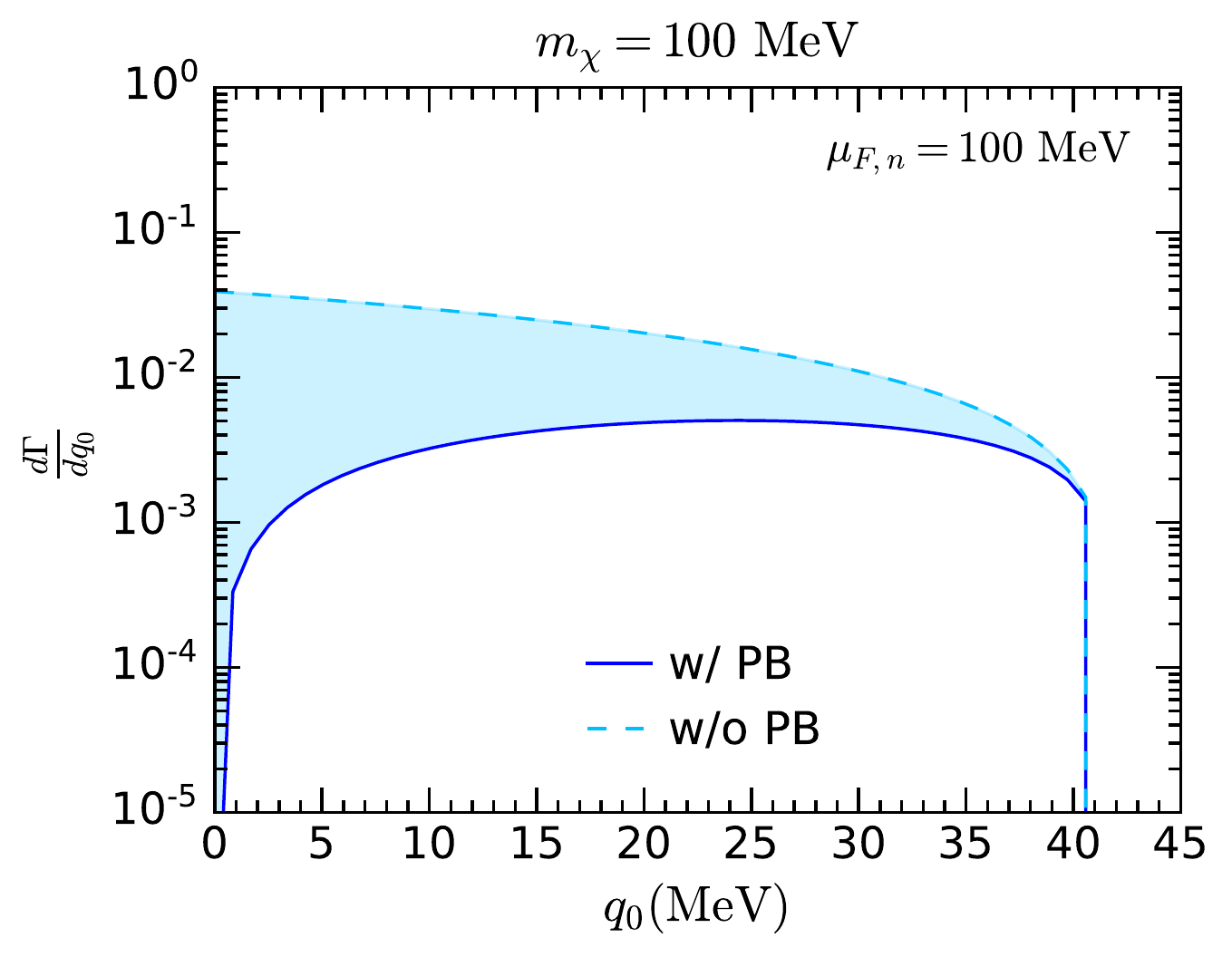}
    \includegraphics[width=.45\textwidth]{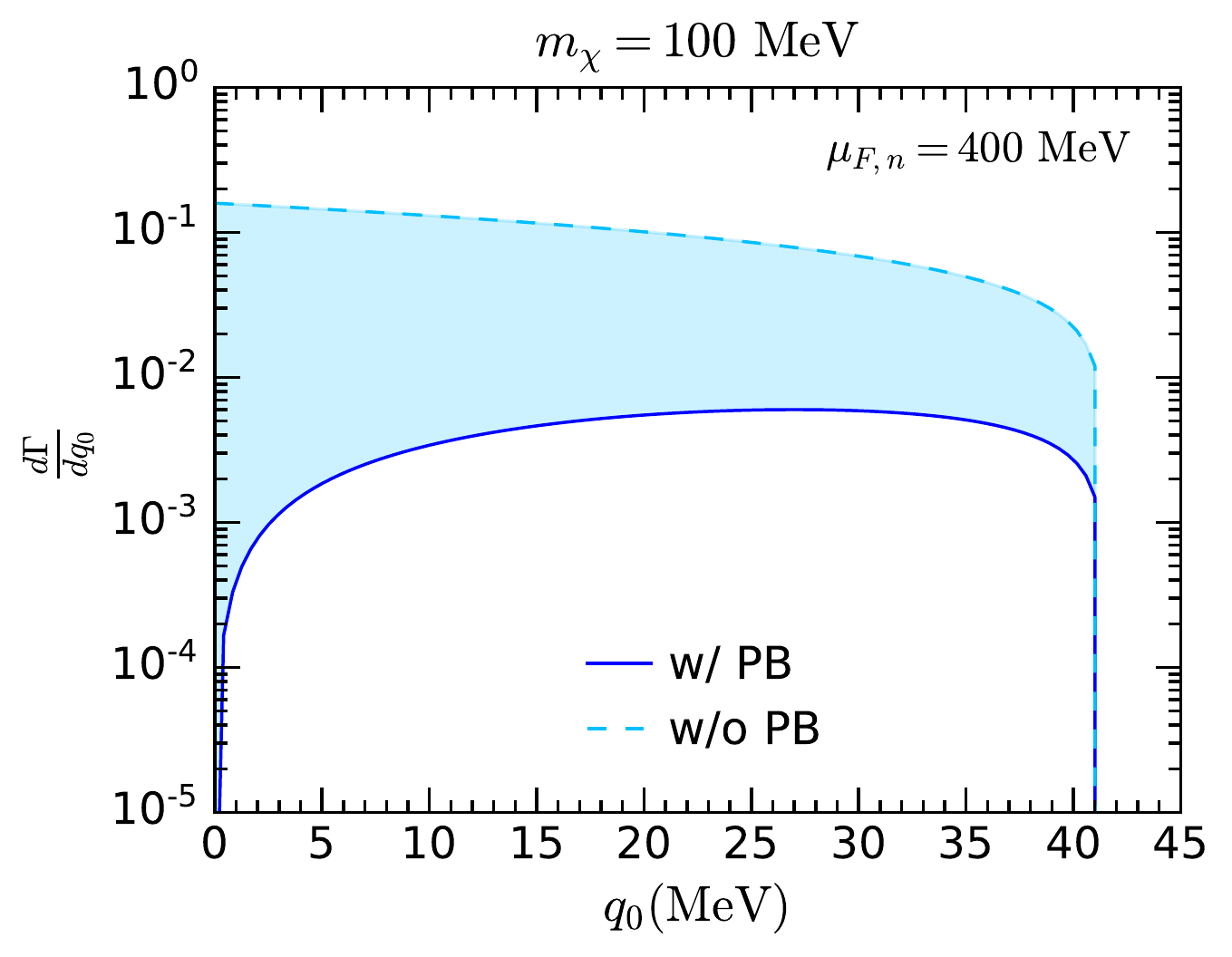}\\
    \includegraphics[width=.45\textwidth]{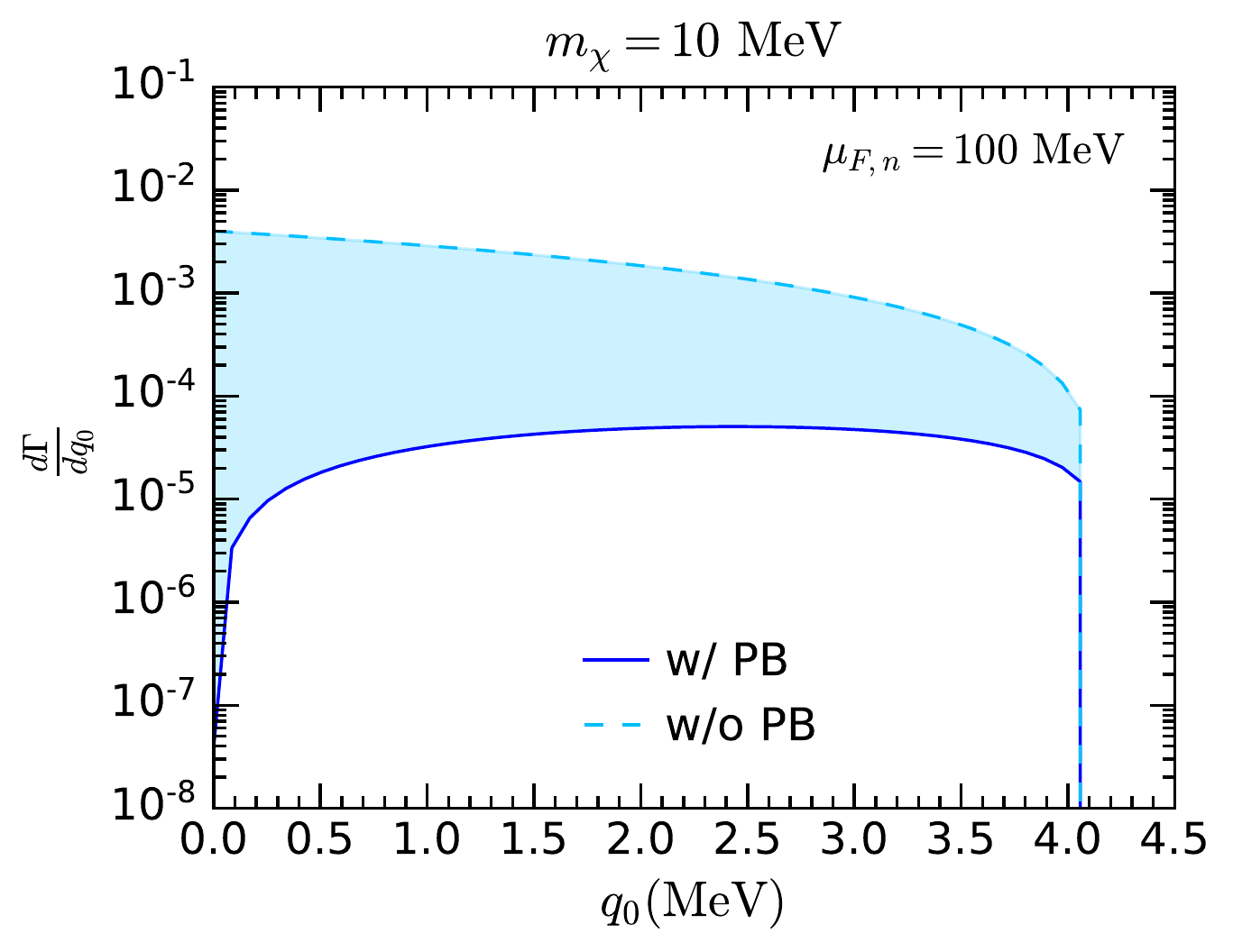}
    \includegraphics[width=.45\textwidth]{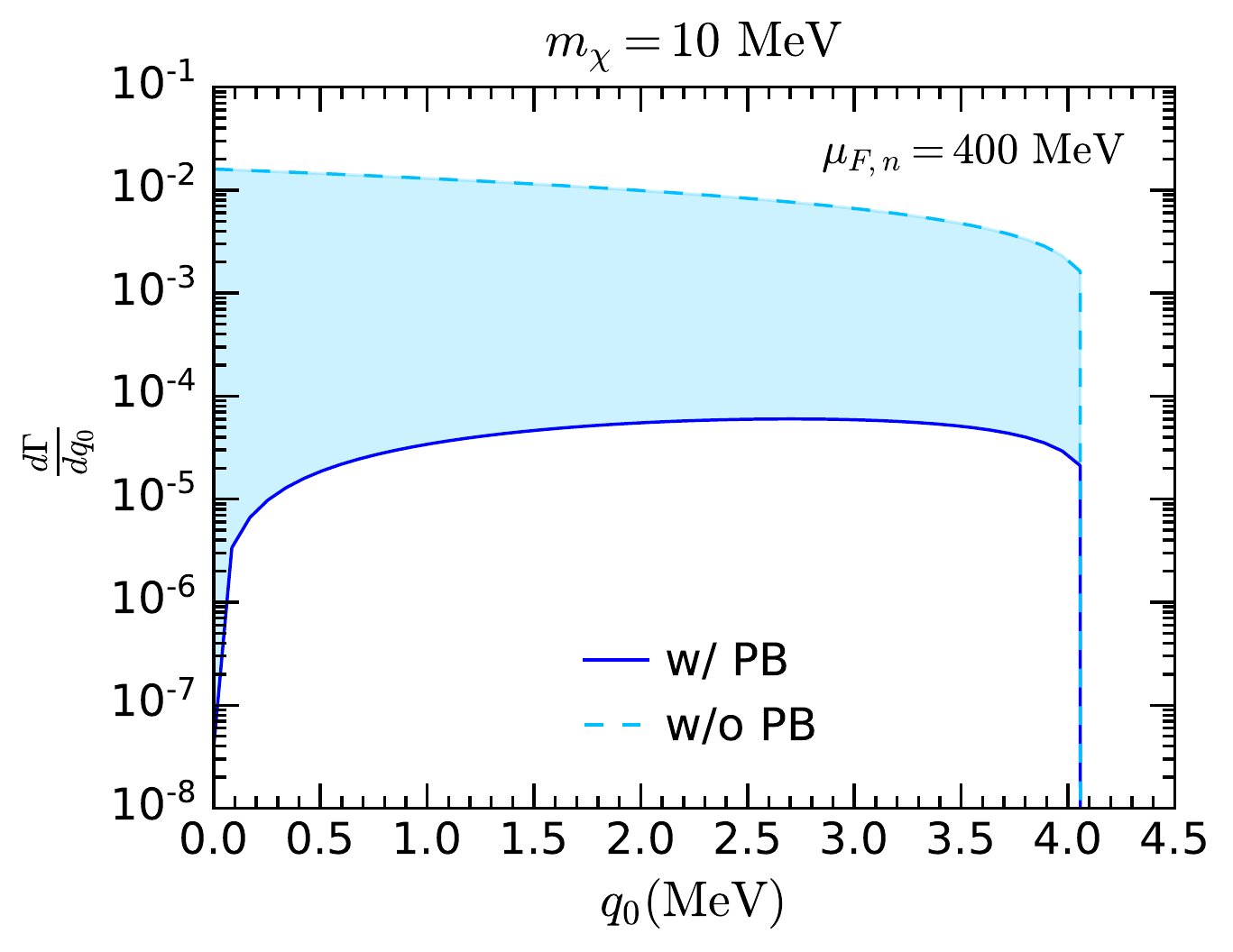}
    \caption{Differential interaction rates $\frac{d\Gamma}{dq_0}$  
    as a function of the energy loss $q_0$ for different values of $m_\chi$ and $\muFn$, constant cross section and $B=0.5$. Blue lines refer to the result that includes Pauli blocking, while the light blue dashed lines refer to the result without PB. Left column:  $\muFn=100\MeV$, right column: $\muFn=400\MeV$. Top: $m_\chi=1\GeV$, middle: $m_\chi=100\MeV$, bottom: $m_\chi=10\MeV$.  }
    \label{fig:gammaNPBmu}
\end{figure}
%%%%%%%%%%%%%%%%%%%%%%%%%%%%%%%%%%%%%%%%%

With the aim of assessing the impact of PB on the DM differential interaction rate, in Fig.~\ref{fig:gammaNPBmu} we compare the calculation 
with (blue solid lines) and without (light blue dashed lines) Pauli blocking, for $B=0.5$ and constant DM-neutron cross section. When Pauli blocking can be neglected, the interaction rate is obtained straightforwardly from Eq.~\ref{eq:scattrate} by stripping away the $(1 - \fFD(E^{'}_n))$ factor. 
The difference between both computations is shaded in light blue. In the top left panel, we can see how the differential rate changes by switching Pauli blocking on or off, for $\muFn=100\MeV$. Indeed, the suppression for $q_0<\muFn$ is evident. The rate calculated without PB is flat for $q_0\lesssim 200\MeV$, while when Pauli suppression is active it undergoes a smooth transition towards $0$ for $q_0<\muFn$. In the top right plot, we increase the neutron chemical potential from $\muFn=100\MeV$ to $\muFn=400\MeV$. Given that in this case $\qomax \sim 0.4 m_\chi \sim 400\MeV$, almost the whole energy range is affected by PB. The higher $\muFn$ changes the spectra (both with and without PB) such that the unsuppressed rate is no longer flat at low $q_0$. The PB suppressed rate reaches a maximum at values of $q_0$ slightly below $\qomax$, and then decreases towards $0$ at lower $q_0$.
In the middle panels, $m_\chi=100\MeV$, and $\qomax\sim40\MeV\ll\muFn$. In this case, it is evident that PB suppression affects the spectrum over the full $q_0=\qomax$ range. In the bottom row, we set $m_\chi=10\MeV$. As expected, for lighter DM,  PB effects are even more pronounced.

\begin{figure}
    \centering
    \includegraphics[width=.85\textwidth]{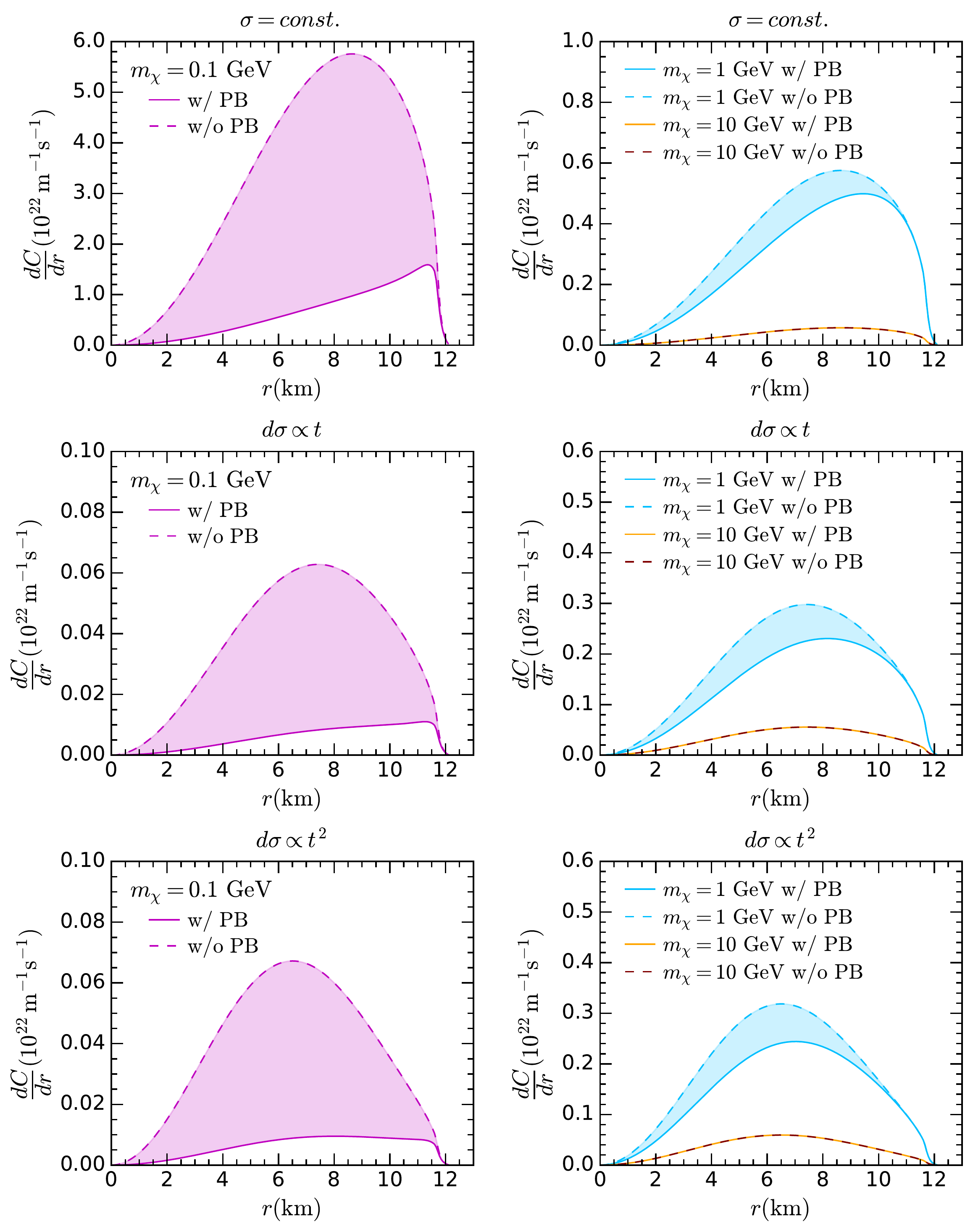}        
    \caption{Differential capture rate as a function of the NS radius $r$, with (solid) and without (dashed) Pauli blocking, for the EoS benchmark BSk24-2. Top: constant cross section, center: $d\sigma\propto t$, bottom: $d\sigma\propto t^2$.}
    \label{fig:diffcap}
\end{figure}

In Fig.~\ref{fig:diffcap} we plot the differential capture rate as a function of the NS radius, with and without Pauli blocking. We see that Pauli blocking is most significant at low DM mass, below about 1~GeV, and insignificant for higher masses. Pauli blocking has a larger impact on the differential capture rate in the deep NS interior, and a negligible effect on the surface. This is particularly apparent in the top left panel of Fig.~\ref{fig:diffcap}. It occurs because the chemical potential is higher in NS interior than it is near the crust, as seen in the radial $\mu_{F,n}$ profile in the bottom left panel of Fig.~\ref{fig:NSradprofs}.

\subsection{Low and intermediate DM mass range}
\label{sec:captureintermediate}

In sections~\ref{sec:caprate} and \ref{sec:intratenumeric}, we have derived a general expression to numerically  calculate the DM   capture and interaction rates,  
Eqs.~\ref{eq:captureclsimplrel} and \ref{eq:omegampaulitext} respectively.   
Using these expressions, we can write 
the complete expression for the capture rate as a function of the differential DM-neutron cross section 
\begin{eqnarray}
C = \frac{2\rho_\chi}{\pi \vstar m_\chi^2} {\rm Erf}\left(\sqrt{\frac{3}{2}}\frac{\vstar}{v_d}\right)\int_0^{\Rstar}  dr  \frac{r^2\zeta(r)}{\sqrt{B(r)}} \int dt dE_n ds \frac{d\sigma}{d\cos\theta_{cm}}\frac{E_n s}{\beta(s)\gamma(s)} \nonumber\\*
\times
\fFD(E_n,r)(1-\fFD(E^{'}_n,r)), \label{eq:capturefinal}
\end{eqnarray}
where the functions $\beta$ and $\gamma$ were given in section~\ref{sec:intratenumeric}. Recall that in the limit $T\rightarrow0$,  $\fFD(E_n,r)$ and $1-\fFD(E^{'}_n,r)$  can be taken to be theta functions,  $\Theta(\mu_{F,n}(r)-E_n)$ and  $\Theta(E^{'}_n-\mu_{F,n}(r))$, respectively.

The differential DM-neutron cross section can be written in terms of the squared matrix element as 
\begin{eqnarray}
 d\sigma &=& \frac{1}{2E_\chi 2E_n |\vec{w}-\vec{u}_n|} d^2\phi |\overline{M}|^2 \\ 
 \frac{d\sigma}{d\cos\theta_{cm}} &=&    \frac{1}{16\pi}\frac{\beta(s)}{2s\beta(s)-\gamma^2(s)} |\overline{M}|^2, \label{eq:difxsec}
\end{eqnarray}
 where $d^2\phi$ is the 2 body phase space, and we have rewritten all quantities in terms of $s$, $\beta$ and $\gamma$. 
Making these substitutions, we obtain an expression for $C$ in terms of $|\overline{M}|^2$, 
\begin{eqnarray}
C =  \frac{\rho_\chi}{8\pi^2 \vstar m_\chi^2} {\rm Erf}\left(\sqrt{\frac{3}{2}}\frac{\vstar}{v_d}\right)\int_0^{\Rstar}   dr  \frac{r^2 \zeta(r) }{\sqrt{B(r)}} \int dt dE_n ds \frac{|\overline{M}|^2 E_n}{2s\beta(s)-\gamma^2(s)} \frac{s}{\gamma(s)} \nonumber \\
\times \fFD(E_n,r)(1-\fFD(E^{'}_n,r)). \label{eq:capturefinalM2}
\end{eqnarray}
This expression can be used to numerically calculate the interaction and capture rates of DM in NSs. 
For nucleon targets and DM masses in the range $1\GeV\lesssim m_\chi\lesssim 10^6\GeV$, i.e. the intermediate mass range, Pauli blocking is negligible, as is the probability that more than one scattering interaction is required for capture. Then, for $|\overline{M}|^2=a t^n$, the previous expression can be simplified to (see Appendix~\ref{sec:capratesimple}) 
\begin{eqnarray}
C &\sim& \frac{4\pi}{\vstar} \frac{\rho_\chi}{m_\chi}  {\rm Erf}\left(\sqrt{\frac{3}{2}}\frac{\vstar}{v_d}\right)\int_0^{\Rstar}  r^2 dr \, n_n(r)  \frac{1-B(r)}{B(r)} \langle\sigma(r)\rangle = C_{approx}\label{eq:csimplelargemtext}, \\
\langle\sigma(r)\rangle &=&\left\langle\int dt \frac{d\sigma}{dt} \right\rangle_s =   \frac{a}{16\pi m_\chi^2 } \frac{\left(\frac{4(1-B(r))m_\chi^2}{B(r)(1+\mu^2)}\right)^{n}}{n+1}. \label{eq:xsecave}
\end{eqnarray}

\begin{figure}
    \centering
    \includegraphics[width=.45\textwidth]{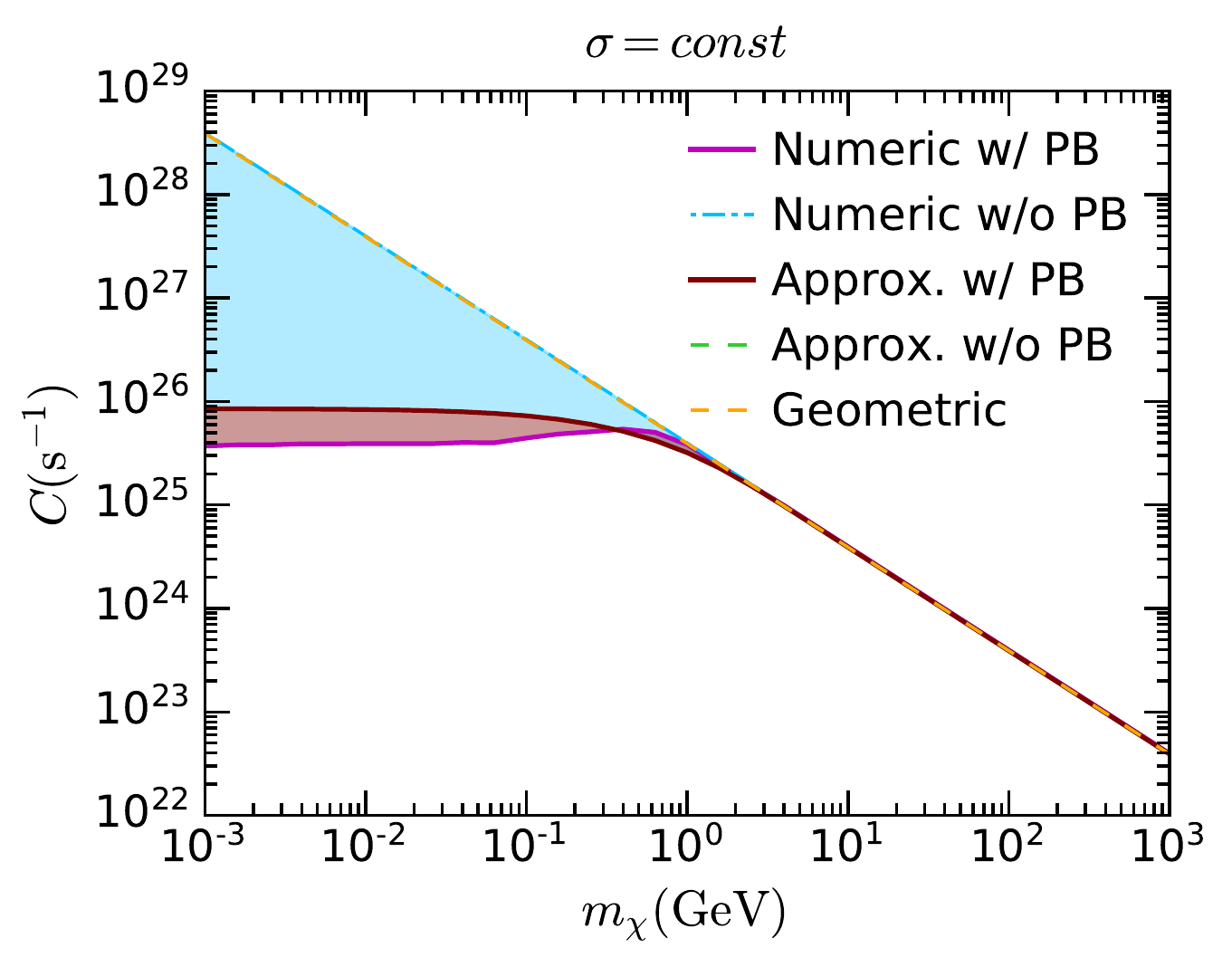}
    \includegraphics[width=.45\textwidth]{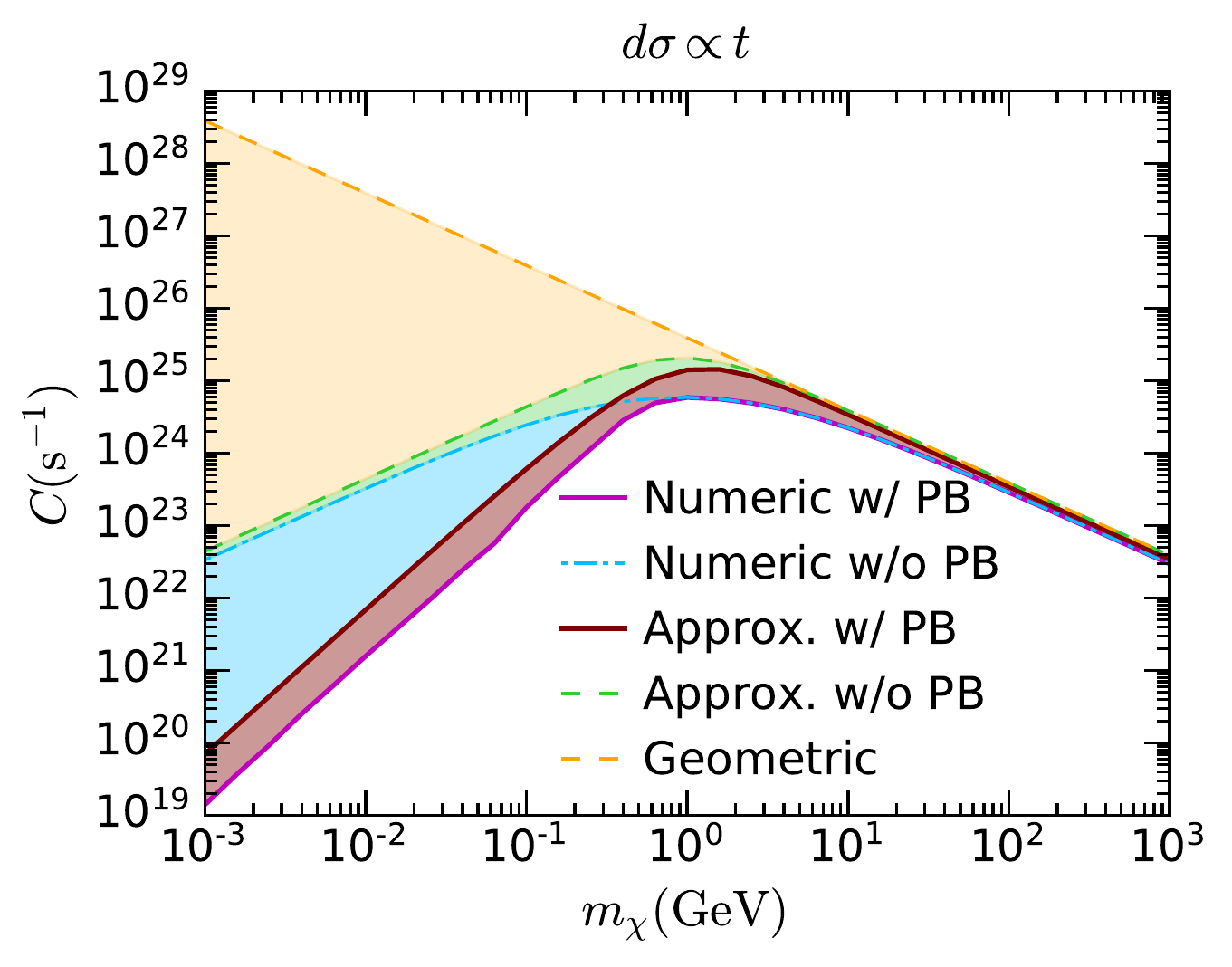}\\
    \includegraphics[width=.45\textwidth]{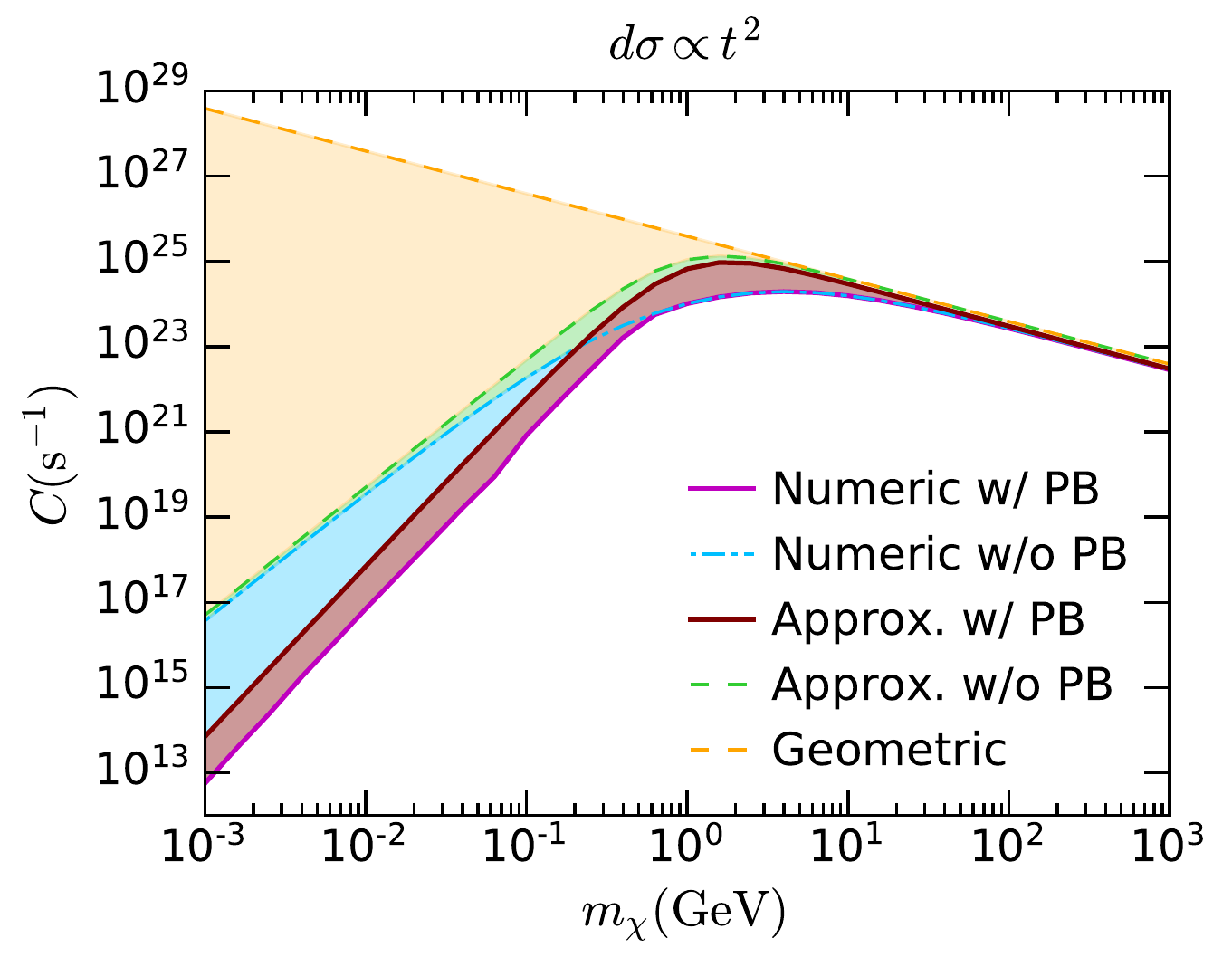}
    \caption{Capture rate as a function of the DM mass for $\sigma=\sigma_{ref}\sim 1.7 \times 10^{-45} \cm^2$ and EoS BSk24-2, calculated with and without Pauli suppression. Top left: constant cross section. Top right: $d\sigma\propto t$, bottom: $d\sigma\propto t^2$, where $t$ is the Mandelstam variable. NS opacity and multiple scattering effects are  neglected, and all rates are normalised to the geometric limit at large DM mass. }
    \label{fig:approxc}
\end{figure}

In Fig.~\ref{fig:approxc}, we show the capture rate as a function of the DM mass for matrix elements proportional to $t^n$ for $n=0,1,2$ and the NS benchmark model BSk24-2. Numerical results obtained using Eq.~\ref{eq:capturefinalM2} are shown in solid magenta; results using the same equation but removing the theta function that enforces Pauli blocking are depicted in light blue; and the approximation for intermediate DM masses, Eq.~\ref{eq:csimplelargemtext}, in green. We show the geometric limit, Eq.~\ref{eq:capturegeom}, in orange for comparison. 
The capture rates were all normalised to the geometric limit at large DM mass. 
No correction for multiple scattering is implemented in these results. 
In the same plots, we also show in brown the result obtained from using a modified version of Eq.~\ref{eq:csimplelargemtext}, where we add inside the integral the ratio between $\Gamma^-$ with and without Pauli blocking, which is obtained in section~\ref{sec:pauliblocking} for several values of $B$ and $\muFn$. 
From Fig.~\ref{fig:approxc}, we can see that Eq.~\ref{eq:csimplelargemtext} is a good approximation to the numerical results obtained without Pauli blocking, and can be safely used for DM masses from a few $\GeV$ up to $m_\chi\sim10^6\GeV$, where multiple scattering becomes relevant. 
On the other hand, for $m_\chi\lesssim 100\MeV$ the brown line is no longer a good approximation to the numerical result with Pauli blocking, Eq.~\ref{eq:capturefinalM2}, as it always overestimates the capture rate by nearly an order of magnitude.  

\begin{figure}
    \centering
    \includegraphics[width=.495\textwidth]{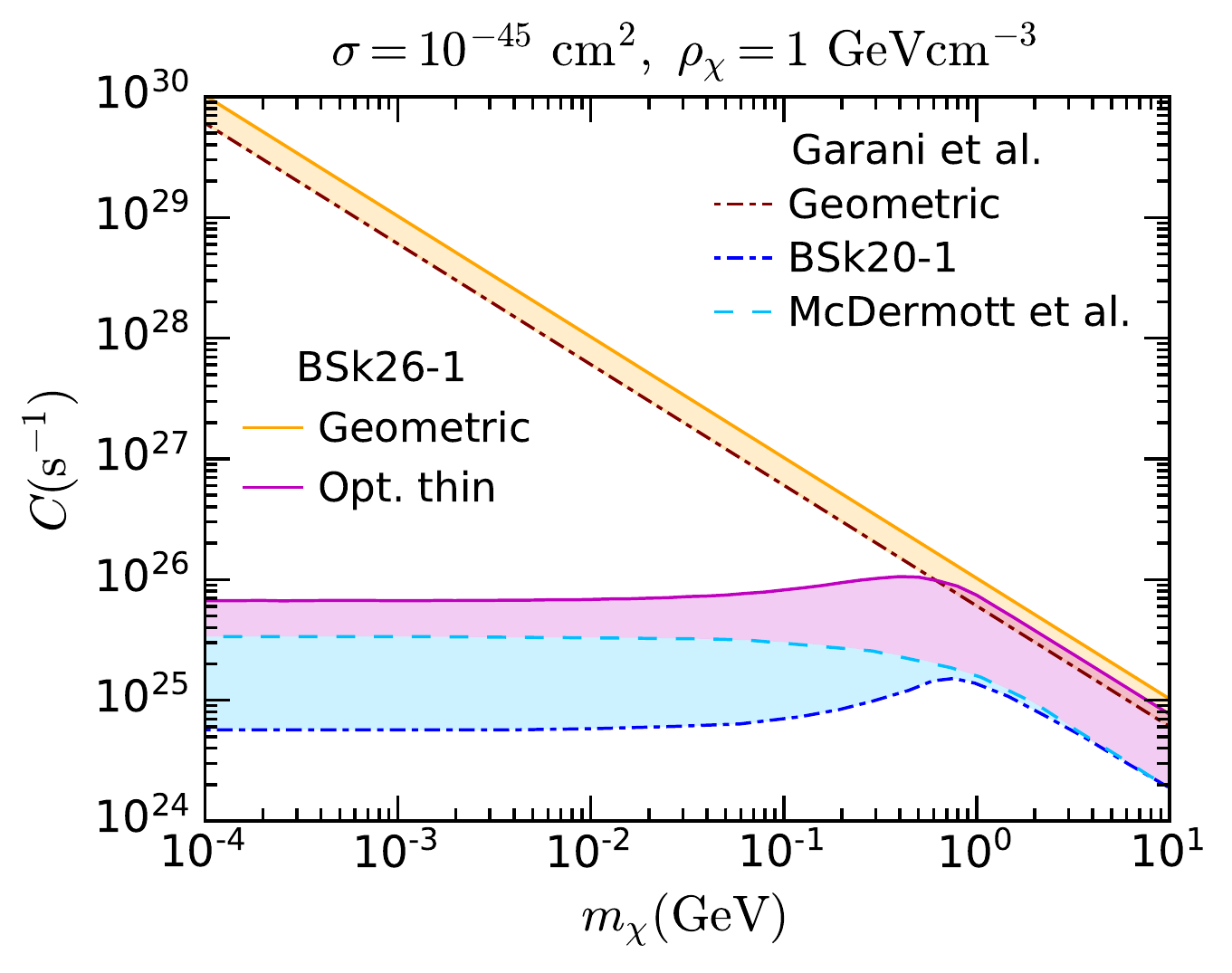}
    \includegraphics[width=.495\textwidth]{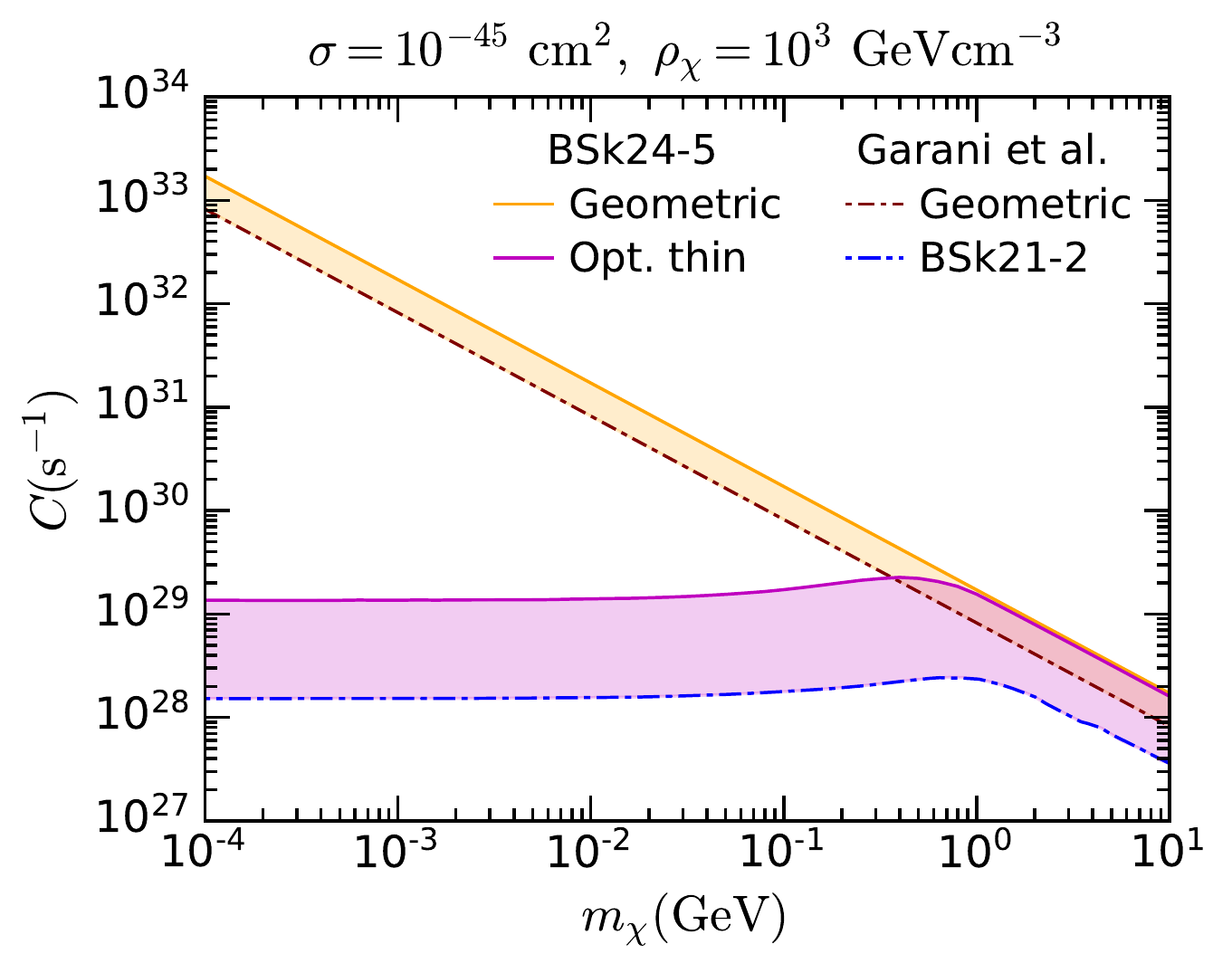}    
    \caption{Left: Capture rate in the optically thin  (magenta) and geometric (orange) limits as a function of the DM mass for constant cross section $\sigma=10^{-45}\cm^2$,  $\rho_\chi=1\GeV\cm^{-3}$ and BSk26 functional for $\Mstar\simeq1.52\Msun$ and $\Rstar\simeq11.6\km$ denoted as BSk26-1. Capture rate calculations from ref.~\cite{Garani:2018kkd} for a NS configuration with EoS BSk20-1~\cite{Potekhin:2013qqa} equivalent to BSk26-1, are shown for comparison. Right: Same as left but for $\rho_\chi=10^3\GeV\cm^{-3}$ and the benchmark model BSk24-5 equivalent to BSk21-2 in ref.~\cite{Garani:2018kkd}: $\Mstar\simeq2.11\Msun$ and  $\Rstar\simeq12.0\km$. 
    }
    \label{fig:Cratecomp}
\end{figure}

In Fig.~\ref{fig:Cratecomp}, we compare our full numerical capture rate calculation  Eq.~\ref{eq:capturefinalM2}, with that of ref.~\cite{Garani:2018kkd}, which includes the stellar structure and Pauli blocking but not general relativity (GR) corrections, for a constant cross section $\sigma=10^{-45}\cm^2$. To that end, we have selected NS configurations that match those of Figs.~1 and 14 of ref.~\cite{Garani:2018kkd},  namely Model A (BSk20-1):  $\Mstar\simeq1.52\Msun$, $\Rstar\simeq11.6\km$ and Model D (BSk21-2): $\Mstar\simeq2.11\Msun$ and $\Rstar\simeq12.0\km$. We denote this new benchmark model as BSk26-1 (left panel of Fig.~\ref{fig:Cratecomp}) and BSk24-5 (right panel). Note that we were not able to use the BSk20 and BSk21 functionals, since there are no public available fits for chemical potentials and particle abundances for those EoS families. However, as discussed earlier in section \ref{sec:NSmodels}, BSk26 (BSk24) yields configurations which are almost indistinguishable from those obtained with BSk20 (BSk21)~\cite{Perot:2019gwl}. It is worth noting that BSk26 is disfavoured by observations \cite{Pearson:2018tkr}.

We can see in the left panel of Fig.~\ref{fig:Cratecomp} that in the non Pauli suppressed region, $m_\chi \gtrsim 1\GeV$, our capture rate calculation in the optical thin limit (solid magenta) exceeds that of ref.~\cite{Garani:2018kkd} (dot-dashed blue) by a factor of $\sim 4$. When Pauli blocking is active, our capture rate calculation is about one order of magnitude  higher than the classical calculation.  Recall that ref.~\cite{Garani:2018kkd}  accounts  for neither gravitational focusing nor relativistic kinematics. 
We also show in dashed light blue the approximation given in ref.~\cite{McDermott:2011jp}, which accounts for Pauli blocking with a suppression factor that depends on the neutron Fermi momentum $\sim m_\chi v_{esc}/p_{F,n}$ for $m_\chi < m_n$. Though this approximation fails to reproduce the capture rate shape due to Pauli blocking in the DM mass range $[0.1\GeV,10\GeV]$, it underestimates the capture rate by only a factor of 2 when the DM mass is below 0.1 GeV. 
Finally, we  compare the geometric limit of Eq.~\ref{eq:capturegeom}  (solid orange) that incorporates GR effects~\cite{Bell:2018pkk} with the non-relativistic expression in ref.~\cite{Garani:2018kkd} (dot-dashed brown). We observe that the former is $\sim 67 \%$ greater than the latter, mostly due to the $1/B(\Rstar)$ GR correction \cite{Goldman:1989nd,Kouvaris:2007ay}. Similar conclusions are obtained when comparing capture rate calculations for Model D of ref.~\cite{Garani:2018kkd} (their Fig.~14) with our approach, as illustrated in the right panel of Fig.~\ref{fig:Cratecomp}.

\section{Multiple Scattering and Star Opacity}
\label{sec:largemassandsigma}

The capture rate expressions obtained in the previous section assume that the cross section is small enough  that the star is in the ``optically thin'' regime, and that DM  capture by multiple scattering is negligible. These assumptions break down if the DM-target cross section is of order the threshold cross section or larger, or if the DM mass exceeds $m_\chi \sim 10^6\GeV$, respectively. In this section, we explain how to modify our previous capture rate expressions to account for the NS optical depth and multiple scattering.  See ref.~\citep{Dasgupta:2019juq} for a recent treatment of multi-scatter capture in white dwarfs.

\subsection{Multiple Scattering}

For DM of mass $m_\chi\gtrsim 10^6\GeV$ scattering on nucleon targets, the capture probability is smaller than $1$ and becomes tiny as we increase the DM mass to large values. To account for this effect, we proceed in the following way. First, instead of setting the DM speed at infinity to $0$, as we did in the previous section, we now assume that the DM particles have a speed $u_\chi\ll 1$ that follows a Maxwell-Boltzmann (MB) distribution, Eq.~\ref{eq:DMveldist}. 
Then, the DM energy at infinity is 
\begin{equation}
    E^\infty_\chi \sim m_\chi \left(1+\frac{1}{2}u_\chi^2\right), 
\end{equation}
and at a distance $r$ from the star it becomes
\begin{equation}
    E_\chi(r) = \frac{m_\chi}{\sqrt{B(r)}} \left(1+\frac{1}{2}u_\chi^2\right).
    \label{eq:Echir}
\end{equation}
The energy that the DM particle should lose in order to be captured is therefore
\begin{equation}
    E_\chi^C(r) =  \frac{1}{2}u_\chi^2 \frac{m_\chi}{\sqrt{B(r)}}. 
\end{equation} 
The probability density function of the energy lost by a DM particle is 
\begin{eqnarray}
\xi(q_0,E_\chi,\muFn) &=& \frac{1}{\Gamma(E_\chi)}\frac{d\Gamma}{dq_0}(q_0,E_\chi,\muFn),\label{eq:normintrate}
\end{eqnarray}
where $\frac{d\Gamma}{dq_0}$ is the DM differential interaction rate, calculated in Appendix~\ref{sec:intrate}. 
The function $\xi$ is defined for any $q_0\ge 0$, however, due to kinematics, the function is non-zero only for $q_0\le\qomax$. $\xi$ depends on $B(r)$ through the ratio $E_\chi/m_\chi$, and for brevity we will simply write $\xi(q_0)$.

%%%%%%%%%%%%%%%%%%%%%%%%%%%%%%%%%%
\begin{figure}[t]
    \centering
    \includegraphics[width=.45\textwidth]{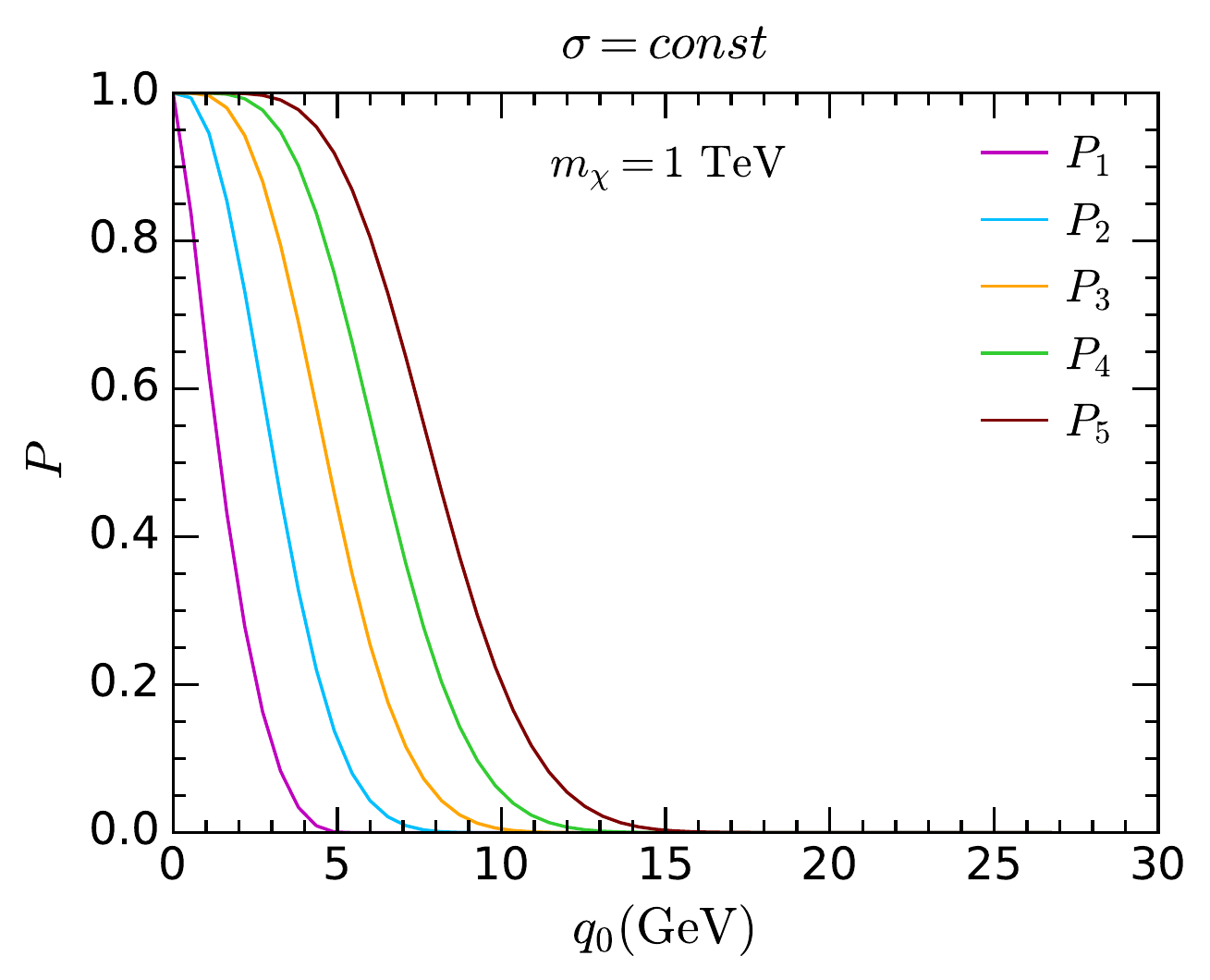}
    \includegraphics[width=.45\textwidth]{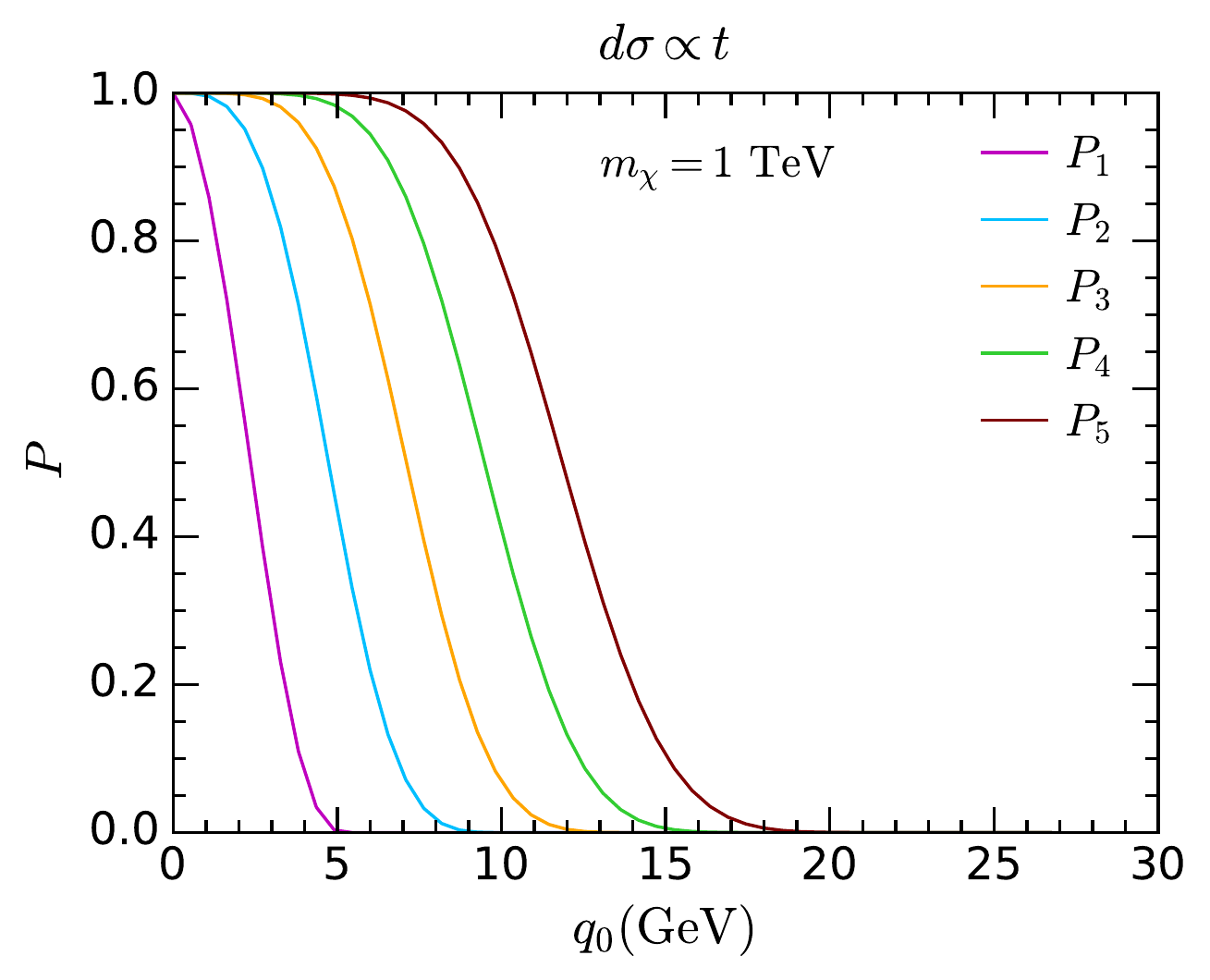}\\
    \includegraphics[width=.45\textwidth]{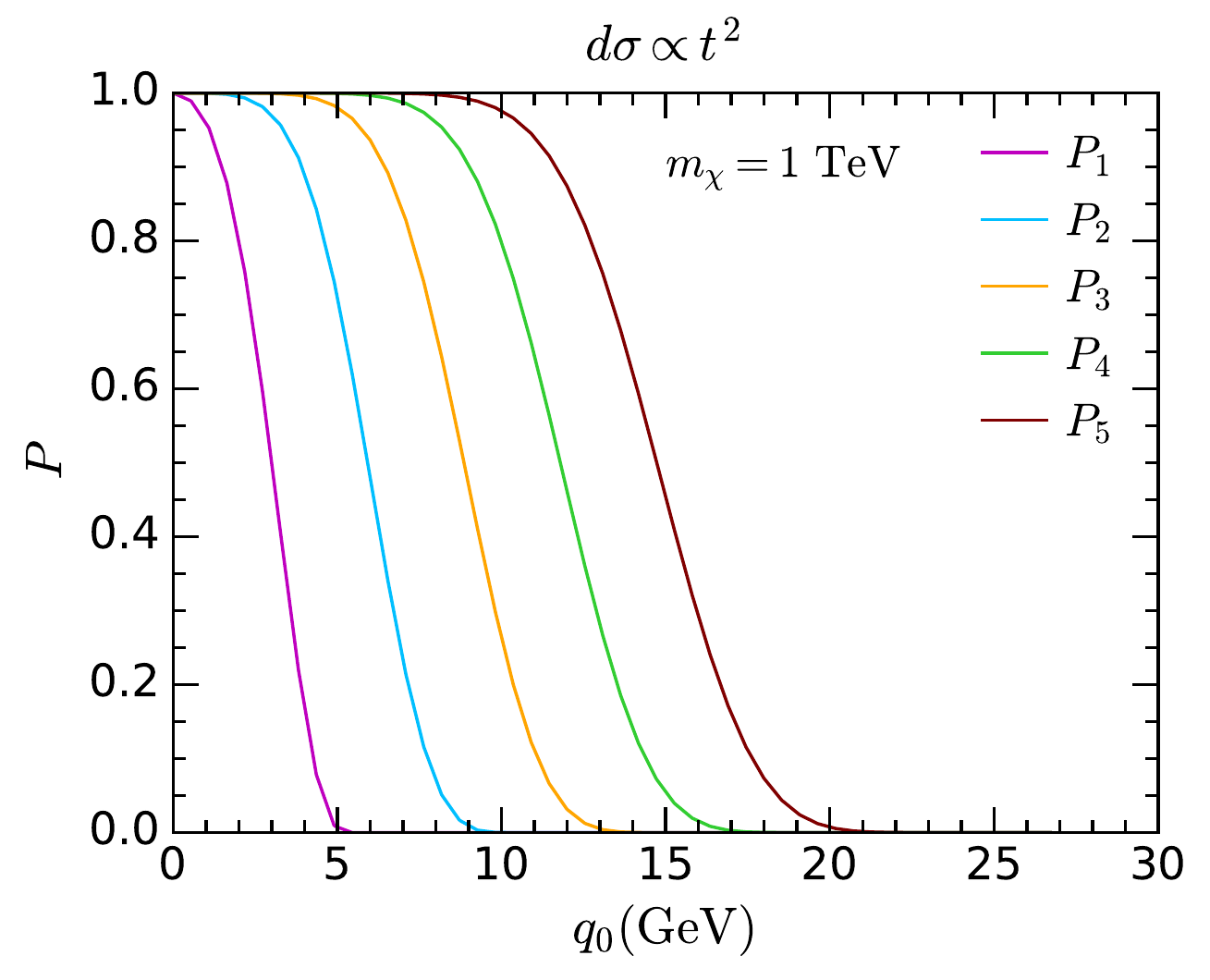}
    \caption{Probabilities to lose an energy $\delta q_0$  after $1,...,5$ scatterings,  $P_1,...,P_5$, as a function of the energy loss $q_0$,  assuming $B=0.5$ and $\muFn=400\MeV$. Results are shown for different dependence on the cross section on the Mandelstam variable $t$: constant DM-neutron cross section (top left), $d\sigma\propto t$ (top right) and $d\sigma\propto t^2$ (bottom). }
    \label{fig:pn}
\end{figure}
%%%%%%%%%%%%%%%%%%%%%%%%%%%%%%%%%%%

We can define the probability to lose an amount of energy of at least $\delta q_0$ in a single collision as
\begin{equation}
    P_1(\delta q_0) = \int_{\delta q_0}^\infty dx \xi(x).
\end{equation}
In the same way, the probability to lose at least the same amount of energy after 2 collisions is
\begin{eqnarray}
 P_2(\delta q_0) &=& P_1(\delta q_0) + \int_{\delta q_0}^\infty dy \int_0^y dx \xi(x)\xi(y-x) = P_1(\delta q_0) + \int_0^{\delta q_0} dz P_1(\delta q_0-z)\xi(z).
\end{eqnarray}
Thus, we obtain the following recursive relation for  $P_N$, 
\begin{eqnarray}
 P_{N+1}(\delta q_0) &=& P_N(\delta q_0) + \int_0^{\delta q_0} dz P_N(\delta q_0-z)\xi(z).\label{eq:pnrecurrent}
\end{eqnarray} 
Fig.~\ref{fig:pn} shows how the probability functions $P_1,...,P_5$ depend on the Mandelstam variable $t$ through the differential cross section. We show results for $\sigma=const.$ (top left), $d\sigma\propto t$ (top right) and $d\sigma\propto t^2$ (bottom), with the values $B=0.5$, $\muFn=400\MeV$.

%%%%%%%%%%%%%%%%%%%%%%%%%%%%%%
\begin{figure}
    \centering
    \includegraphics[width=.6\textwidth]{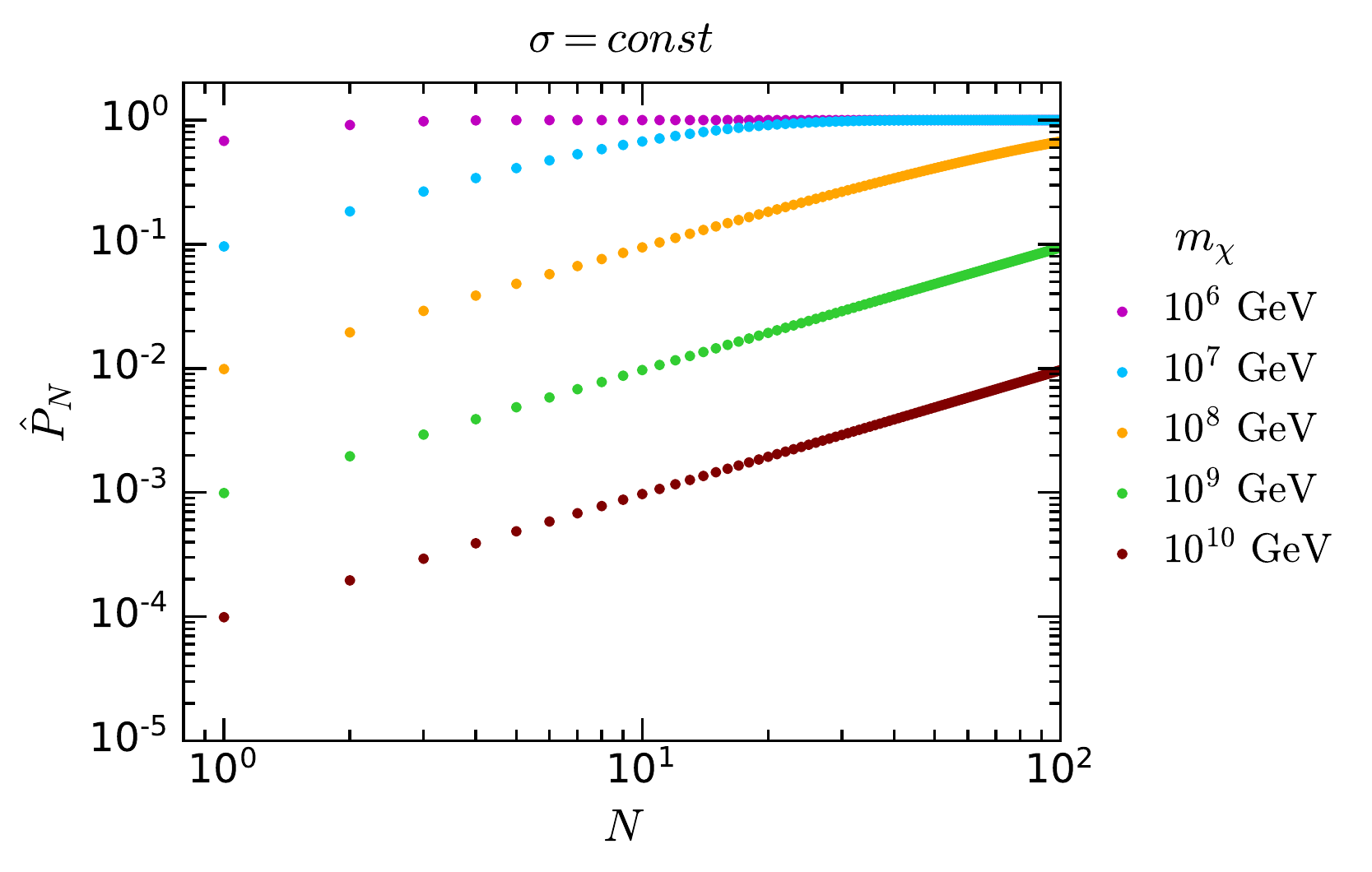}
    \caption{Cumulative probability $\hat{P}_N$ for $B=0.5$,  $\muFn=400\MeV$ for constant $|\overline{M}|^2$ as a function of the number of scatterings $N$ for several DM masses. }
    \label{fig:pnofn}
\end{figure}
%%%%%%%%%%%%%%%%%%%%%%%%%%%%%%

\begin{figure}
    \centering
    \includegraphics[width=.45\textwidth]{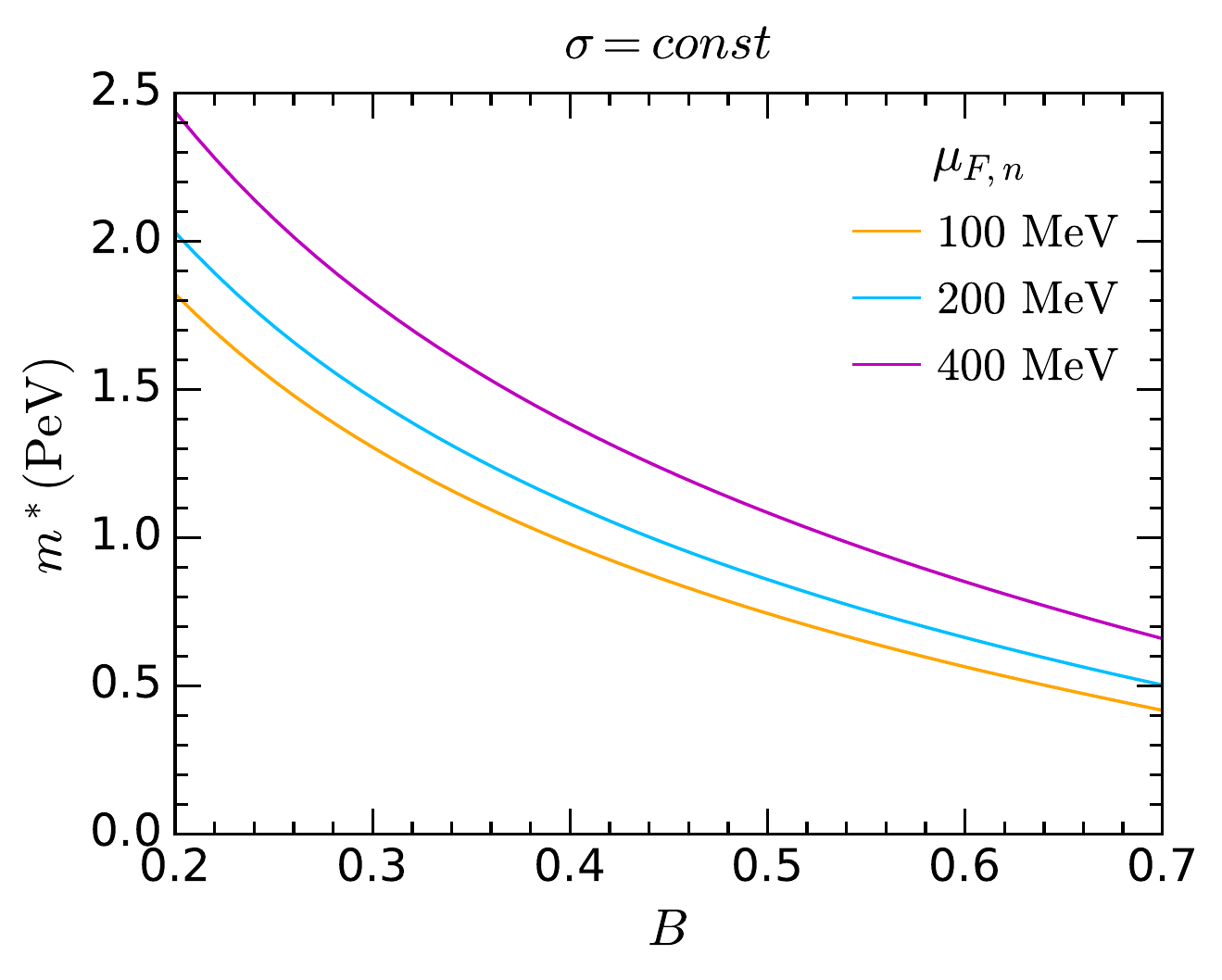}
    \includegraphics[width=.45\textwidth]{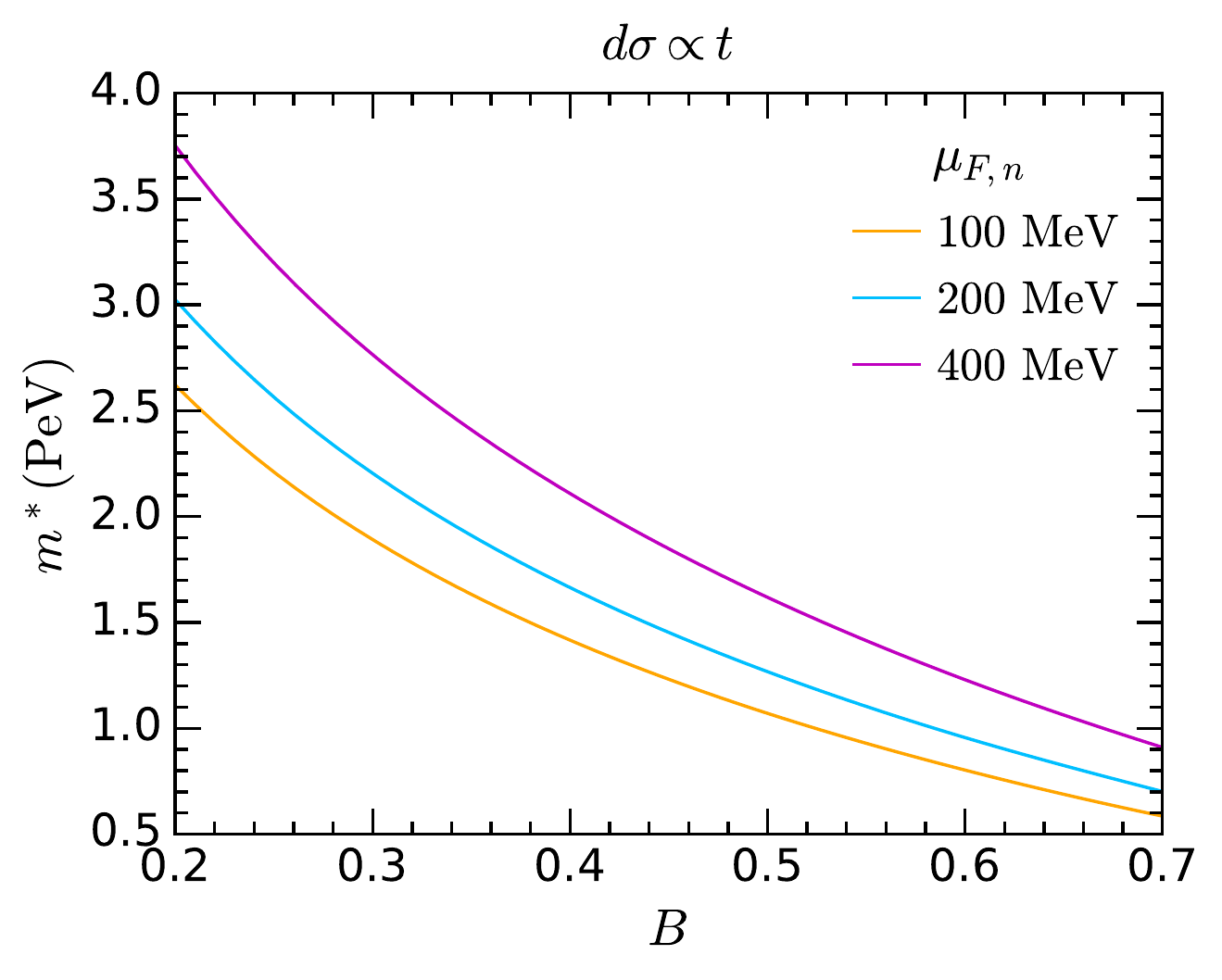}    
    \includegraphics[width=.45\textwidth]{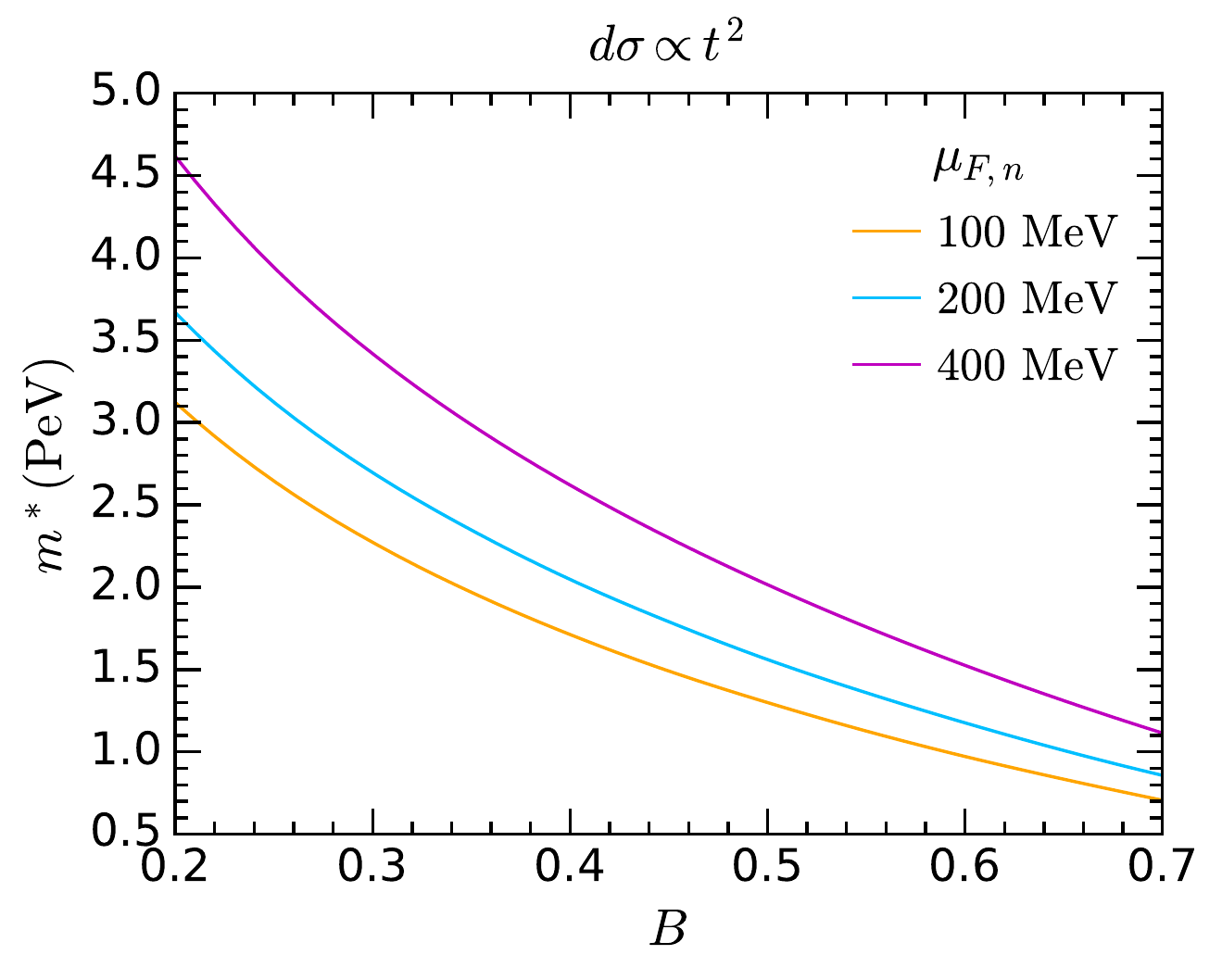}      
    \caption{Value of $m^*$ as a function of $B$ for different values of $\muFn$, $\sigma=const.$ (top left), $d\sigma\propto t$ (top right) and $d\sigma\propto t^2$ (bottom).}
    \label{fig:mstarbmu}
\end{figure}

We  define the probability of a DM particle to be captured after exactly $N$ scatterings by averaging over the MB energy distribution   
\begin{eqnarray}
c_N(r) &=& \frac{1}{\int_0^\infty \frac{\fMB(u_\chi)}{u_\chi}du_\chi}\int_0^\infty \frac{\fMB(u_\chi)}{u_\chi}du_\chi \left[P_N\left(\frac{1}{2}\frac{m_\chi u_\chi^2}{\sqrt{B(r)}}\right)-P_{N-1}\left(\frac{1}{2}\frac{m_\chi u_\chi^2}{\sqrt{B(r)}}\right)\right], 
\end{eqnarray}
where $c_N$ depends on $r$ through the dependence  of $P_1$  on $B(r)$. Note that although our results will assume a Maxwell-Boltzmann velocity distribution, it is straightforward to repeat the calculations with any other DM velocity distribution. The cumulative probability 
 $\hat{P}_N$ that a DM particle is captured after $N$ interactions with a total energy loss  $\delta q_0=E_\chi^C$ is 
\begin{eqnarray}
\hat{P}_N(r) = \sum_{i=1}^N c_i = \frac{1}{\int_0^\infty \frac{\fMB(u_\chi)}{u_\chi}du_\chi}\int_0^\infty \frac{\fMB(u_\chi)}{u_\chi}du_\chi P_N\left(\frac{1}{2}\frac{m_\chi}{\sqrt{B(r)}} u_\chi^2\right).
\end{eqnarray}
The resulting cumulative probability is shown as a function of the number of scatterings $N$ in Fig.~\ref{fig:pnofn}, for constant cross section and several DM masses. 
The cumulative probability $\hat{P}_N$ for the above values of $B,\muFn$ is well approximated by the function
\begin{eqnarray}
\hat{P}_N \sim  1-e^{-\frac{N m^*}{m_\chi}}.\label{eq:pncum}
\end{eqnarray}
In particular, for single scattering
\begin{equation}
c_1=\hat{P}_1 \sim  1-e^{-\frac{m^*}{m_\chi}}.\label{eq:c1}
\end{equation}
Further discussion of the multi-scattering regime, and justification of this fitting function, can be found in Appendix~\ref{sec:opticalfactors}.
For the values $B=0.5$ and $\muFn=400$~MeV, we find
\begin{eqnarray}
m^{*} &=&1.08 \times 10^6 \GeV,\quad |\overline{M}|^2\propto t^0,\\
m^{*} &=&1.62 \times 10^6 \GeV,\quad |\overline{M}|^2\propto t^1,\\
m^{*} &=&2.01 \times 10^6 \GeV,\quad |\overline{M}|^2\propto t^2.
\end{eqnarray}
We illustrate how $m^*$ varies with $B$ and $\muFn$ in Fig.~\ref{fig:mstarbmu}.

%%%%%%%%%%%%%%%%%%%%%%%%%%%%%%
%%%%%%%%%%%%%%%%%%%%%%%%%%%%%%

\subsection{Neutron Star Opacity}

If the DM-neutron cross section is of order  the threshold value or larger, i.e., large enough that the flux of DM particles passing through the NS is significantly attenuated over the path length, we should consider the star opacity in the capture rate calculation. We outline our calculation of the star opacity below, with further details presented in Appendix~\ref{sec:opticalfactors}. 

\begin{figure}
    \centering
    \includegraphics[width=.45\textwidth]{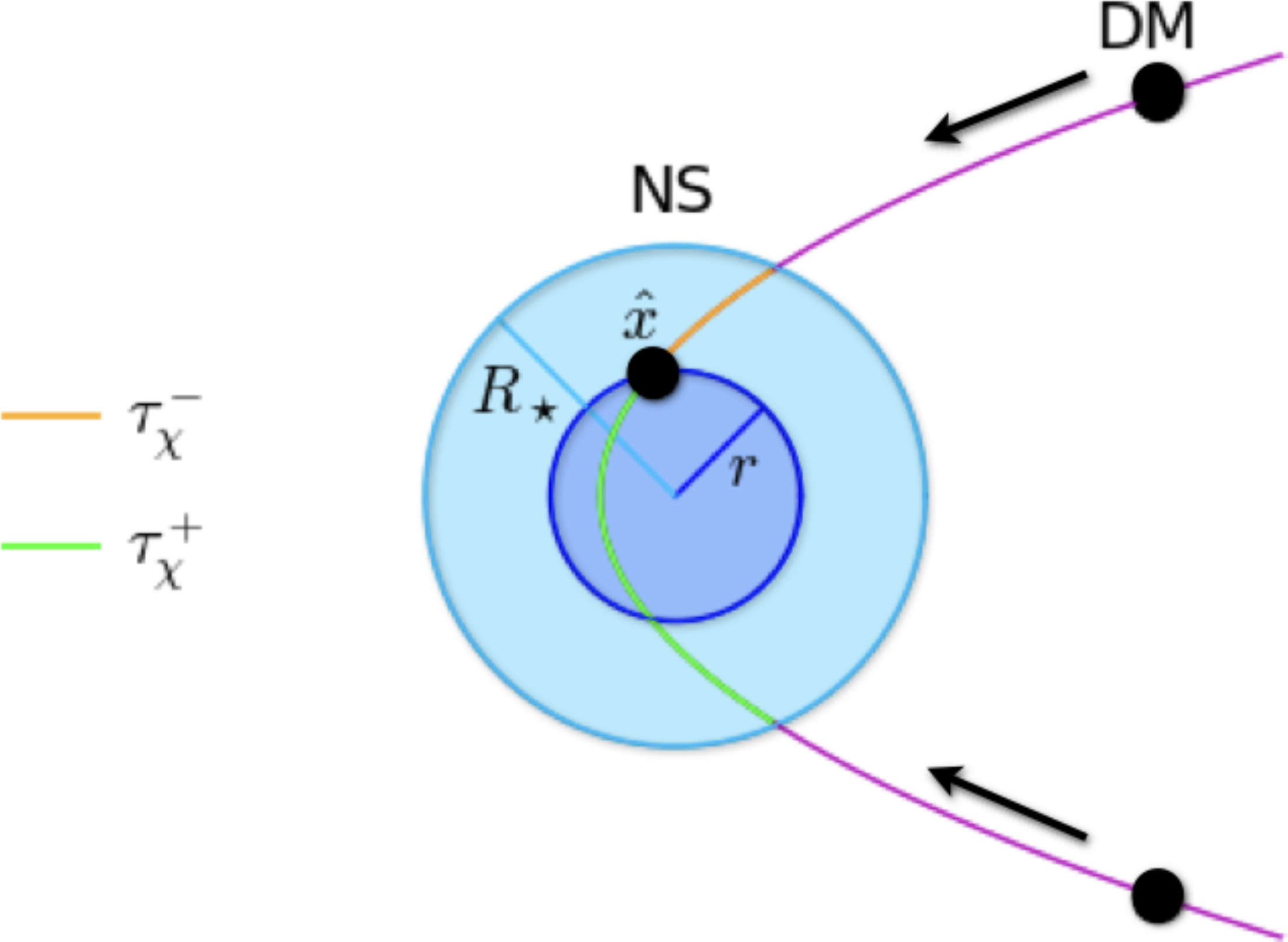}
    \caption{Schematic representation of the  orbit of a DM particle around a NS and the possible trajectories it follows within the NS until it reaches $\hat{x}$. The shortest (longest) path shown in orange (green) is used to calculate $\optdepth^-$ ($\optdepth^+$). }
    \label{fig:orbits}
\end{figure}

The NS opacity can be quantified with the optical factor $\eta$,
\begin{equation}
\eta(\optdepth)=e^{-\optdepth}, 
\label{eq:optfactor}
\end{equation}
where $\optdepth$ is the optical depth seen by a DM particle as it traverses the stellar interior until it is finally captured.
Then, to account for the effect of the opacity on the capture rate, we insert the optical factor $\eta$ in  Eq.~\ref{eq:captureclsimplrel}, 
\begin{eqnarray}
C_{opt} &=& \frac{4\pi}{\vstar}\frac{\rho_\chi}{m_\chi}{\rm Erf}\left(\sqrt{\frac{3}{2}}\frac{\vstar}{v_d}\right) \int_0^{\Rstar}  r^2 dr  \frac{\sqrt{1-B(r)}}{B(r)}\Omega^{-}(r) \eta(r).  \label{eq:captureopticaldepth}
\end{eqnarray}
The optical factor $\eta$ essentially acts as an extinction factor, removing the DM particle from the incoming DM flux after it interacts (and is captured) the first time, thus avoiding double-counting interactions.
The optical depth, $\optdepth$ is determined using
\begin{eqnarray} 
\optdepth(r,\gamma,J) = \int_{\gamma}\Omega^{-}(r)\dfrac{d\tau'}{dl}dl=  \int_{\gamma} dx  \frac{\sigma(x)\zeta(x)n_{free}(r)}{\sqrt{1-\frac{J^2}{J_{max}^2(x)}}}, 
\end{eqnarray} 
where $\tau'$ is the proper time and $\gamma$ is the path followed by the DM particle within the NS to reach the radial distance $r$. Note that for every point $\hat{x}$ on the NS shell of radius $r$, there are two possible trajectories that a DM particle could follow. The shortest path goes from the surface to $\hat{x}$ without passing the perihelion and the longest starts at the surface, reaches the perihelion and then goes to $\hat{x}$, as depicted in Fig.~\ref{fig:orbits}. These trajectories lead to two equally probable optical depths that we must average over
\begin{eqnarray}
\optdepth^-(r,J) &=&  \int_{\Rstar}^r dx \frac{\sigma(x)n_n(r)}{\sqrt{1-\frac{J^2}{J_{max}^2(x)}}},\\
\optdepth^+(r,J) 
&=& \optdepth^{-}(r,J) + 2 \int_{r_{min}}^r dx  \frac{\sigma(x)n_n(r)}{\sqrt{1-\frac{J^2}{J_{max}^2(x)}}},  
\end{eqnarray}
where $r_{min}$ is determined by the angular momentum of the DM particle. For further details, see Appendix~\ref{sec:opticalfactorsingle}.

\begin{figure}
    \centering
    \includegraphics[width=.5\textwidth]{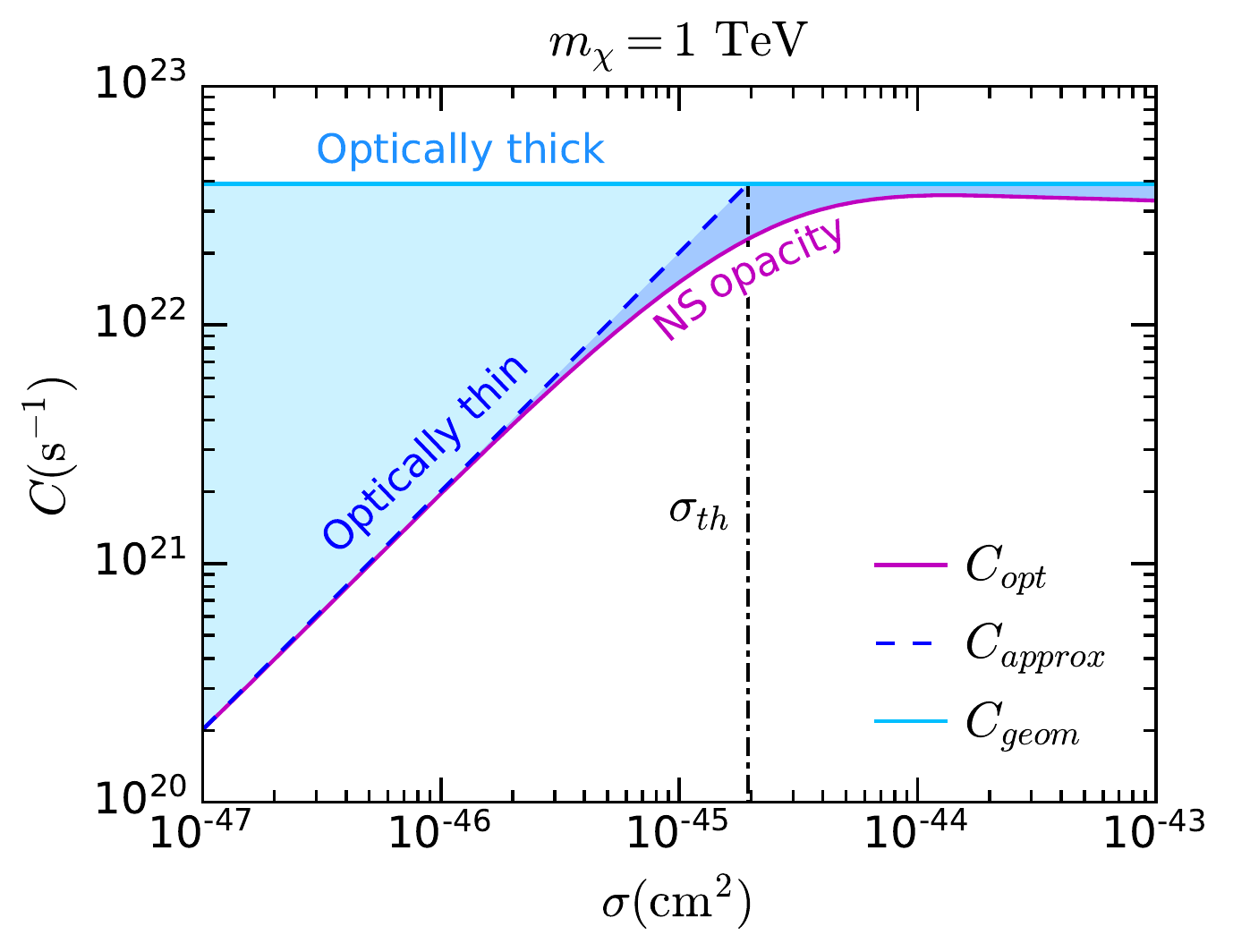}
    \caption{Capture rate as a function of the cross section for $m_\chi=1\TeV$ and the NS model BSk24-2. 
The purple solid  line indicates the capture rate $C_{opt}$ calculated  taking into account the optical factor $\eta$, Eq.~\ref{eq:optfactor}. The light blue solid line denotes the geometric limit $C_{geom}$ and the blue dashed line  the capture rate $C_{approx}$ calculated in the optically thin regime, i.e. without the optical factor $\eta$.}
    \label{fig:geomlim0}
\end{figure}

In Fig.~\ref{fig:geomlim0}, we show the transition from the optically thin regime to the geometric limit for $m_\chi\lesssim10^6\GeV$, by plotting the capture rate as a function of cross section for  $m_\chi=1\TeV$. The blue dashed line indicates the capture rate $C_{approx}$ (Eq.~\ref{eq:csimplelargemtext}), calculated in the optically thin approximation, i.e. neglecting the optical factor $\eta$, and is therefore proportional to the DM-neutron cross section. The light blue line denotes the geometric limit $C_{geom}$ (Eq.~\ref{eq:capturegeom}). The intersection  of these two lines gives the value of the threshold cross section, $\sigma_{th}$ (black dashed  line). 
The  capture rate, calculated including the optical depth factor $\eta$ (Eq.~\ref{eq:optfactor}), $C_{opt}$ (Eq.~\ref{eq:captureopticaldepth}) is depicted in purple. This calculation is well approximated by $C_{approx}$ (blue dashed line) until $\sigma\sim {\rm few} \times 10^{-46}\cm^2$. For larger cross sections, the optical depth factor suppresses the capture rate, such that it asymptotes to the geometric limit when $\sigma \gtrsim 10^{-44}\cm^2$. 

\begin{figure}
    \centering
    \includegraphics[width=.5\textwidth]{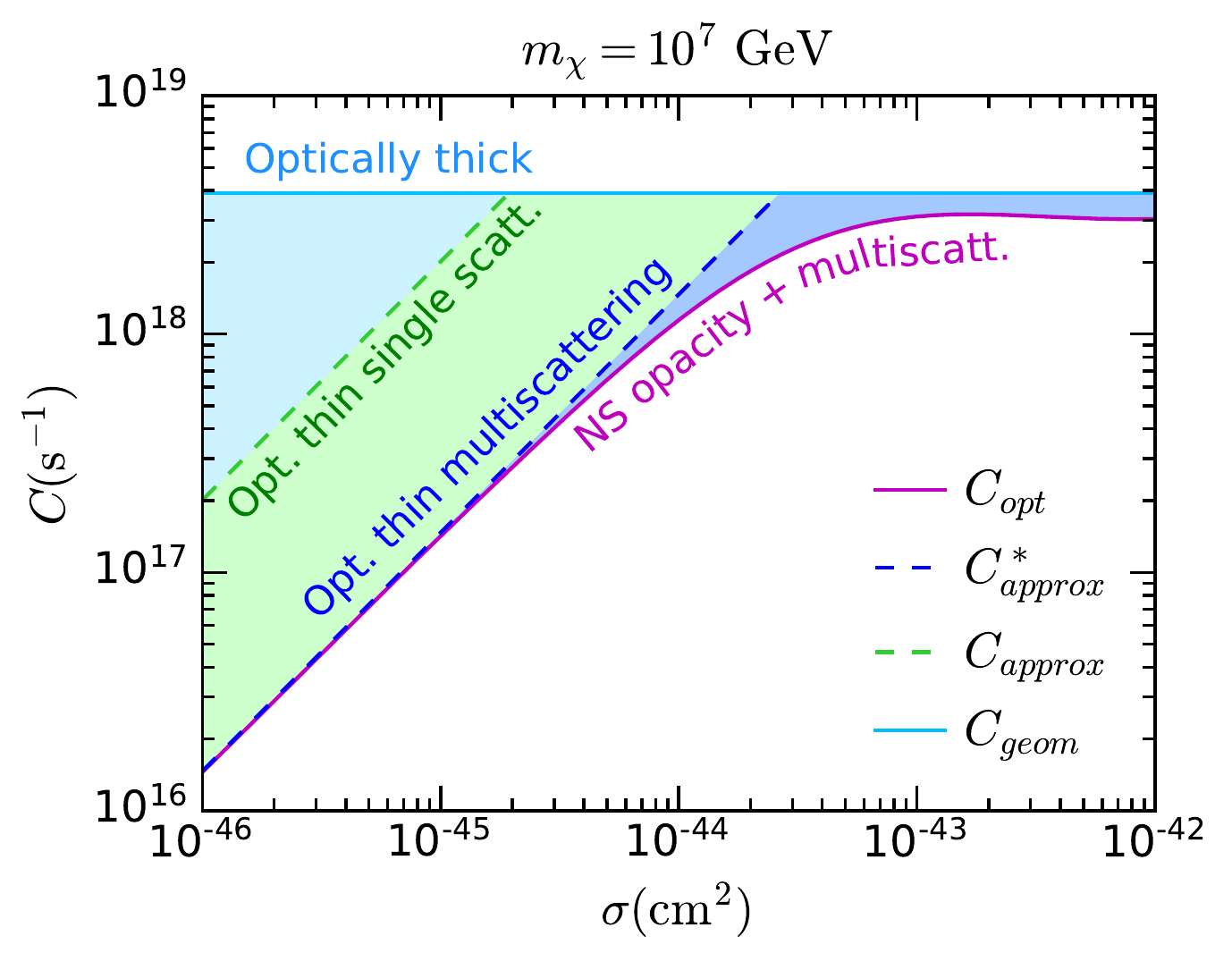}
    \caption{Capture rate as a function of the cross section for $m_\chi=10^7\GeV$ and the NS benchmark model BSk24-2. 
The purple solid  line indicates the capture rate $C_{opt}$, calculated taking into account the optical factor $\eta$, given by Eq.~\ref{eq:optfactorlargem}, and considering multiscattering. The light blue solid line denotes the geometric limit $C_{geom}$.     
   The blue dashed line indicates the capture rate $C_{approx}^*$, calculated without the optical factor $\eta$ but including the suppression factor due to a capture probability $c_1\ll1$. The green dashed line denotes the capture rate $C_{approx}$, calculated without the optical factor $\eta$ and with   $c_1=1$.  }
    \label{fig:multlim0}
\end{figure}

\subsection{Opacity and Multi-Scattering}

In the case that the DM-neutron cross section and DM mass are both large, we have to simultaneously account for opacity and multiscattering effects, which are both highly nonlinear (see Appendix~\ref{sec:opticalfactormulti}). We can properly incorporate this by adding to the integrand of Eq.~\ref{eq:csimplelargemtext} the optical factor $\eta(\optdepth(r))$ calculated in Appendix~\ref{sec:opticalfactormulti}. 
In Appendix~\ref{sec:largexslimit}, we show that this formalism correctly reproduces the geometric limit in the case where the cross section is sufficiently large.

For $m_\chi\gtrsim10^6\GeV$, we obtain an optical factor of 
\begin{equation}
\eta(r) = \frac{1}{n^*(r)} e^{-\optdepth(r)/n^*(r)}, 
\label{eq:optfactorlargem}
\end{equation}
where $\frac{1}{n^*}=c_1$ is the capture probability defined in Eq.~\ref{eq:c1}. See Appendix~\ref{sec:opticalfactormulti} for the full derivation. The modified optical factor in Eq.~\ref{eq:optfactorlargem} can be interpreted  similarly to that in Eq.~\ref{eq:optfactor}, with the difference being that the capture probability for a single interaction is no longer $1$, but rather $c_1=\frac{1}{n^*}$, and thus an average of $n^*$ interactions is required to remove the particle from the incoming flux.  This expression for the optical factor should be used in Eq.~(\ref{eq:captureopticaldepth}) whenever $m_\chi\gtrsim10^6\GeV$ and $\sigma \sim \sigma_{th}$.

For large mass and small cross sections, $m_\chi\gg10^6\GeV, \sigma\ll\sigma_{th}$, the capture probability is significantly smaller than $1$, which should thus be accounted for, while the probability of subsequent scatterings beyond the first is negligible, and hence the use of an optical factor is unnecessary. Therefore, neglecting the factor that depends on the optical depth $\optdepth$, we obtain a suitable approximation that accounts only for multiple scattering,
\begin{eqnarray}
C_{approx}^* =\frac{4\pi}{\vstar}\frac{\rho_\chi}{m_\chi}{\rm Erf}\left(\sqrt{\frac{3}{2}}\frac{\vstar}{v_d}\right)  \int  r^2 dr  \frac{\sqrt{1-B(r)}}{B(r)}\Omega^{-}(r) \frac{1}{n^*(r)} 
\label{eq:capturesimplelargem}. 
\end{eqnarray}

The transition from the optically thin regime to the geometric limit in the case of large mass, $m_\chi>10^6\GeV$, is illustrated in Fig.~\ref{fig:multlim0} for $m_\chi=10^7\GeV$. The purple solid line corresponds to $C_{opt}$, which correctly accounts for both the NS opacity and multiple scattering by using the expression for $\eta$ given by Eq.~\ref{eq:optfactorlargem} in Eq.~\ref{eq:captureopticaldepth}. The blue dashed line shows $C_{approx}^*$ (Eq.~\ref{eq:capturesimplelargem}), which is a good approximation of the capture rate at small cross sections, and includes the capture probability $c_1$, but not the optical factor $\eta$. 
The green dashed line corresponds to $C_{approx}$ (Eq.~\ref{eq:csimplelargemtext}), which includes neither $\eta$ nor $c_1$ and therefore overestimates $C$ by a factor $n^*=1/c_1$ at small cross sections; see the green shaded area. Compared to the intermediate DM mass range, the key difference in this heavy mass regime is that we require significantly larger cross sections to saturate the geometric limit.

\section{Results}
\label{sec:results}

%%%%%%%%%%%%%%%%%%%%%%%%%%%%%%%%%%%%%%%%%%%%%%%%%%%%%%%%%%
\begin{table}[tb]\centering
\begin{tabular}{|c|c|c|c|}
\hline
 $\sigma$ & $m_\chi\lesssim 1\GeV$ & $1\GeV\lesssim m_\chi\lesssim 10^6\GeV$ & $m_\chi\gtrsim 10^6\GeV$
\\ \hline
$\sigma\ll\sigma_{th}$ & $C_{PB}$ (\ref{eq:captureclsimplrel},\ref{eq:omegampaulitext}) & $C_{approx}$ (\ref{eq:csimplelargemtext})  & $C_{approx}^*$ (\ref{eq:capturesimplelargem})
\\
$\sigma\sim\sigma_{th}$ & $C_{PB+opt^*}$ & $C_{opt}$ (\ref{eq:captureopticaldepth}) with $\eta$ (\ref{eq:optfactor}) & $C_{opt}$ (\ref{eq:captureopticaldepth}) with $\eta$ (\ref{eq:optfactorlargem}) 
\\
$\sigma\gg\sigma_{th}$ & $C_{geom}$ (\ref{eq:capturegeom}) & $C_{geom}$ (\ref{eq:capturegeom}) & $C_{geom}$ (\ref{eq:capturegeom})
\\ \hline
\end{tabular}
\caption{Different regimes for the DM capture rate in NSs The DM mass ranges specified are valid for nucleon targets.}\label{tab:caprecap}
\end{table}
%%%%%%%%%%%%%%%%%%%%%%%%%%%%%%%%%%%%%%%%%%%%%%%%%%%%%%%%%%

In this section, we present our results for the capture rate of fermionic DM scattering from neutrons within a NS in the zero temperature approximation, which is in fact valid for NS temperatures $T\lesssim 10^6\K$ for the DM mass range considered here. We calculate the capture rate only for scalar/pseudoscalar-scalar/pseudoscalar interactions between DM and neutrons, i.e. effective operators D1-D4 in Table~\ref{tab:operatorshe}, whose differential cross sections depend only on the Mandelstam variable $t$ but not on $s$. 
We assume realistic radial profiles for the neutron number density, chemical potential and relativistic corrections encoded in $B(r)$ as explained in section~\ref{sec:NSmodels} for the configurations of the EoS BSk24 in Table~\ref{tab:eos}.

Table~\ref{tab:caprecap} summarises the various kinematic regimes identified in the previous sections, and the relevant approximations needed to accurately calculate the capture rate.
Note that we have not given an explicit equation for $C_{PB+opt^*}$. This label refers to Eqs.~\ref{eq:captureclsimplrel} and \ref{eq:omegampaulitext} with the optical depth factor $\eta$ included, using the proper calculation of optical depth $\optdepth$ that correctly accounts for Pauli blocking. In any case, very large DM-neutron cross sections are required to saturate the geometric limit in the low mass regime, i.e. $\sigma_{th} \gg 10^{-45}\cm^2$ when $m_\chi\ll 1\GeV$.

The mass regimes specified in Table~\ref{tab:caprecap}
are valid for nucleon targets, and we have focused on operators D1-D4 to illustrate our results. However, most of our results are applicable generally to other operators or to other targets (with the mass ranges adjusted appropriately).  Specifically,
Eqs.~\ref{eq:captureclsimplrel},~\ref{eq:omegampaulitext}, which are to be evaluated numerically, are applicable to all operators and targets, and work until multiple scattering becomes relevant, when $m_\chi\gtrsim \qomax/v_{\star}^2$.
The optical factor of Eq.~\ref{eq:optfactor} for the intermediate mass range is also applicable to all operators and targets. The optical factor of Eq.~\ref{eq:optfactorlargem} and the value of $m^*$, which are used to include multiple scattering effects in the large mass range, $m_\chi\gtrsim \qomax/v_{\star}^2$, can be easily computed for operators D1-D4 (or any other operator that depends only on $t$) for all targets. For other operators it can be used only by numerically solving the shape of the differential rate, a task that may be computationally intensive to achieve with high precision.
Our approximated formulas, Eqs.~\ref{eq:csimplelargemtext},~\ref{eq:capturesimplelargem} and \ref{eq:captureopticaldepth} have been checked to be accurate only for nucleon targets, but can be applied to any operator (for $s$-dependent ones, see Appendix~\ref{sec:capratesimple} on how to remove the $s$ dependence). In any case, one can substitute the relevant factors ($\eta,m^*$) into Eqs.~\ref{eq:captureclsimplrel},~\ref{eq:omegampaulitext} to calculate the capture rate, in the appropriate mass range, for other targets.

\begin{figure}
    \centering
    \includegraphics[width=.45\textwidth]{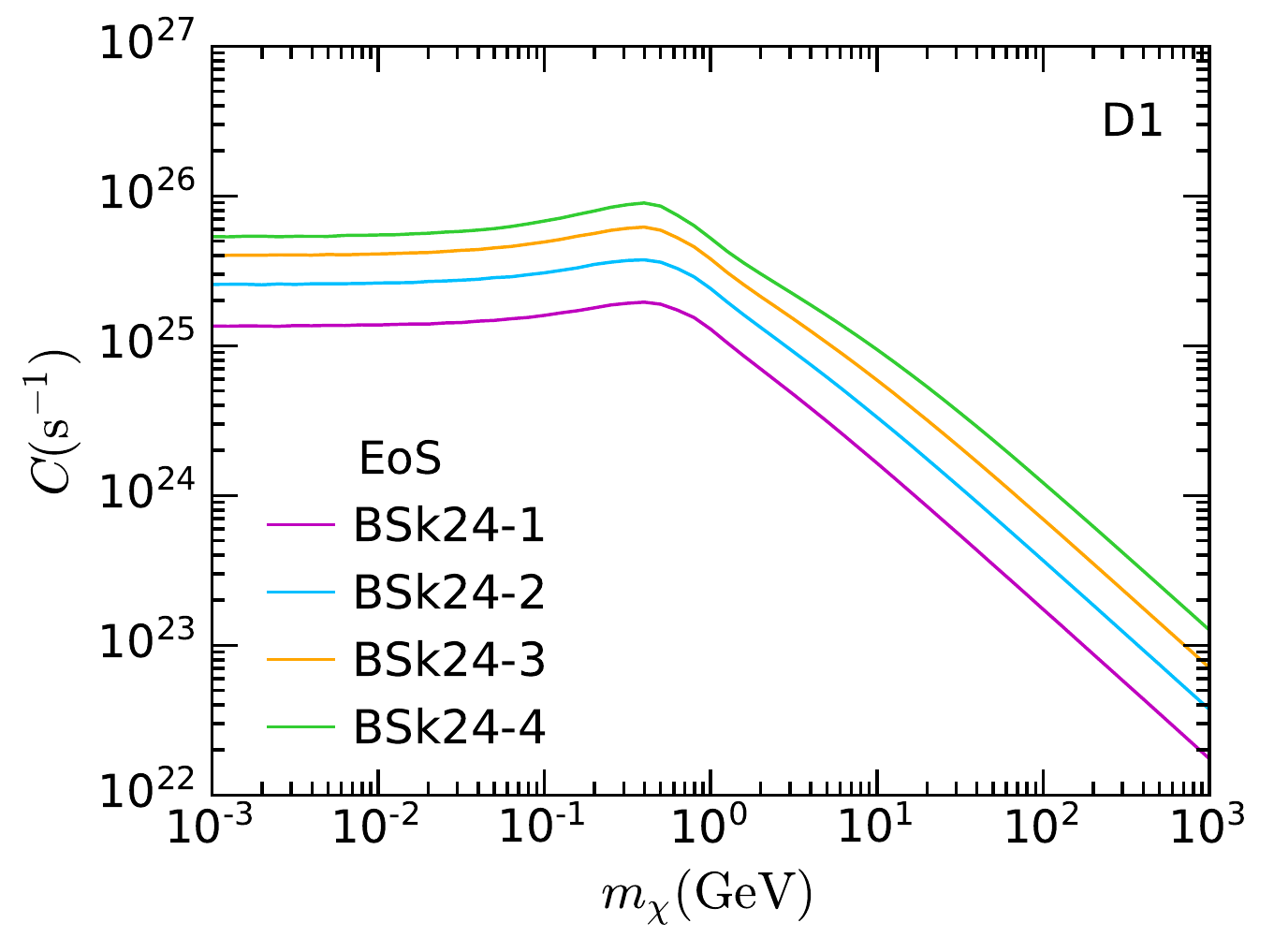}
    \includegraphics[width=.45\textwidth]{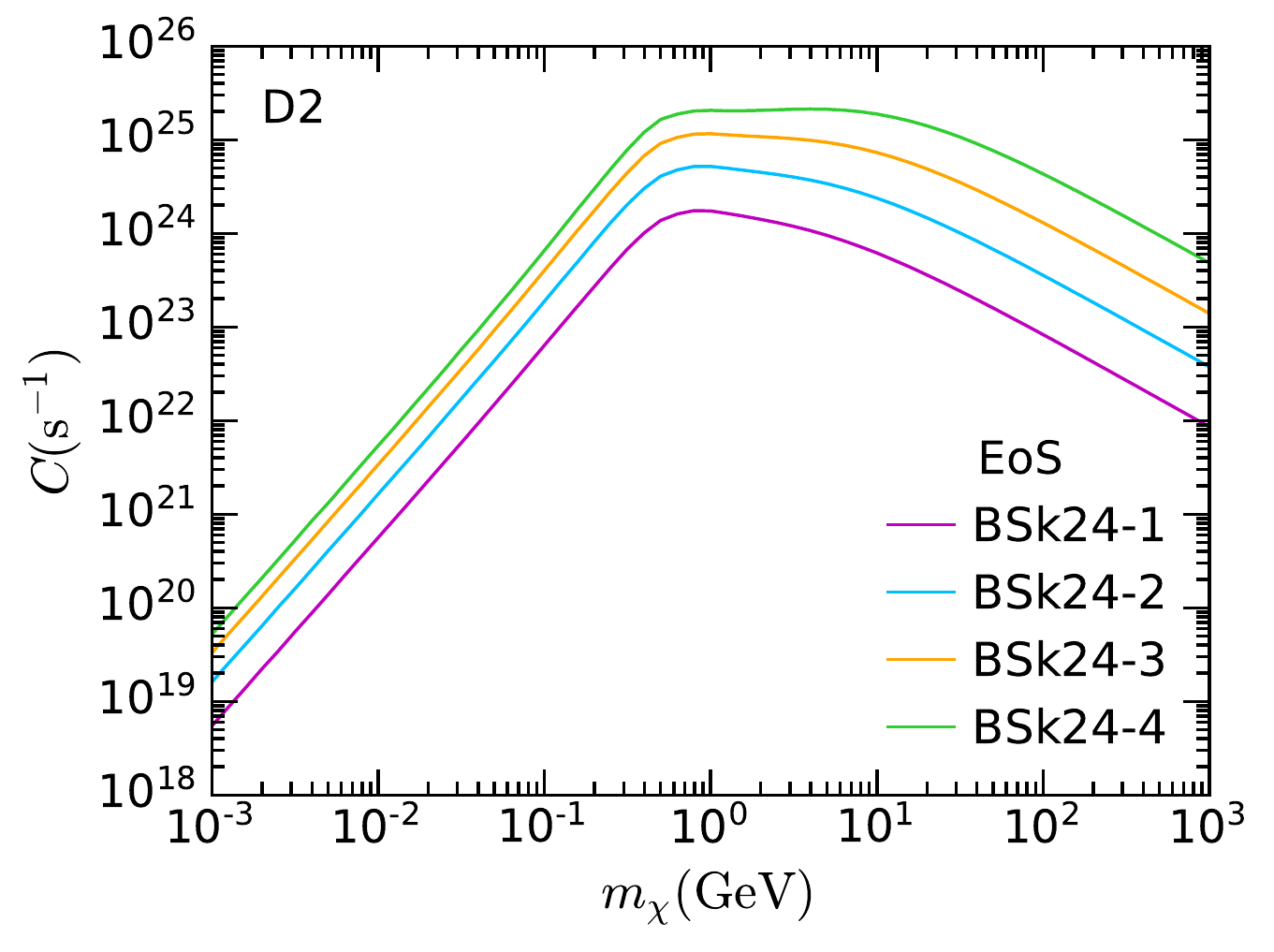}\\
    \includegraphics[width=.45\textwidth]{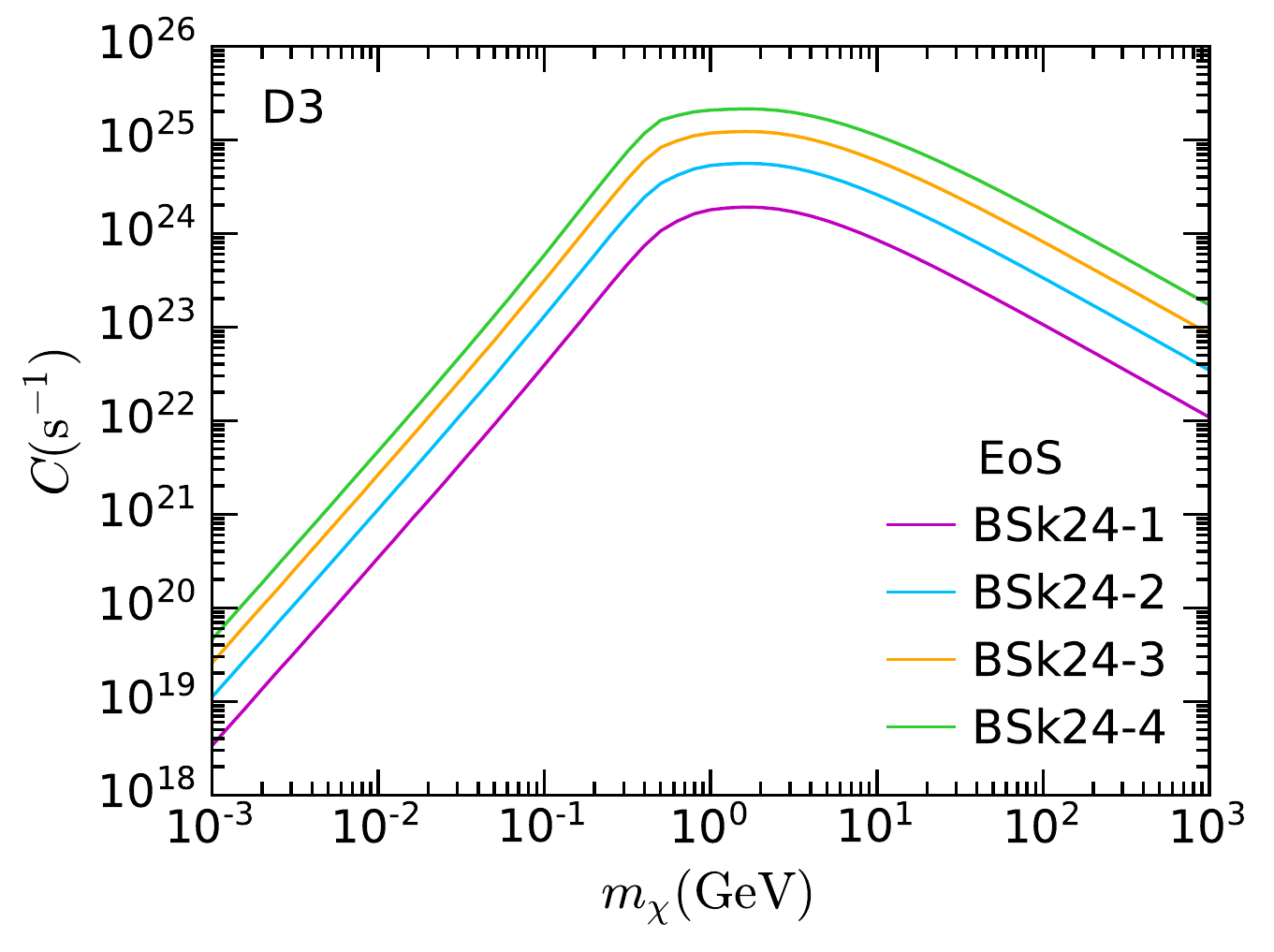}
    \includegraphics[width=.45\textwidth]{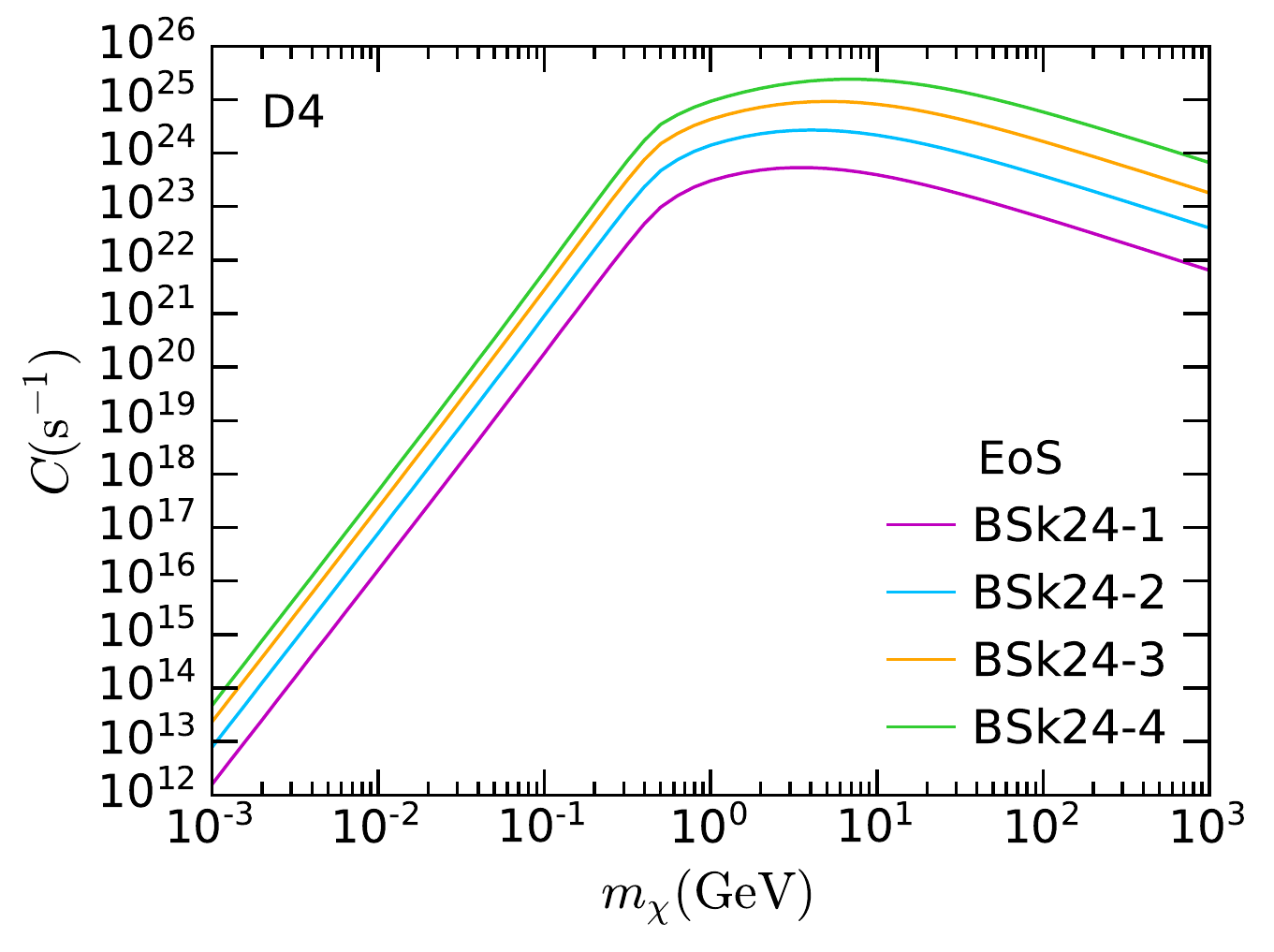}
    \caption{Capture rate in the optically thin limit as a function of the DM mass for $\sigma=\sigma_{ref}\sim 1.7 \times 10^{-45} \cm^2$ and the  configurations of the EoS BSk24 given in Table~\ref{tab:eos}. Rate calculated using the 4-dimensional integral in Eq.~\ref{eq:capturefinal}, which includes Pauli blocking and neglects the  NS opacity and multiple scattering for the EFT operators D1 (top left), D2 (top right), D3 (bottom left) and D4 (bottom right) in Table~\ref{tab:operatorshe}.}
    \label{fig:captureEOS}
\end{figure}

In order to estimate the NS EoS impact on the DM capture rate computation,  we numerically calculate it using the exact expression in the optically thin limit, Eq.~\ref{eq:capturefinal}, that properly accounts for gravitational focusing and Pauli blocking but neglects the star opacity.  In this approximation, the capture rate is proportional to the differential DM-neutron cross section. 
Fig.~\ref{fig:captureEOS} shows how this rate varies with the NS  EoS  for operators D1-D4 and the EoS configurations given in Table~\ref{tab:eos}, and in turn with the NS mass and radius. 
The value of the cross section was chosen so that at large DM mass the capture rate is equal to the geometric limit. 
Note that properly including the optical depth factor $\eta$ would have given a lower value of $C$ (see section~\ref{sec:largemassandsigma}). It is worth remarking that we  should not use larger values of the cross section in the optically thin approximation, as this would lead to capture rates exceeding the geometric limit.
Depending on the operator considered, going from the lightest to the heaviest NS can change the capture rate by a minimum of one order of magnitude, such as in the case of operators D1, D2 and D3 (at low DM mass), and up to 2 orders of magnitude, as in the case of operators D2 (only at large DM mass) and D4. 

%%%%%%%%%%%%%%%%%%%%%%%%%%%%%%
%%%%%%%%%%%%%%%%%%%%%%%%%%%%%%

At large DM mass, all operators  show the same scaling with the DM mass. At  $m_\chi\lesssim1\GeV$, a different picture arises as Pauli blocking leads to different suppressions of the capture rate for the different operators. However, we observe that the four operators give very similar results to those of Fig.~\ref{fig:approxc}, where we analysed the dependence of the capture rate on the momentum transfer $t$.  We note that  operator D1, which contains in its squared matrix element, $|\overline{M}|^2$, a term independent of $t$, gives a result that is very similar to that of $\sigma=const$. Operators D2 and D3, for which $|\overline{M}|^2$ does not include terms independent of $t$, but rather terms proportional to $t$ and $t^2$, yield  very similar results to that of $d\sigma\propto t$. 
Overall, we conclude that the lowest power of the transferred momentum determines the mass scaling of the capture rate at low DM  mass.

%%%%%%%%%%%%%%%%%%%%%%%%%%%%%%
\begin{figure}
    \centering
    \includegraphics[width=.45\textwidth]{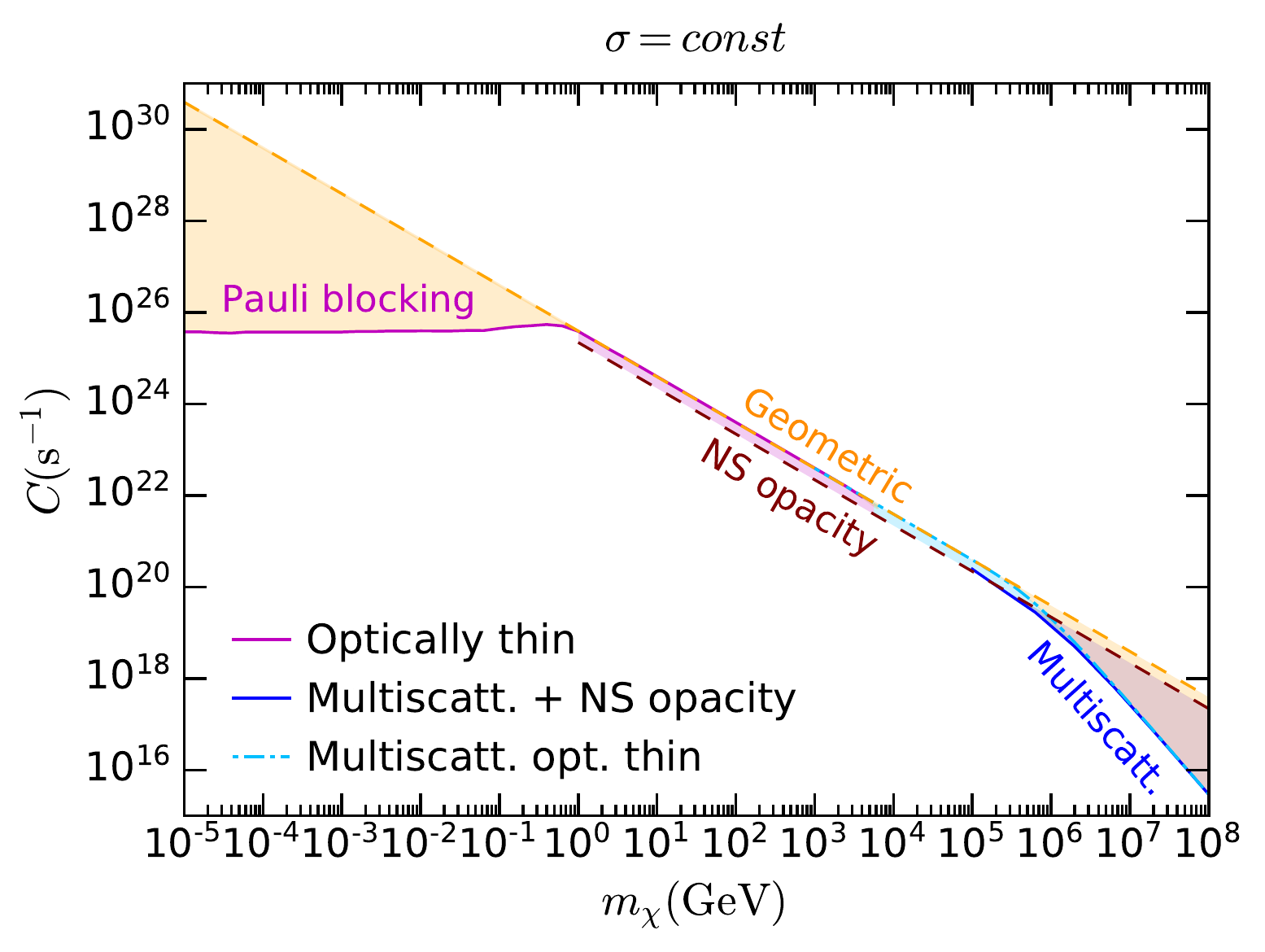}    
    \includegraphics[width=.45\textwidth]{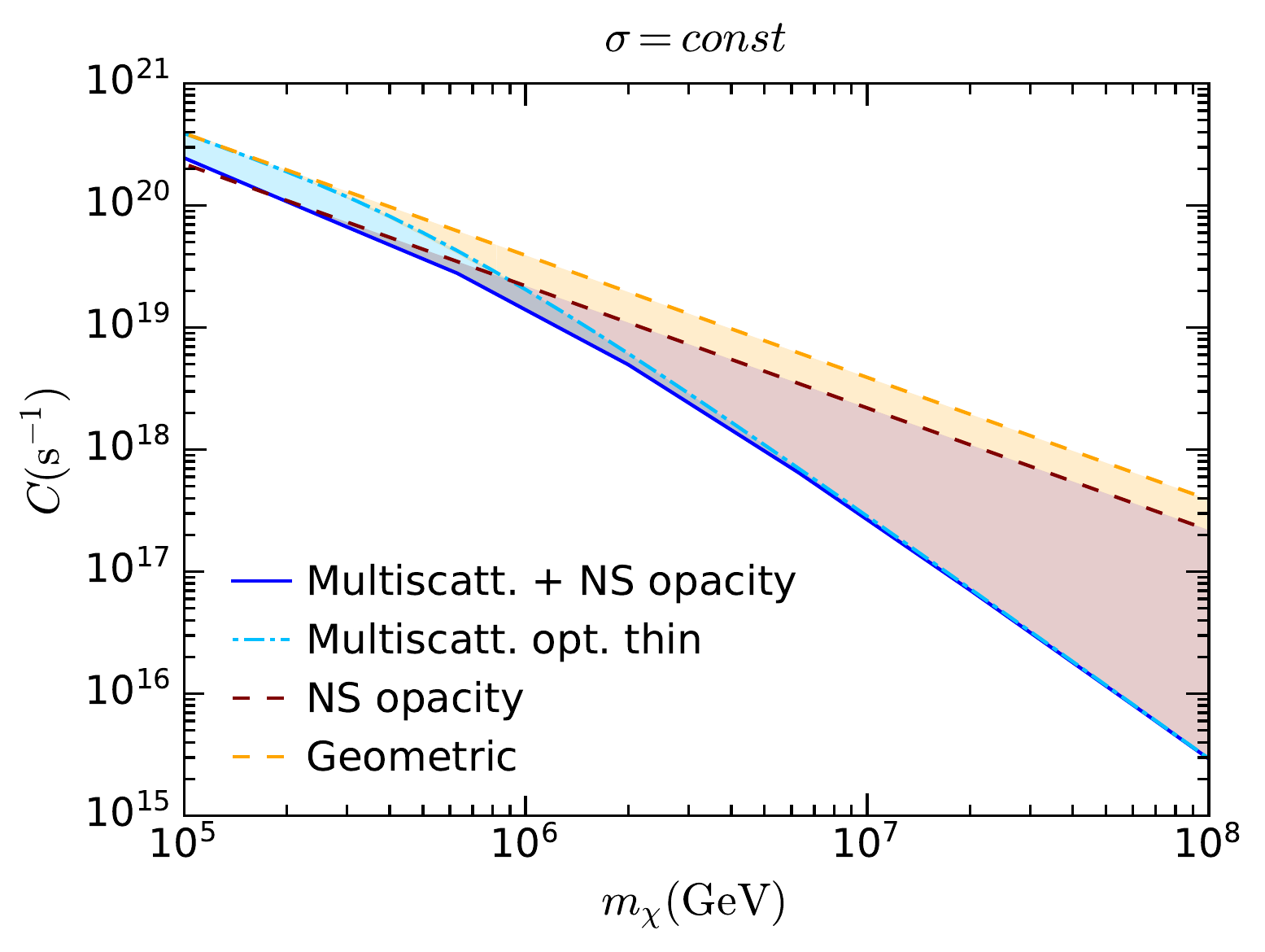} %\\
    \includegraphics[width=.45\textwidth]{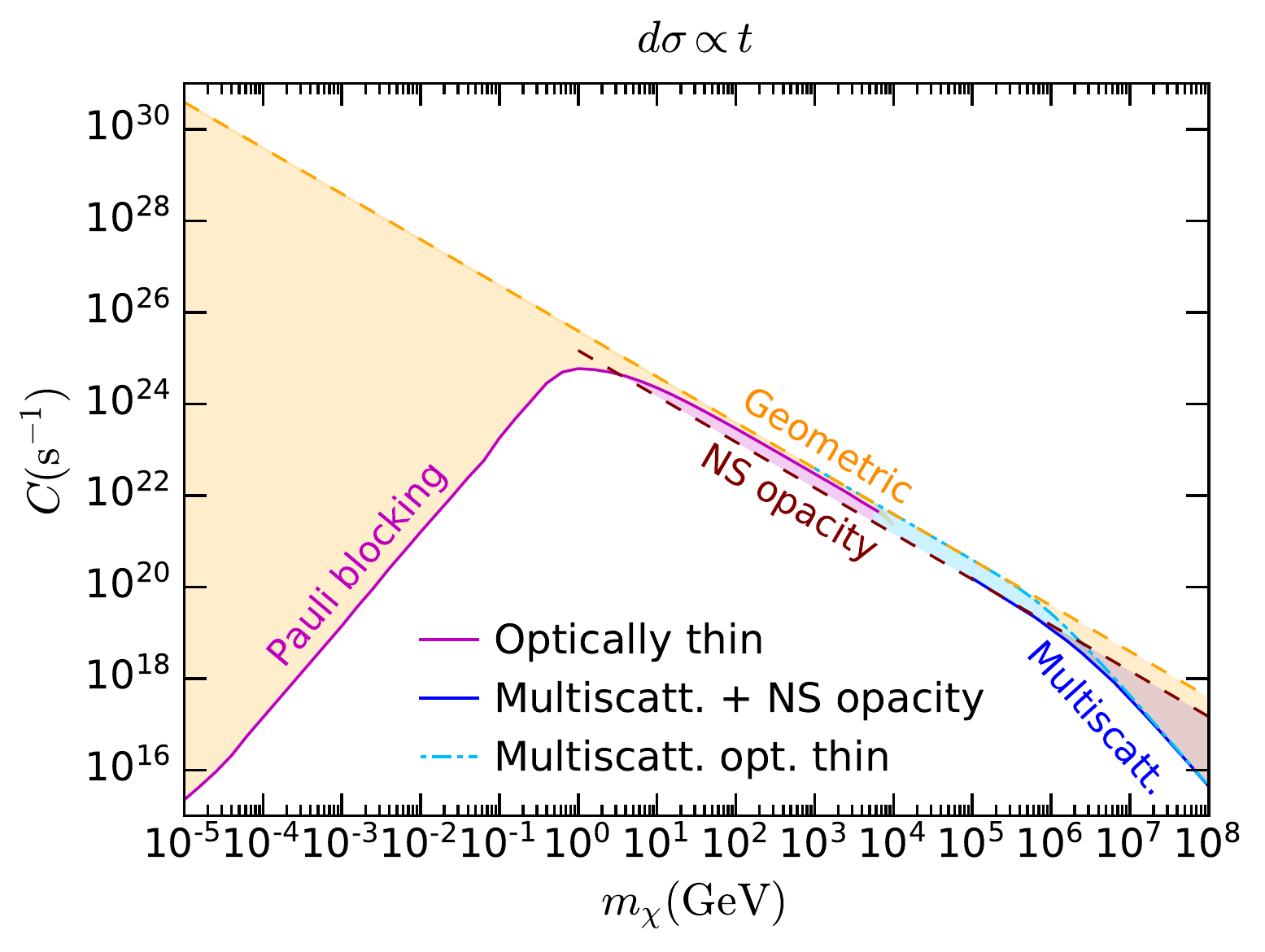}       
    \includegraphics[width=.45\textwidth]{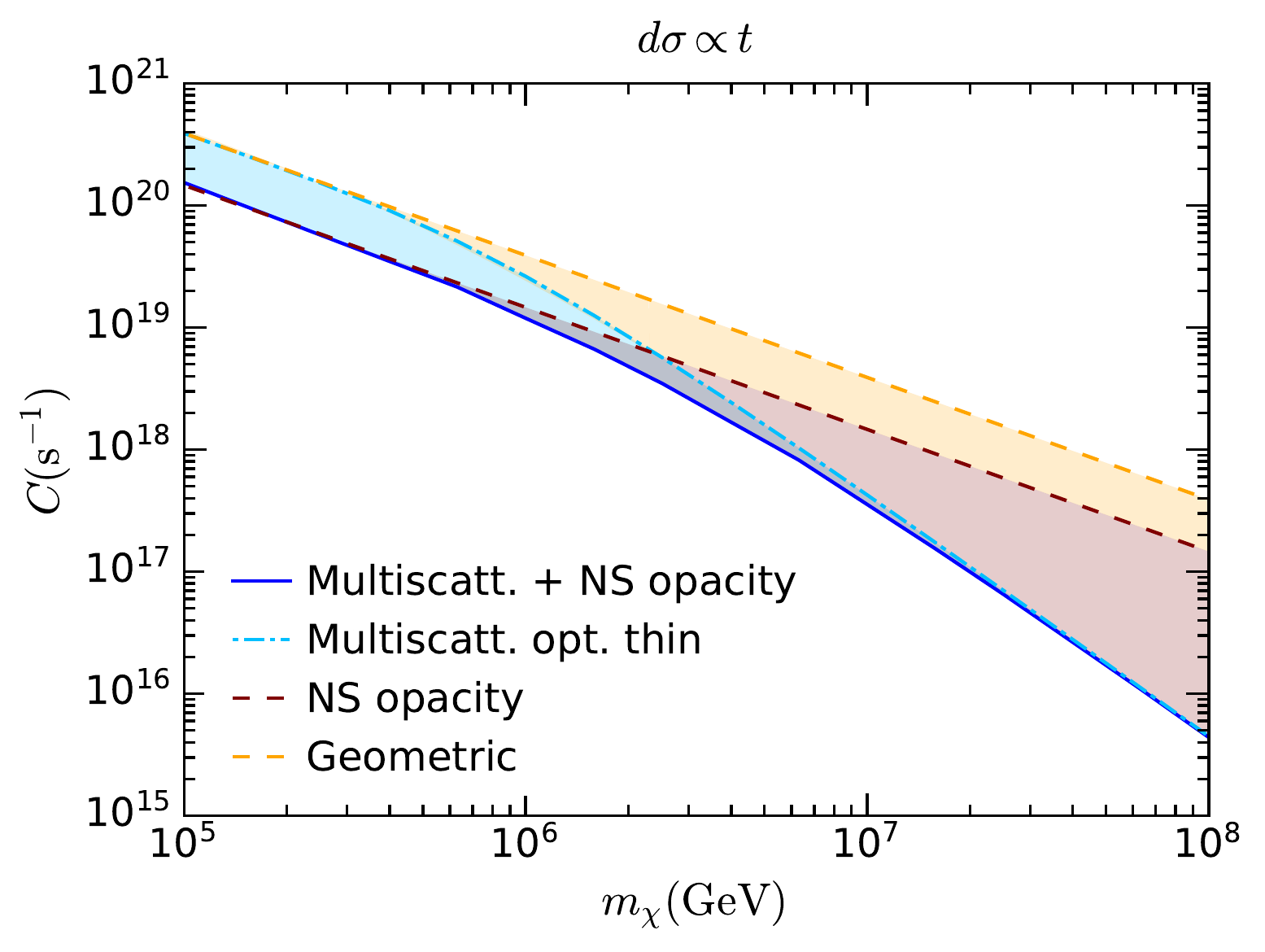} \\
    \includegraphics[width=.45\textwidth]{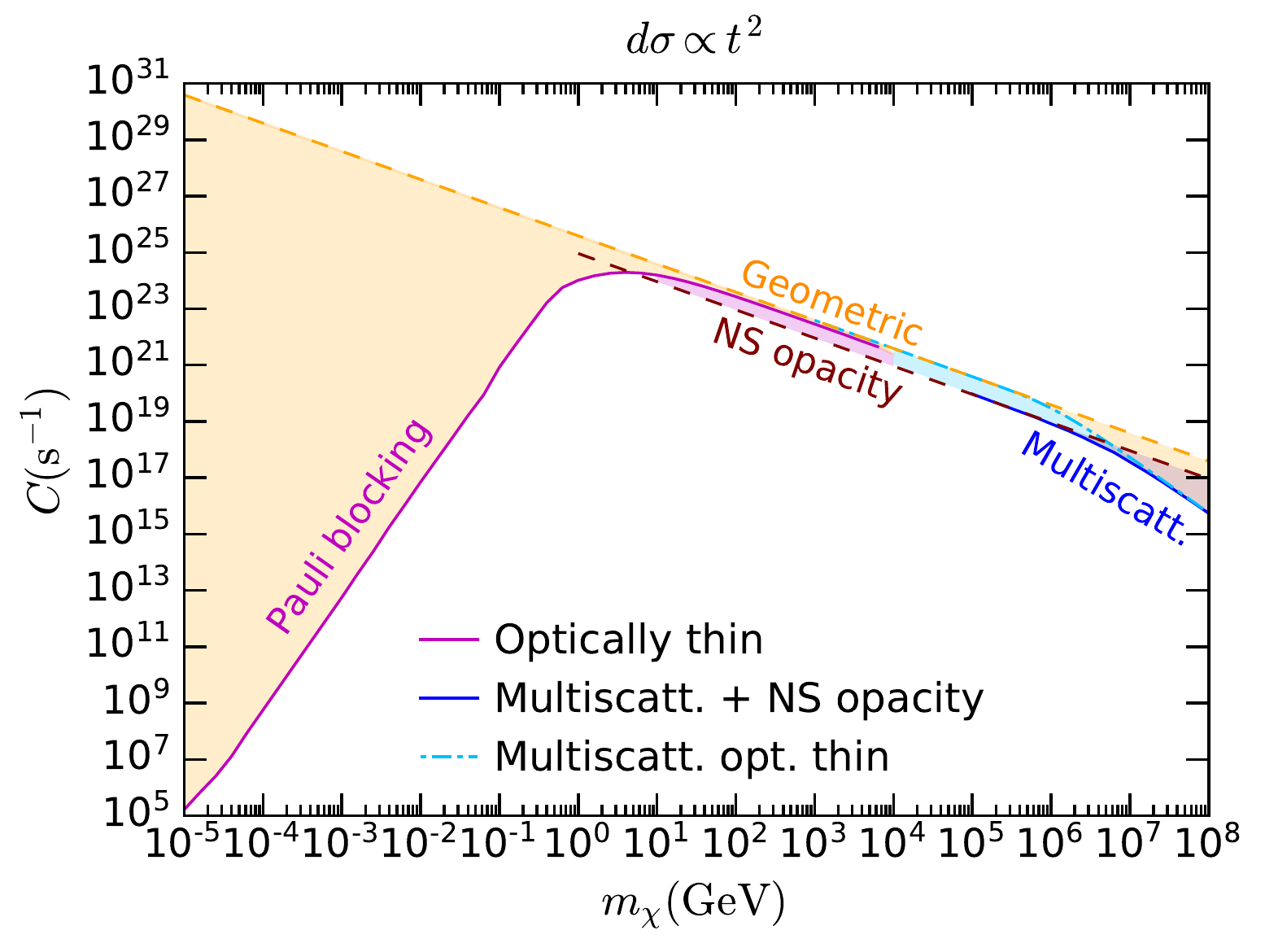}       
    \includegraphics[width=.45\textwidth]{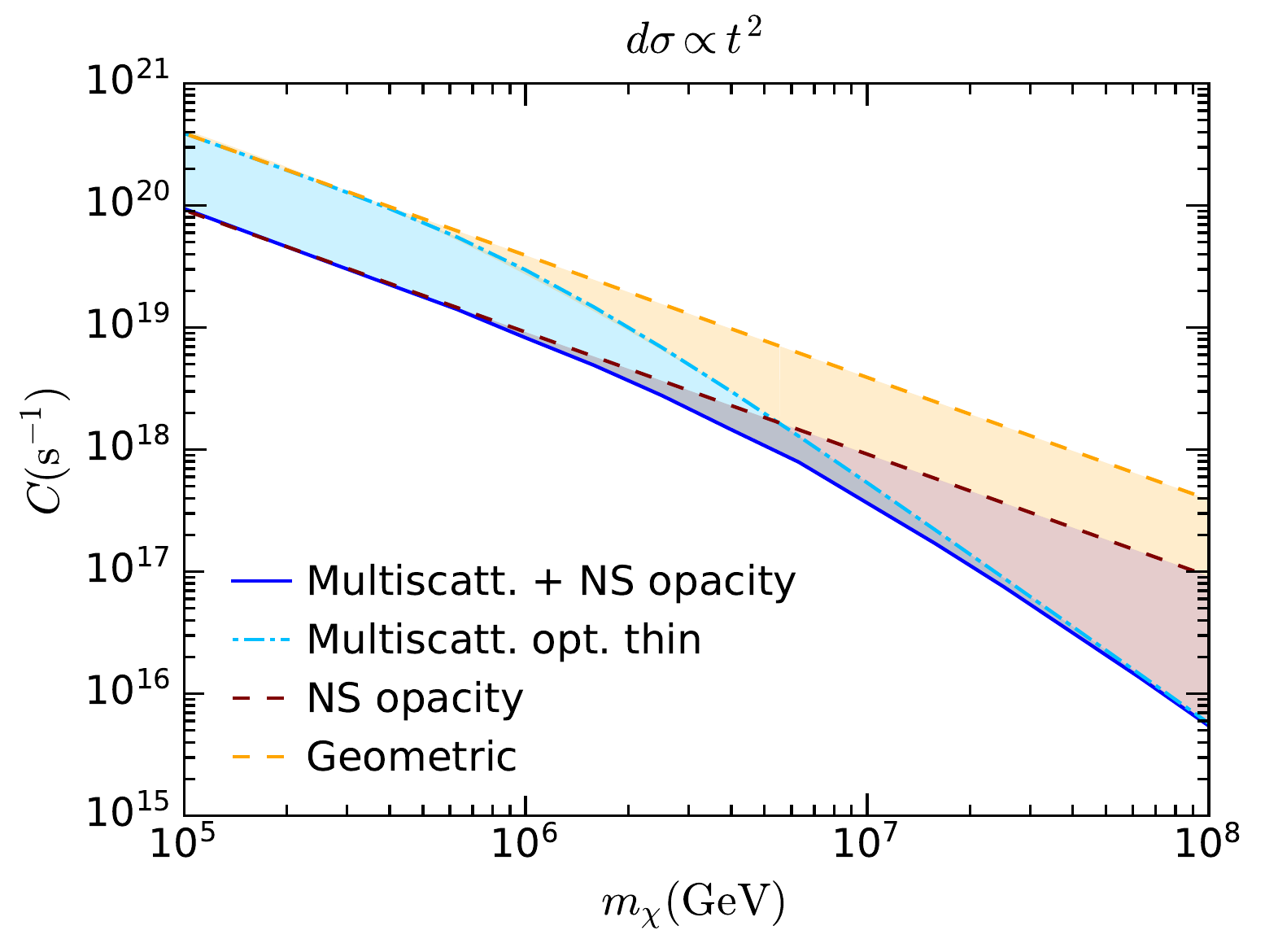}        
    \caption{Capture rate for constant cross section (top row), $d\sigma\propto t$ (middle row) and $d\sigma\propto t^2$ (bottom row),  for  $\sigma=\sigma_{ref}\sim 1.7 \times 10^{-45}\cm^2$ and NS EoS configuration BSk24-2. We extend the plot in the top left panel of Fig.~\ref{fig:approxc} to large DM masses. Left: Full mass range. Right: Same as before but only for large DM mass range. 
    }
    \label{fig:capturelargemass}
\end{figure}
%%%%%%%%%%%%%%%%%%%%%%%%%%%%%%

In Fig.~\ref{fig:capturelargemass}, we show the capture rate for a broad DM mass range, spanning 13 orders of magnitude from $m_\chi=10\keV$ to $m_\chi=10^8\GeV$, including all the regimes identified in Table~\ref{tab:caprecap}, for $d\sigma\propto const.$ (first row), $t^1$ (second row) and $t^2$ (third row).
In the left panels, we show the full mass range.  As in previous figures, the magenta line indicates the capture rate calculated in the optically thin limit using the 4-dimensional integration in Eq.~\ref{eq:captureclsimplrel} that accounts for Pauli blocking. 
At large DM masses, Pauli suppression plays no role and the capture rate approaches the geometric limit (dashed orange line). 
We also show in Fig.~\ref{fig:capturelargemass} three new lines, portraying the effect of the inclusion of the NS optical depth and multiple scattering, which become relevant at $m_\chi\sim10^6\GeV$. 
The difference among these calculations is better shown in the right panels, where only the large DM mass range is considered. The brown dashed line indicates the result that includes the optical depth factor $\eta$ but neglects multiple scattering, obtained using Eq.~\ref{eq:optfactor} in Eq.~\ref{eq:captureopticaldepth}.   
As we can see, for $\sigma=\sigma_{ref}$ this causes a small suppression of the capture rate, when compared to the result where the optical depth factor is ignored (light blue dot dashed line). For larger $\sigma$, neglecting the optical depth would result in a capture rate that exceeds the geometric limit (orange dashed line), while the inclusion of the optical depth factor $\eta$ causes the capture rate to saturate, tending to  $C_{geom}$ for large cross sections. 
The light blue dot dashed line indicates the capture rate calculated by neglecting the optical depth factor, but including multiple scattering, given by Eq.~\ref{eq:capturesimplelargem}. At $m_\chi\sim 10^5\GeV$ that line matches the geometric limit, due to the chosen value of the cross section $\sigma=\sigma_{ref}$. On the other hand, at larger DM masses $m_\chi\gtrsim10^6\GeV$, multiple scattering is required to capture DM particles, hence an additional suppression factor of $1/m_\chi$ arises, as given in Eq.~\ref{eq:capturesimplelargem}. Therefore the capture rate becomes increasingly smaller than $C_{geom}$ (orange and brown shaded areas). Finally, the capture rate calculated including both effects is depicted in blue. At $m_\chi\sim10^5\GeV$, we can observe the suppression produced by the optical depth factor $\eta$ (light blue shaded region), while at larger DM masses the proper additional suppression $1/m_\chi$ emerges. 

Comparing the plots for different $t^n$ dependence, we can see that increasing the power of $n$ has a small effect on the mass scale where the various suppressions become relevant. For example, comparing the blue and light blue lines, which both include multiple scattering effects, we see that the change of slope moves further to the right for larger $n$. This is a consequence of the fact that larger powers of $n$ result in larger energy transfer (see, for example, Fig.~\ref{fig:diffgamma}) and therefore a larger capture probability $c_1$ and larger $m^*$.
However, the qualitative behaviour is the same for all choices of $d\sigma$: the suppression of the capture rate is primarily due to Pauli blocking at low mass, opacity effects in the $1$--$10^6$~GeV mass range, and multiscattering effects (i.e. a low capture probability) at the largest masses.

\section{Conclusions}
\label{sec:conclusions}

Neutron stars (NSs) are relativistic objects by nature, in the sense that their structure and evolution should be studied in the context of General Relativity. For instance, their luminosity and observed radius are affected by gravitational redshift.  In adddition, dark matter particles are accelerated to quasi-relativistic speeds upon infall to a NS.
Therefore, NSs have greater reach to probe dark matter (DM) interactions than any other known stellar objects, and most Earth-based direct detection experiments, where the identification of DM interactions is hampered by small momenta, velocities or recoil energies. Consequently, NSs have gained increasing attention in the last few years, especially  in light of upcoming infrared telescopes that might be able to detect a signal in the wavelength range in which an old, faint, nearby NS would emit radiation. This signal can be interpreted in terms of heating induced by DM interactions with the particle species present in the NS, and used to constrain the strength of DM interactions. 

A key ingredient in any scenario involving the accumulation of DM in a NS is the capture rate. Its proper derivation beyond the geometric limit, however, implies knowledge of the NS internal structure, which is still an open problem in nuclear astrophysics. We are compelled then to assume an equation of state (EoS) that relates pressure to other fundamental parameters, including those required for the capture rate computation, namely, the number density, abundance and chemical potential of each particle species present in the star. 
For the purpose of numerical computation, we have therefore selected a family of NS EoS, specifically, the unified equation of state with Brussels-Montreal functional BSk24. We have presented results for a set of benchmark configurations of that functional, although it is of course possible to repeat the capture rate computation  using any other EoS. 
We find that the choice of EoS can lead to variations in the capture rates by one or two orders of magnitude.

In this paper, we have improved and extended the existing framework to calculate the DM capture rate, relaxing the simplifying assumptions that have previously been made. Specifically, we have derived exact expressions for the capture rate that correctly incorporate relativistic kinematics, gravitational focusing, Pauli blocking, the opacity of the star and multiple-scattering effects. We also properly incorporate the NS internal structure, consistently calculating the radial profiles of the EoS dependent parameters and the general relativistic corrections, by solving the Tolman-Oppenheimer-Volkoff equations.

Neutron stars are composed of strongly degenerate matter, which results in significant Pauli blocking of scattering interactions when the dark mater is light, $m_\chi\lesssim1\GeV$, suppressing the capture rate by several orders of magnitude. By including the radial dependence of the chemical potential in our calculations, we correctly account for Pauli suppression at any NS layer. However, note that the chemical potential is dependent on the EoS assumption.

For very large DM masses, $m_\chi\gtrsim10^6\GeV$, the energy lost in a single collision is less than the DM initial energy. In this regime, a single scattering interactions is insufficient for capture and hence further scattering must be considered. To correctly compute the DM capture probability due to multiple scattering, we have derived, for the first time, an exact equation for the DM interaction rate in degenerate matter, and used that result to compute the differential capture rate as a function of the DM energy loss. This enables us to compute the cumulative probability that a DM particle is captured after multiple interactions, averaging over the initial DM velocity distribution.

Our framework correctly incorporates the NS opacity in  the DM capture probability. For  $m_\chi\lesssim10^6\GeV$, only one scattering is required. Then, to avoid double-counting interactions, it is sufficient to modify the differential capture rate by including an optical depth factor, calculated along all possible DM trajectories within the star. 
On the other hand, for  $m_\chi\gtrsim10^6\GeV$, the star opacity and multiple scattering, both of which are non-linear phenomena, must be treated at the same time to ensure a consistent calculation. As a result, the proper optical factor includes the capture probability for an arbitrary number of scatterings. 

In addition to obtaining an expression to numerically calculate the capture rate for any arbitrary differential cross section, parametrized  in terms of the Mandelstam variables $s$ and $t$, we also derived simplified expressions, valid at large DM mass, for differential cross sections that depend only on $t$ and not on $s$. These approximations greatly improve the computation speed.

Finally, although we have framed our results in terms of the scattering of DM from neutron targets, it is straightforward to obtain the capture rate for DM scattering from any other NS species, simply just by replacing the neutron number density and chemical potential radial profiles with those of the relevant target. Moreover, our framework can be applied to DM capture in other compact objects.

%\bigskip
\bigskip

\noindent{\it Note added: Ref~\citep{Joglekar:2020} appeared during the final stages of preparation of our manuscript. This paper has some similarities with the material we present here. Specifically, both calculations use a relativistic description of the scattering, but differ in other aspects.}

\section*{Acknowledgements}
NFB and SR were supported by the Australian Research Council and MV by the Commonwealth of Australia.

%\newpage
\bigskip

\appendix
%%%%%%%%%%%%%%%%%%%%%%%%%%%%%%%%%%%%%%%%%%
%%%%%%%%%%%%%%%%%%%%%%%%%%%%%%%%%%%%%%%%%%
%%%%%%%%%%%%%%%%%%%%%%%%%%%%%%%%%%%%%%%%%%
%%%%%%%%%%%%%%%%%%%%%%%%%%%%%%%%%%%%%%%%%%

\section{Interaction rate in the optically thin limit}
\label{sec:capratethin}
%%%%%%%%%%%%%%%%%%%%%%%%%%%%%%%%%%%%%%%%%%
%%%%%%%%%%%%%%%%%%%%%%%%%%%%%%%%%%%%%%%%%%

%%%%%%%%%%%%%%%%%%%%%%%%
\subsection{Non-degenerate weak field limit}
\label{sec:weakfieldlimit}
%%%%%%%%%%%%%%%%%%%%%%%%

When setting up the centre  of mass energy interval in section~\ref{sec:intratenumeric}, we have set the DM energy to 0 at infinity. This means that when taking the classical non-relativistic limit, the interaction rate would approach 
\begin{eqnarray}
\Omega^{-}(r) &\rightarrow& n_n(r) v_{esc}(r) \sigma, \label{eq:OmegaClass}
\end{eqnarray}
in the simple case of constant cross section. Taking Eq.~\ref{eq:omegampauli}, one can first strip out the Pauli blocking term $(1-\fFD)$, and then the integration in $t$ and $s$ can be performed analytically. Then, first taking the limit $u_n\rightarrow0, \quad E_n=m_n/\sqrt{1-u_n^2}$, and then the weak field approximation $B(r)\rightarrow1-v_{esc}^2(r)$, for a constant cross section $\frac{d\sigma}{d\cos\theta}=\frac{\sigma}{2}$, we find 
\begin{eqnarray}
\Omega^{-}(r) &\rightarrow& 
m_n^2 \frac{\sigma}{2}\frac{2u_n v_{esc}(r) \fFD(E_n,r)}{\pi^2}dE_n = 
m_n^3 \frac{\sigma}{2} \frac{2u_n v_{esc}(r)  \fFD(E_n,r)}{\pi^2}u_n du_n  \\ &=& m_n^3 \frac{\sigma}{2}\frac{v_{esc}(r)  \fFD(E_n,r)}{2\pi^3} d^3u_n 
= \frac{\sigma}{2} \frac{v_{esc}(r)  \fFD(E_n,r)}{2\pi^3} d^3p. 
\end{eqnarray}
Cases with $\sigma\propto t^n$ give similar results.
Recall that
\begin{equation}
    \frac{2\fFD(E_n)}{(2\pi)^3} d^3p,
\end{equation}
is the number density of neutron states. Then, following expression in \ref{eq:MBtoFD} we substitute it with the classical number density $n_n(r)$, to obtain the expected classical limit given by Eq.~\ref{eq:OmegaClass}.

%%%%%%%%%%%%%%%%%%%%%%%%%%%%%%%%%%%%%%%%%%
\subsection{Intermediate DM mass range}
\label{sec:capratesimple}
%%%%%%%%%%%%%%%%%%%%%%%%%%%%%%%%%%%%%%%%%%
The interaction rate in Eq.~\ref{eq:omegampaulitext} can be rewritten in terms of the DM momentum $p_\chi$, such that
\begin{equation} 
\Omega^{-}(r) = \frac{\zeta(r)}{32\pi^3}\int dt dE_n ds |\overline{M}|^2 \frac{E_n}{2s\beta(s)-\gamma^2(s)} \frac{1}{p_\chi}\frac{s}{\gamma(s)}\fFD(E_n,r)(1-\fFD(E_n^{'},r)),
\end{equation}
where we have also used Eq.~\ref{eq:difxsec}.  Assuming  that the squared matrix element depends only on $t$, i.e. $|\overline{M}|^2 = \bar{g}(s) t^n$, 
we can straightforwardly  perform the integral over $t$, 
\begin{equation}
\Omega^{-}(r) = \frac{\zeta(r)}{32\pi^3}\int dE_n ds \bar{g}(s) \frac{E_n\gamma(s)}{2s\beta(s)-\gamma^2(s)} \frac{1}{p_\chi}\fFD(E_n,r)(1-\fFD(E_n^{'},r)) \frac{1}{n+1}\left(\frac{\gamma^2(s)}{s}\right)^n.
\end{equation}

We now assume that either $\mu\gg1$ or $\mu\ll1$. In both cases, the integration range for $s$ shrinks to $[s_0-\delta s, s_0+\delta s]$, with $\delta s \ll s_0$, and the following simplifications can be made;
\begin{eqnarray}
s_0 &=& m_n^2+m_\chi^2 + 2\frac{E_n m_\chi}{\sqrt{B(r)}} = m_n^2+m_\chi^2 + 2E_n E_\chi,\\
\delta s &=& 2\sqrt{\frac{1-B(r)}{B(r)}}m_\chi\sqrt{E_n^2-m_n^2} = 2p_\chi\sqrt{E_n^2-m_n^2}, \\ 
\frac{\gamma(s)}{2s\beta(s)-\gamma^2(s)} &\rightarrow& \frac{\sqrt{1-B(r)}}{2\left(m_n^2+m_\chi^2\right)} =\frac{p_\chi}{2E_\chi\left(m_n^2+m_\chi^2\right)},\\
\frac{\gamma^2(s)}{s} &\rightarrow& \frac{4(1-B(r))m_\chi^2}{B(r)\left(1+\mu^2\right)} =\frac{4p_\chi^2}{1+\mu^2}. 
\end{eqnarray}
If $g(s)$ is regular in $s_0$, we can estimate the integral in $s$ to be $\bar{g}(s_0) 2\delta s$, approximating the integrand as constant in that range, which gives
\begin{align}
\Omega^{-}(r) \sim \zeta(r)\frac{\bar{g}(s_0)}{16\pi^3} \frac{\sqrt{E_\chi^2-m_\chi^2}}{E_\chi\left(m_n^2+m_\chi^2\right)}  \frac{\left[\frac{4\left(E_\chi^2-m_\chi^2\right)}{1+\mu^2}\right]^n}{n+1}\int dE_n E_n \sqrt{E_n^2-m_n^2} \nonumber \\
\times \fFD(E_n,r)(1-\fFD(E_n^{'},r)). 
\end{align}
To perform the integral in $E_n$, we have to potentially deal with Pauli blocking. However, for $\mu\gg1$, Pauli blocking is not effective and we can drop the $1-\fFD$ term to obtain
\begin{eqnarray}
 \int_{m_n}^{m_n+\mu_{F,n}(r)} dE_n E_n \sqrt{E_n^2-m_n^2}  \fFD(E_n,r) = \frac{\left[\mu_{F,n}(r)(2m_n+\mu_{F,n}(r))\right]^{3/2}}{3} = \pi^2 n_{free}(r).
\end{eqnarray}
This, together with $\zeta(r)$, result in an overall factor of $\pi^2 n_n(r)$, leaving 
\begin{eqnarray}
\Omega^{-}(r) &\sim& \frac{n_n(r)}{16\pi} \frac{\sqrt{E_\chi^2-m_\chi^2}}{   m_\chi^2 E_\chi} \frac{\bar{g}(s_0)}{n+1} 
\left[\frac{4(1-B(r))m_\chi^2}{B(r) (1+\mu^2)}\right]^n,   
\end{eqnarray}
and the capture rate reads, 
\begin{equation}
C \sim \frac{1}{4 \vstar} \frac{\rho_\chi}{m_\chi^3}  {\rm Erf}\left(\sqrt{\frac{3}{2}}\frac{\vstar}{v_d}\right)\int_0^{\Rstar}  r^2 dr \, n_n(r)  \frac{1-B(r)}{B(r)} \frac{\bar{g}(s_0)}{n+1} \left[\frac{4(1-B(r))m_\chi^2}{B(r) (1+\mu^2)}\right]^n.    
\end{equation}
We can now rewrite these expressions in terms of the cross section which has been averaged over $s$, 
\begin{eqnarray}
\langle\sigma(r)\rangle=\left\langle\int dt \frac{d\sigma}{dt} \right\rangle_s &=& \frac{1}{2\delta s}\int_{s_0-\delta s}^{s_0+\delta s} ds \int dt \frac{d\sigma}{dt} = \frac{1}{64\pi m_\chi^2 m_n^2} \frac{B(r)}{(1-B(r))} \bar{g}(s_0) \int dt \, t^n \\
&=& \frac{1}{64\pi m_\chi^2 m_n^2} \frac{B(r)}{(1-B(r))} \frac{\bar{g}(s_0)}{(n+1)} \left[\frac{4(1-B(r))m_\chi^2}{B(r)(1+\mu^2)}\right]^{n+1}\\
&=& \frac{1}{16\pi \left(m_n^2+m_\chi^2\right)}  \frac{\bar{g}(s_0)}{(n+1)} \left[\frac{4(1-B(r))m_\chi^2}{B(r)(1+\mu^2)}\right]^{n}, 
\end{eqnarray}
which leads to, 
\begin{eqnarray}
\Omega^{-}(r) &\sim& n_n(r) \langle\sigma(r)\rangle \frac{\sqrt{E_\chi^2-m_\chi^2}}{E_\chi}, \\
C &\sim& \frac{4\pi}{\vstar} \frac{\rho_\chi}{m_\chi}  {\rm Erf}\left(\sqrt{\frac{3}{2}}\frac{\vstar}{v_d}\right)\int_0^{\Rstar}  r^2 dr \, n_n(r)  \frac{1-B(r)}{B(r)} \langle\sigma(r)\rangle\label{eq:csimplelargem}. 
\end{eqnarray}
From Eq.~\ref{eq:csimplelargem}, we can identify the typical $1/m_\chi$ scaling. This equation also looks very similar to the non-relativistic case, with $1-B(r)$ playing the role of the escape velocity $v_{esc}^2(r)$ and $1/B(r)$ being a relativistic correction.

%%%%%%%%%%%%%%%%%%%%%%%%%%%%%%%%%%%%%%%%%%
%%%%%%%%%%%%%%%%%%%%%%%%%%%%%%%%%%%%%%%%%%
\section{Interaction rate including Pauli Blocking}
\label{sec:intrate}
%%%%%%%%%%%%%%%%%%%%%%%%%%%%%%%%%%%%%%%%%%
%%%%%%%%%%%%%%%%%%%%%%%%%%%%%%%%%%%%%%%%%%

The DM scattering rate is defined in ref.~\citep{Bertoni:2013bsa} as
\begin{eqnarray}
\Gamma = 2\int \frac{d^3k^{'}}{(2\pi)^3}\int \frac{d^3p}{(2\pi)^3}\int \frac{d^3p^{'}}{(2\pi)^3} \frac{|\overline{M}|^2}{(2E_\chi)(2E^{'}_\chi)(2E_n)(2E^{'}_n)}(2\pi)^4\delta^4\left(k_\mu+p_\mu-k_\mu^{'}-p_\mu^{'}\right) \nonumber \\
\times \fFD(E_n)(1-\fFD(E^{'}_n))
\end{eqnarray}
where $k^\mu=(E_\chi,\vec{k})$, $k^{'\mu}=(E^{'}_\chi,\vec{k'})$ are the DM initial and final momenta, and $p^\mu=(E_n,\vec{p})$,
$p^{'\mu}=(E^{'}_n,\vec{p'})$ are the target particle initial and final momenta.
We will now  calculate this rate analytically, making no approximations for as long as possible. Following refs.~\citep{Bertoni:2013bsa,Reddy:1997yr}, we write this in terms of the neutron response function, $S(q_0, q)$\footnote{Note that we will factorise some terms differently, and multiply and divide by some masses to keep the same energy dimension for each terms.}
\begin{eqnarray}
\Gamma &=& \int \frac{d^3k^{'}}{(2\pi)^3} \frac{|\overline{M}|^2}{(2E_\chi)(2E^{'}_\chi)(2m_n)(2m_n)}S(q_0,q), \label{eq:intrate}\\
S(q_0,q) &=& 2\int \frac{d^3p}{(2\pi)^3}\int \frac{d^3p^{'}}{(2\pi)^3} \frac{m_n^2}{E_n E^{'}_n} (2\pi)^4\delta^4\left(k_\mu+p_\mu-k_\mu^{'}-p_\mu^{'}\right)\fFD(E_n)(1-\fFD(E^{'}_n)),
\end{eqnarray}
where we have assumed that $|\overline{M}|^2$ can be written only in terms of $k$ and $k^{'}$, i.e. $|\overline{M}|^2\propto t^n$, where $t=(k^\mu-k^{'\mu})^2$ is the Mandelstam variable.
A few $\Theta$ functions are missing in the amplitude, namely $\Theta(E^{'}_\chi)\Theta(E_n)\Theta(E^{'}_n)$. We can actually demand that such energies should not only be positive, but also higher than their respective masses, so we multiply Eq.~\ref{eq:intrate} by 
 $\Theta(E^{'}_\chi-m_\chi)\Theta(E_n-m_n)\Theta(E^{'}_n-m_n)$.
 
In addition, as we are only interested in exothermic scattering, i.e. $q_0>0$, we also include a $\Theta(q_0)$ factor. The scattering rate and response function then become
\begin{align}
\Gamma &= \int \frac{d^3k^{'}}{(2\pi)^3} \frac{|\overline{M}|^2}{(2E_\chi)(2E^{'}_\chi)(2m_n)(2m_n)}\Theta(E^{'}_\chi-m_\chi)\Theta(q_0)S(q_0,q),\\
S(q_0,q) &= \frac{1}{2\pi^2}\int d^3p \frac{m_n^2}{E_n E^{'}_n} \delta\left(q_0+E_n-E^{'}_n\right)\fFD(E_n)(1-\fFD(E^{'}_n))\Theta(E_n-m_n)\Theta(E^{'}_n-m_n),     
\end{align}
where we have integrated over $d^3p^{'}$. 
After that  $E^{'}_n$ is fixed to
\begin{equation}
E^{'}_n(E_n,q,\theta) = \sqrt{m_n^2+(\vec{p}+\vec{q})^2} = \sqrt{E_n^2+q^2+2qp\cos\theta} > m_n, \quad\forall p,q,\theta, |\cos\theta|<1,
\end{equation} 
where $\theta$ is the angle between $\vec{p}$ and $\vec{q}$. The integral over $d^3p$ can be performed by changing to $d^3p = 2\pi p E_n\,dE_n\,d\cos\theta$ and, following refs.~\citep{Bertoni:2013bsa,Reddy:1997yr}, using the delta function to integrate over $\theta$. However, as noted in ref.~\citep{Reddy:1997yr}, we should remember that this gives rise to a $\Theta$ function, namely $\Theta(1-\cos^2\theta)$. First, we calculate the derivative of the argument of the delta function
\begin{equation}
\left|\frac{d}{d\cos\theta}\left(q_0+E_n-E^{'}_n(E_n,q,\theta)\right)\right|=\left|\frac{dE^{'}_n}{d\cos\theta}(E_n,q,\theta)\right| = \frac{qp}{E^{'}_n}, 
\end{equation}
and then integrate over $\theta$ to obtain
\begin{eqnarray}
S(q_0,q) &=& \frac{m_n^2}{\pi q}\int dE_n  \fFD(E_n)(1-\fFD(E_n+q_0))\Theta(E_n)  \Theta(1-\cos^2\theta(q,q_0,E_n)).
\end{eqnarray}
Using 
\begin{equation}
\cos\theta(q,q_0,E_n) =  \frac{q_0^2-q^2+2E_nq_0}{2q\sqrt{E_n^2-m_n^2}},
\end{equation}
we can determine the integration interval for $E_n$. 
For the $q^2>q_0^2$ case, $q^\mu$ is expected to be space-like, $t=q_\mu q^\mu <0$, and the response function becomes
\begin{eqnarray}
S^{-}(q_0,q) &=& \frac{m_n^2}{\pi q}\int_{E_n^{\, t^{-}}}^\infty dE_n  \fFD(E_n)(1-\fFD(E_n+q_0)),
\end{eqnarray}
where $E_n^{t^-}$ is the minimum energy of the neutron before the collision, which is obtained from kinematics and given by 
\begin{equation}
E_n^{\, t^{-}} = -\left(m_n+\frac{q_0}{2}\right) + \sqrt{\left(m_n+\frac{q_0}{2}\right)^2+\left(\frac{\sqrt{q^2-q_0^2}}{2}-\frac{m_n q_0}{\sqrt{q^2-q_0^2}}\right)^2}. 
\end{equation} 
For  the $t>0$ case, we instead have
\begin{eqnarray}
S^{+}(q_0,q) &=& \frac{m_n^2}{\pi q}\int_0^{E_n^{\, t^{+}}} dE_n  \fFD(E_n)(1-\fFD(E_n+q_0)), 
\end{eqnarray}
where
\begin{eqnarray}
E_n^{\, t^{+}} &=& -\left(m_n+\frac{q_0}{2}\right) + \sqrt{\left(m_n+\frac{q_0}{2}\right)^2-\left(\frac{\sqrt{q_0^2-q^2}}{2}+\frac{m_n q_0}{\sqrt{q_0^2-q^2}}\right)^2}. 
\end{eqnarray}
We note that
\begin{eqnarray}
1-\fFD(E_n+q_0) = \fFD(-E_n-q_0), 
\end{eqnarray}
and use the following result for FD integrals,  
\begin{eqnarray}
F(x,z) = \int dx \fFD(x)\fFD(-x-z) = \frac{e^z \left[\log \left(e^{x+z}+1\right)-\log \left(e^x+1\right)\right]}{e^z-1}.
\end{eqnarray}
In addition, for finite non-zero values of $E_n$, we can identify 3 distinct regimes, 
\begin{eqnarray}
&E_n&>\mu_{F,n},\\
\mu_{F,n}-q_0<&E_n&<\mu_{F,n},\\
&E_n&<\mu_{F,n}-q_0.
\end{eqnarray}
To calculate the response function in the $T\rightarrow0$ limit, we use the following results for each of the above $E_n$ intervals, 
\begin{eqnarray}
\lim_{T\rightarrow0} T F(E_n/T,q_0/T) &=& q_0, \quad E_n>\mu_{F,n},\\
\lim_{T\rightarrow0} T F(E_n/T,q_0/T) &=& E_n+q_0-\mu_{F,n}, \quad \mu_{F,n}-q_0<E_n<\mu_{F,n},\\
\lim_{T\rightarrow0} T F(E_n/T,q_0/T) &=& 0, \quad E_n<\mu_{F,n}-q_0. 
\end{eqnarray}
The expressions above look like a step function with a smooth passage from $0$ to $q_0$ for $E_n\in[\mu_{F,n}-q_0,\mu_{F,n}]$. Note that the middle case was absent in refs.~\citep{Bertoni:2013bsa,Reddy:1997yr}. Then, we can define
\begin{eqnarray}
\lim_{T\rightarrow0} T F(E_n/T,q_0/T) &=& q_0 \,  g\left(\frac{E_n-\mu_{F,n}}{q_0}\right),
\end{eqnarray}
where $g(x)=1$ for $x>0$, $g(x)=0$ for $x<-1$, and $g(x)=1+x$ for $-1<x<0$, giving the response function in the $T\rightarrow0$ limit as
\begin{eqnarray}
S^{-}(q_0,q) &=&  \frac{m_n^2q_0}{\pi q}\left[1-g\left(\frac{E_n^{\,t^{-}}-\mu_{F,n}}{q_0}\right)\right]=\frac{m_n^2q_0}{\pi q}h\left(\frac{E_n^{\,t^{-}}-\mu_{F,n}}{q_0}\right),\label{eq:Sminus}\\
S^{+}(q_0,q) &=&  \frac{m_n^2q_0}{\pi q}\left[g\left(\frac{E_n^{\,t^{+}}-\mu_{F,n}}{q_0}\right)-1\right]=-\frac{m_n^2q_0}{\pi q}h\left(\frac{E_n^{t^{+}}-\mu_{F,n}}{q_0}\right),
\end{eqnarray}
where $h(x)=1-g(x)$ is a function such that $h(x)=1$ for $x<-1$, $h(x)=0$ for $x>0$ and $h(x)=-x$ for $-1<x<0$. Note that $S^{-}\ge0$, while $S^{+}\le0$. We drop the discussion of $S^{+}$ from now on, as it is not required for elastic scattering\footnote{It would be necessary for inelastic scattering, for example.}. Comparing to refs.~\citep{Bertoni:2013bsa,Reddy:1997yr}, our result has a factor $h(x)$, which encodes the smooth transition, while they instead use $\Theta(x)$.

Returning to the scattering rate expression 
\begin{eqnarray}
\Gamma^- 
&=& \int \frac{d\cos\theta k^{'2}dk^{'}}{64\pi^2E_\chi E^{'}_\chi m_n^2} |\overline{M}|^2 \Theta(E_\chi-q_0-m_\chi)\Theta(q_0)S^-(q_0,q), 
\end{eqnarray}
we change variables from $k^{'},\cos\theta$ to $q_0,q$, 
\begin{eqnarray}
q_0 &=& E_\chi-\sqrt{k^{'2}+m_\chi^2},\label{eq:q0}\\
q^2 &=& k^2+k^{'2}-2k k^{'}\cos\theta, \label{eq:q}
\end{eqnarray}
and plug in the result for $S^-$, Eq.~\ref{eq:Sminus}, to obtain, 
\begin{eqnarray}
\Gamma^{-} 
&=& \frac{1}{64\pi^3E_\chi k }\int dq q_0dq_0 |\overline{M}|^2  h\left(\frac{E_n^{\, t^{-}}-\mu_{F,n}}{q_0}\right) \Theta(E_\chi-q_0-m_\chi)\Theta(q_0). 
\end{eqnarray}
To simplify the integration for $t$-dependent matrix elements, we define $t_E=-t=q^2-q_0^2$, then 
\begin{eqnarray}
\Gamma^{-} 
&=&\frac{1}{2^7\pi^3E_\chi k }\int_0^{E_\chi-m_\chi}q_0 dq_0 \int \frac{dt_E }{\sqrt{q_0^2+t_E}} |\overline{M}|^2  h\left(\frac{E_n^{\,t^{-}}-\mu_{F,n}}{q_0}\right).
\end{eqnarray}
Next, we will assume that $|\overline{M}|^2\propto t_E^n$, with $n=0,1,2$.
\begin{eqnarray}
\Gamma^{-}(E_\chi) &\propto& \frac{1}{2^7\pi^3E_\chi k }\int_0^{E_\chi-m_\chi}q_0 dq_0 \int \frac{t_E^n dt_E }{\sqrt{q_0^2+t_E}}  h\left(\frac{E_n^{\,t^{-}}-\mu_{F,n}}{q_0}\right).
\end{eqnarray}
To find  the integration interval for $t_E$ from the range of $\cos\theta\in[-1,1]$, we combine Eqs.~\ref{eq:q0}, \ref{eq:q} and the expression for $t_E$, obtaining a result in terms of $q_0$, $E_\chi$ and $m_\chi$, 
\begin{equation}
    t_E^\pm = 2 \left[ E_\chi(E_\chi-q_0)-m_\chi^2\pm k\sqrt{(E_\chi-q_0)^2-m_\chi^2}\right].
\end{equation}
These roots need to be compared with the ranges of the $h(x)$ function above. Therefore, there are three possible intervals, one for  $h(x)=0$, another for  $h(x)=x$ and the remaining one for  $h(x)=1$. 

To be in the case $h(x)=0$, we require $E_n^{\,t^{-}}-\mu_{F,n}>0$, such that
\begin{eqnarray}
\frac{m_n^2q_0^2}{t_E}+\frac{t_E}{4}-\mu_{F,n}(q_0+\mu_{F,n})-m_n(q_0+2\mu_{F,n})&>&0.\label{eq:tEineq}
\end{eqnarray}
As $t_E>0$, this is true for values of $t_E$ that are not between the two roots of Eq.~\ref{eq:tEineq}, which we denote $t_{\mu^{+}}^\pm$, 
\begin{equation}
t_{\mu^{+}}^\pm = 2\left[\mu_{F,n}(\mu_{F,n}+q_0)+m_n(2\mu_{F,n}+q_0)  \pm \sqrt{\left(\mu_{F,n}(\mu_{F,n}+q_0)+m_n(2\mu_{F,n}+q_0)\right)^2-m_n^2q_0^2}\right].
\end{equation}

For $0<h(x)<1$, we require both
\begin{eqnarray}
E_n^{\,t^{-}}-\mu_{F,n}&<&0,\\
E_n^{\,t^{-}}-\mu_{F,n}+q_0&>&0.
\end{eqnarray}
In this case, we need to consider values in between $t_{\mu^{+}}^\pm$ but not in between $t_{\mu^{-}}^\pm$, where $t_{\mu^{-}}^\pm$ is obtained directly from $t_{\mu^{+}}^\pm$ by substituting $\mu_{F,n}$ with $\mu_{F,n}-q_0$, 
\begin{equation}
t_{\mu^{-}}^\pm =  2\left[\mu_{F,n}(\mu_{F,n}-q_0)+m_n(2\mu_{F,n}-q_0)\pm\sqrt{\left(\mu_{F,n}(\mu_{F,n}-q_0)+m_n(2\mu_{F,n}-q_0)\right)^2-m_n^2q_0^2}\right].
\end{equation}
These four roots are, for $0\leq q_0<\mu_{F,n}$, always in the order
$t_{\mu^{+}}^{+} \ge t_{\mu^{-}}^{+} \ge t_{\mu^{-}}^{-} \ge t_{\mu^{+}}^{-} \ge 0$. 
For $\mu_{F,n}<q_0<2m_n+\mu_{F,n}$  the $t_{\mu^-}^\pm$ roots do not exist, and for $q_0> 2m_n+\mu_{F,n}$ they become negative, and the order of the remaining roots is $ t_{\mu^{+}}^{+} \ge t_{\mu^{+}}^{-} \ge 0$. 

The final case $h(x)=1$ can be obtained as the complementary to the first two as  $t_{\mu^{-}}^{+} \ge t_E \ge t_{\mu^{-}}^{-}$, 
which means that this case does not exist when the $t_{\mu^-}^\pm$ do not exist or are negative. 

We define an operator that encodes the $t_E$ integral over the aforementioned intervals, 
\begin{align}
    \mathcal{I}(\tilde{f}(t),t_1^+,t_2^+,t_1^-,t_2^-) =& \sum_{i=1,2}\sum_{j=1,2} \left(F(t_i^+)-F(t_j^-)\right)\Theta\left(t_{3-i}^+-t_i^+\right)\Theta\left(t_i^+-t_j^-\right) \nonumber\\
    &\times\Theta\left(t_j^--t_{3-j}^-\right),\\
    F(t) =& \int dt \, \tilde{f}(t), 
\end{align}
and then rewrite $\Gamma^{-}$ as
\begin{align}
\Gamma^{-}(E_\chi) \propto \,& \frac{1}{2^7\pi^3E_\chi k } \left[\int_0^{E_\chi-m_\chi}q_0 dq_0 \,\, \mathcal{I}\left(\frac{t_E^n  }{\sqrt{q_0^2+t_E}},t_E^+,t_{\mu^-}^+,t_E^-,t_{\mu^-}^-\right)\Theta(\mu_{F,n}-q_0)\right. \nonumber\\
&- \int_0^{E_\chi-m_\chi} dq_0 \,\, \mathcal{I}\left(\frac{t_E^n  }{\sqrt{q_0^2+t_E}} \left(E_n^{\,t^{-}}-\mu_{F,n}\right),t_E^+,t_{\mu^+}^+,t_E^-,t_{\mu^-}^+\right)\Theta(\mu_{F,n}-q_0)  \nonumber\\
&- \int_0^{E_\chi-m_\chi} dq_0 \,\, \mathcal{I}\left(\frac{t_E^n  }{\sqrt{q_0^2+t_E}} \left(E_n^{\,t^{-}}-\mu_{F,n}\right),t_E^+,t_{\mu^-}^-,t_E^-,t_{\mu^+}^-\right)\Theta(\mu_{F,n}-q_0)  \nonumber\\
&\left. - \int_0^{E_\chi-m_\chi} dq_0 \,\, \mathcal{I}\left(\frac{t_E^n  }{\sqrt{q_0^2+t_E}} \left(E_n^{\,t^{-}}-\mu_{F,n}\right),t_E^+,t_{\mu^+}^+,t_E^-,t_{\mu^+}^-\right)\Theta(q_0-\mu_{F,n}) \right] .\label{eq:gammafinal}
\end{align}
Note that all these contributions are positive as $\left(E_n^{\,t^{-}}-\mu_{F,n}\right)$ is negative in the given integration ranges.
The value of the primitives of these functions are listed below for $n=0,1,2$, 
\begin{eqnarray}
&&\int \frac{t_E^n dt_E }{\sqrt{q_0^2+t_E}} = 2D_n(q_0^2,t_E)\sqrt{q_0^2+t_E},\label{eq:Dn} \\
&&\int \frac{t_E^n dt_E }{\sqrt{q_0^2+t_E}}\left(E_n^{\,t^{-}}-\mu_{F,n}\right) = R_n(q_0^2,t_E),
\end{eqnarray}
\begin{eqnarray}
D_0(x,y) &=&1,\\
D_1(x,y) &=&\frac{y-2x}{3},\\
D_2(x,y) &=&\frac{3y^2-4xy+8x^2}{15},\\
R_n(x,y) &=& \frac{\tilde{H}_n(x,y)}{2\sqrt{x+y}}+(-1)^{n+1}\Gamma(n+1)m_n^{2(1+n)}\log\left(x+y\right) \nonumber \\ 
&&+(-1)^n\Gamma(n+1)m_n^{2(1+n)}\log \tilde{g}(x,y),\\
\tilde{H}_0(x,y) &=& y \left(\sqrt{\frac{\left(4 m_n^2+y\right) (x+y)}{y}}-4 \mu_{F,n}\right)-4 m_n (x+y)-2 x^{3/2}-4 \mu_{F,n} x-2 \sqrt{x} y,\\
\tilde{H}_1(x,y) &=&\frac{1}{2} y \left(2 m_n^2+y\right) \sqrt{\frac{\left(4 m_n^2+y\right) (x+y)}{y}}-\frac{2}{3} (y-2 x) (x+y) \left(2 \mu_{F,n}+2 m_n+\sqrt{x}\right),\\
\tilde{H}_2(x,y) &=&\frac{1}{3} y \left(-6 m_n^4+m_n^2 y+y^2\right) \sqrt{\frac{\left(4 m_n^2+y\right) (x+y)}{y}}-\frac{2}{15} (x+y) \left(8 x^2-4 x y+3
   y^2\right)\nonumber\\
   &&\times\left(2 \mu_{F,n}+2 m_n+\sqrt{x}\right),\\
\tilde{g}(x,y) &=& \left(\sqrt{y \left(4 m_n^2+y\right)}+2 m_n^2+y\right) (x+y), 
\end{eqnarray}
where $\Gamma(n+1)$ is the Gamma function. 

All interaction rate spectra will have an endpoint at $q_0 = \qomax$, the maximum amount of energy that can be lost in a single interaction.
The value of $\qomax$ is shown in left panel of Fig.~\ref{fig:q0max} as a function of $B$ in the case of large DM mass ($m_\chi = 1\TeV$), for several values of $\muFn$. The endpoint can be found as the minimum between the DM kinetic energy and the root of one of the following two equations (only one of them, at most, has a positive root for each choice of $m_\chi$, $\muFn$ and $E_\chi$) 
\begin{eqnarray}
    t_E^- = t_{\mu^+}^+,\\
    t_E^+ = t_{\mu^+}^-.
\end{eqnarray}
For $m_\chi\gg m_n$, the second equation never has a solution, and the solution of the first equation is always much lower than the kinetic energy. This results in the value of $\qomax$ to become independent of $m_\chi$ in this mass range.

\begin{figure}
    \centering
    \includegraphics[width=.8\textwidth]{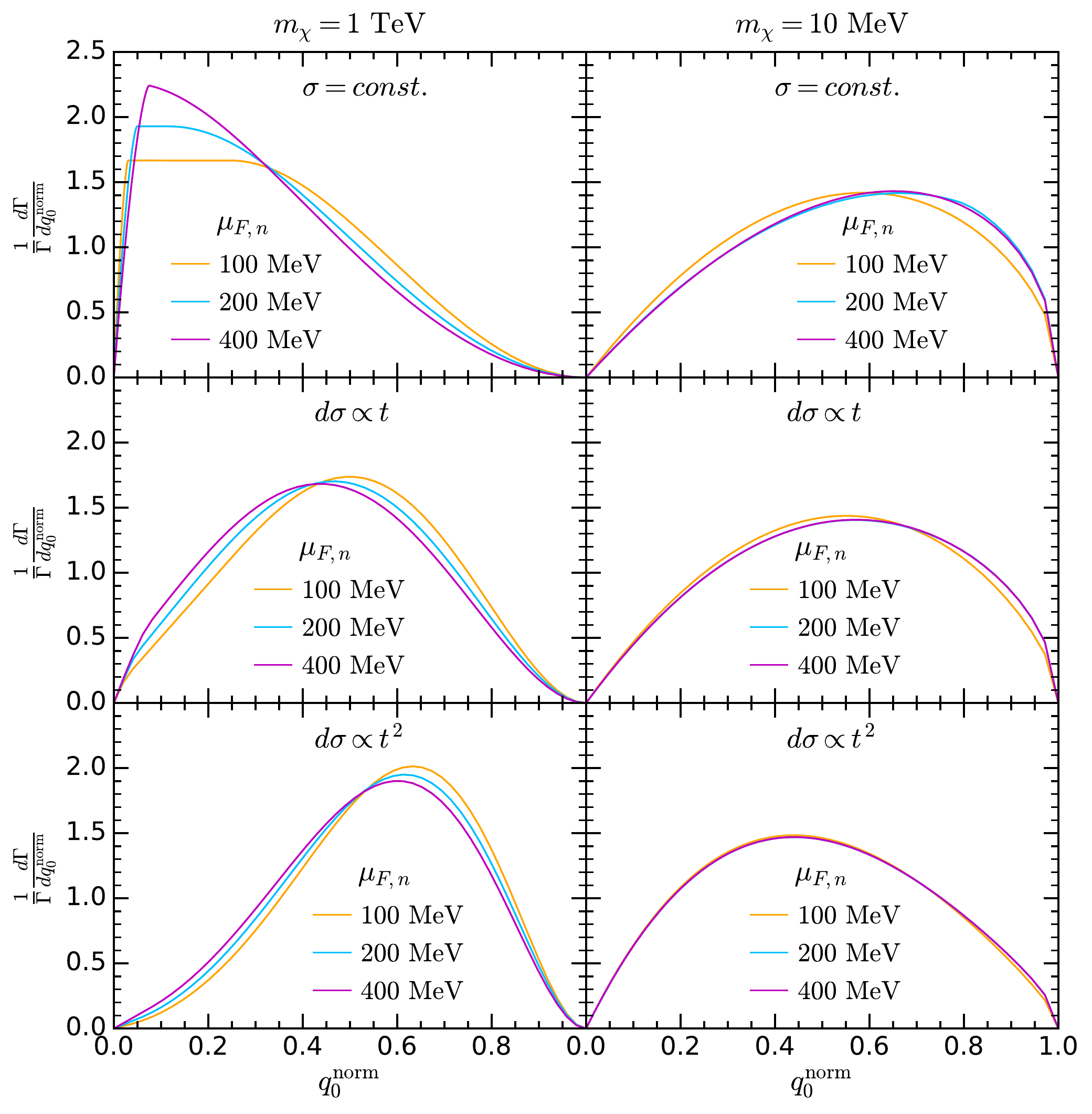}
    \caption{Normalised differential interaction rates, $\frac{1}{\Gamma}\frac{d\Gamma}{d\qonorm}$, as a function of $\qonorm$ for different values of $\muFn$, with $m_\chi=1\TeV$ (left panels) $m_\chi=10\MeV$ (right panels) and $B=0.5$. Top row: $n=0$, middle row: $n=1$, bottom row: $n=2$.  }
    \label{fig:diffgamma}
\end{figure}

The shape of the differential interaction rate  depends very weakly on $m_\chi$ and $B$ for $m_\chi\gg m_n$ and $m_\chi \ll m_n$, as seen by plotting it as a function of $\qonorm=q_0/\qomax$. Therefore, we use as a reference $m_\chi=1\TeV$ (left) and $m_\chi=10\MeV$ (right),  $B=0.5$, and show the normalised differential interaction rates in Fig.~\ref{fig:diffgamma} for $n=0,1,2$.
We observe in the left panels that for $n=0$ interaction rates are flat (or peaked, depending on $\muFn$) at low energy and suppressed at high energies, while for $n=1,2$ the profiles become peaked at higher and  higher energies. Conversely, for $m_\chi=10\MeV$ the peak of the spectrum is shifted to lower energies with increasing power of t ($d\sigma\propto t^n$).

%%%%%%%%%%%%%%%%%%%%%%%%%%%%%%%%%%%%%%%%%%
%%%%%%%%%%%%%%%%%%%%%%%%%%%%%%%%%%%%%%%%%%

%%%%%%%%%%%%%%%%%%%%%%%%%%%%%%%%%%%%%%%%%%
\section{Probability to scatter from a NS target}
\label{sec:opticalfactors}
%%%%%%%%%%%%%%%%%%%%%%%%%%%%%%%%%%%%%%%%%%
In this section, we calculate the probability for a DM particle to scatter off a neutron or any other NS species exactly N times, which we denote by $Q_N$. This is done for large DM mass, $m_\chi\gg 1\GeV$ for nucleon targets, which allows for Pauli blocking to be neglected. We will later discuss how to calculate it for low mass DM. 

\subsection{Single Scattering}
\label{sec:opticalfactorsingle}

We start from $Q_0$, the probability that a DM particle has never interacted with a neutron target within the star. This probability is $1$ until the DM particle enters the star. After that, the probability evolves as 
\begin{eqnarray}
\frac{dQ_0}{dt}(t) &=&-\Omega^{-}(t)Q_0(t),\\
Q_0(r) &=& e^{-\int_{\gamma}\Omega^{-}(r)dr\frac{d\tau'}{dr}} = e^{-\optdepth(r,\gamma,J)},
\end{eqnarray} 
where $\optdepth$ is the optical depth, $\tau'$ is the proper time, and $\gamma$ is the path followed by the DM particle. The optical depth for $|\overline{M}|^2\propto t^n$ is defined as
 \begin{eqnarray} 
\optdepth(r,\gamma,J) &=& \int_{\gamma} dx  \frac{\sigma_{surf}\zeta(x) n_{free}(r)\sqrt{1-B(x)}\left(\frac{1-B(x)}{B(x)}\right)^n}{\sqrt{1-B(x)\left(1+\frac{J^2}{m_\chi^2x^2}\right)}\left(\frac{1-B(\Rstar)}{B(\Rstar)}\right)^{n}}\label{eq:optdepth}\\
&=& \int_{\gamma} dx  \frac{\sigma_{surf}\zeta(x)n_{free}(r)\left(\frac{1-B(x)}{B(x)}\right)^n}{\sqrt{1-\frac{J^2}{J_{max}^2(x)}}\left(\frac{1-B(\Rstar)}{B(\Rstar)}\right)^{n}}, 
\end{eqnarray} 
where $\sigma_{surf}$ is the cross section on the surface of the star.  
As usual $A(r)$ factors  were reabsorbed in the neutron number density, 
and we have used that for $|\overline{M}|^2\propto t^n$,
 \begin{equation}
    \sigma(r) \left[\frac{1-B(r)}{B(r)}\right]^{-n} = const., 
\end{equation}
to define the cross section in Eq.~\ref{eq:optdepth} at a given NS radius $\sigma(r)$ as a function of $\sigma_{surf}$. 
For each value of angular momentum $J$ and position  $\hat{x}$  at a radius $r$ within the NS, there is a single orbit with two possible paths for a DM particle to reach $\hat{x}$. The shortest path, which has optical depth $\optdepth^{-}$, goes from the surface to $\hat{x}$ without passing  the perihelion. On the other hand, the longest path will have optical depth $\optdepth^+$, and  goes from the surface to the perihelion and then to $\hat{x}$ instead. These paths are depicted in Fig.~\ref{fig:orbits}. The optical depths are calculated as 
\begin{eqnarray}
\optdepth^-(r,J) &=&  \int_{\Rstar}^r dx  \frac{\sigma_{surf}n_n(r)\left(\frac{1-B(x)}{B(x)}\right)^n}{\sqrt{1-\frac{J^2}{J_{max}^2(x)}}\left(\frac{1-B(\Rstar)}{B(\Rstar)}\right)^{n}},\\
\optdepth^+(r,J) &=&  \int_{\Rstar}^{r_{min}}+\int_{r_{min}}^r dx  \frac{\sigma_{surf}n_n(r)\left(\frac{1-B(x)}{B(x)}\right)^n}{\sqrt{1-\frac{J^2}{J_{max}^2(x)}}\left(\frac{1-B(\Rstar)}{B(\Rstar)}\right)^{n}}\\
&=& \optdepth^{-}(r,J) + 2 \int_{r_{min}}^r dx  \frac{\sigma_{surf}n_n(r)\left(\frac{1-B(x)}{B(x)}\right)^n}{\sqrt{1-\frac{J^2}{J_{max}^2(x)}}\left(\frac{1-B(\Rstar)}{B(\Rstar)}\right)^{n}},
\end{eqnarray}
where we can find $r_{min}(J)$ by solving $J = J_{max}(r_{min})$, 
\begin{equation}
J = m_\chi r_{min} \sqrt{\frac{1-B(r_{min})}{B(r_{min})}}.
\end{equation} 
Substituting $J$ with $y J_{max}(r)$, 
\begin{eqnarray}
\optdepth^-(r,y) &=&  \int_r^{\Rstar} dx  \frac{\sigma_{surf}n_n(x)\left(\frac{1-B(x)}{B(x)}\right)^n}{\sqrt{1-y^2\frac{J_{max}^2(r)}{J_{max}^2(x)}}\left(\frac{1-B(\Rstar)}{B(\Rstar)}\right)^{n}},\label{eq:taum}\\
\optdepth^+(r,y) &=&  \optdepth^{-}(r_{min},y) + 2\int_{r_{min}}^r dx  \frac{\sigma_{surf}n_n(x)\left(\frac{1-B(x)}{B(x)}\right)^n}{\sqrt{1-y^2\frac{J_{max}^2(r)}{J_{max}^2(x)}}\left(\frac{1-B(\Rstar)}{B(\Rstar)}\right)^{n}}, \label{eq:taup}
\end{eqnarray}
allowing $Q_0$ to be calculated as
\begin{equation}
    Q_0(r,y) = \frac{1}{2}\left[e^{-\optdepth^{-}(r,y)}+  e^{-\optdepth^{+}(r,y)}\right].
\end{equation}
Nevertheless, recall that when calculating the capture rate in section~\ref{sec:caprate}, we integrated over $J$. Hence, we will need to average $Q_0$ over the $J$ distribution, given by 
\begin{eqnarray}
f_J(y)dy = \frac{y dy}{\sqrt{1-y^2}}.
\end{eqnarray}
The final expression for $Q_0$ is then
\begin{eqnarray}
Q_0(r) &=& \frac{1}{2}\int_0^1 \frac{y dy}{\sqrt{1-y^2}} \left[e^{-\optdepth^{-}(r,y)} + e^{-\optdepth^{+}(r,y)}\right].
\end{eqnarray}

\subsection{Multiple Scattering}
\label{sec:opticalfactormulti}

\begin{figure}
    \centering
    \includegraphics[width=.45\textwidth]{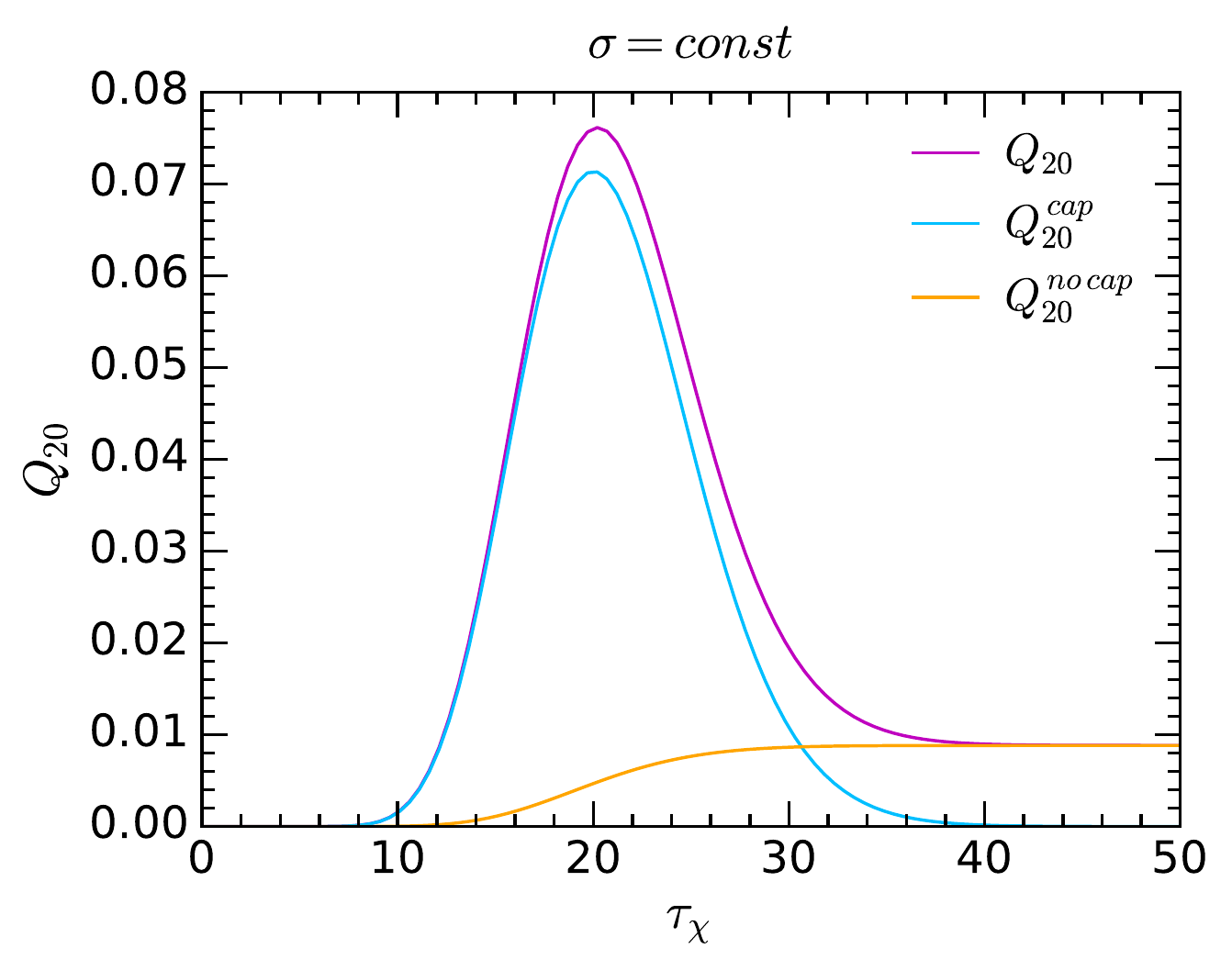}
    \caption{Probabilities $Q_{20}$ (purple), $Q_{20}^{cap}$ (blue) and  $Q_{20}^{no \, cap}$ (orange)  as a function of the optical depth, $\optdepth$, for constant cross section, and  $n^*=92$ corresponding e.g. to $m_\chi=10^8\GeV$ and $m^*=1.08\times 10^6\GeV$. }
    \label{fig:q10}
\end{figure}

Let $Q_1(r)$ be the probability that a DM particle scatters off a NS target exactly one time. We will assume that it depends on $r$ only through the optical depth $\optdepth(r)$. 
We will also assume that the total energy loss is much lower than the total energy. This assumption is always true as the DM speed far away from the NS is not relativistic. The probabilities of exactly 2 and $N$ scatterings occurring will then be
\begin{eqnarray}
 Q_2(\optdepth) &=& \int_0^{\optdepth} Q_1(\optdepth-x) f(x)dx,\\
 Q_{N}(\optdepth) &=& \int_0^{\optdepth} Q_{N-1}(\optdepth-x) f(x)dx,
\end{eqnarray}
where $f(x)$ is an unknown  probability distribution function. This problem can be solved using Laplace transforms, 
\begin{eqnarray}
 \tilde{Q}_{N}(s) &=& \tilde{Q}_{N-1}(s)\tilde{f}(s),\\
 \tilde{Q}_{N}(s) &=& \tilde{Q}_1(s) \left[\tilde{f}(s)\right]^{N-1}.
\end{eqnarray}
The cumulative probability is
\begin{eqnarray}
    \sum_{N=1}^\infty \tilde{Q}_1(s) \left[\tilde{f}(s)\right]^N = \tilde{Q}_1(s) \frac{1}{1-\tilde{f}(s)}, \label{eq:Qcummlap}
\end{eqnarray}
where we assumed $\tilde{f}(s)<1, s>0$, and $\tilde{f}(0)\le1$. The probability not to scatter for an optical depth $\optdepth$ is
\begin{eqnarray}
Q_0(\optdepth) &=& e^{-\optdepth},\\
\tilde{Q}_0(s) &=& \int_0^\infty d\optdepth e^{-\optdepth} e^{-\optdepth s} = \frac{1}{s+1}.\label{eq:Q0tilde}
\end{eqnarray}
The sum of all probabilities must be one, whose Laplace transform is $1/s$. Thus, from the sum of Eq.~\ref{eq:Q0tilde} and Eq.~\ref{eq:Qcummlap}, we obtain
\begin{eqnarray}
   \tilde{Q}_1(s) 
   &=&\frac{1-\tilde{f}(s)}{s(s+1)},\\
   \tilde{Q}_{N}(s) &=& \frac{1-\tilde{f}(s)}{s(s+1)} \left[\tilde{f}(s)\right]^{N-1}, 
\end{eqnarray}
which implies  that $Q_1(\optdepth)$ has the following form of
\begin{eqnarray}
    Q_1(\optdepth) &=& 1-e^{-\optdepth} +\int_0^{\optdepth} \left(1 - e^{-(\optdepth-x)}\right)f(x)dx.
\end{eqnarray}
For a given $f$, one can now find the single scattering probability as the inverse Laplace transform. 
The kernel $f$ needs to be of the form
\begin{equation}
    \tilde{f}(s) = \frac{A}{1+s},
\end{equation}
where $A$ is a constant.
The quantity $\tilde{f}(0) = A$ 
sets the survival probability after one scattering, i.e. the probability that after a single  scatter the particle is not ``removed'' by the medium and continues to propagate, being able scatter again.  On the other hand,  $1-\tilde{f}(0) = 1-A$ 
sets the absorption probability after one scattering, i.e. that after a single scatter the particle is ``absorbed'', which in our case means captured. This quantity should match the DM capture probability, which is 
\begin{eqnarray}
    c_1 &=&\hat{P}_1 = \frac{1}{n^*} =1 - e^{-\frac{m^*}{m_\chi}},\\
    c_1 &=& \frac{1}{n^*}\rightarrow \frac{m^*}{m_\chi},  \quad m_\chi\gg m^*,
\end{eqnarray}
where the last approximation holds if $n^*\gg1$, i.e. $\hat{P}_1\ll1$. This sets $A=1-\hat{P}_1$. 
For $m_\chi\lesssim 10^6\GeV$, the probability $\hat{P}_1\sim1$, leading to 
\begin{eqnarray}
    \tilde{Q}_1(s) &=& \frac{1}{s(s+1)},\\
    \tilde{Q}_N(s) &=& 0, \quad N\ge 2,\\
    Q_1(\optdepth) &=& 1-e^{-\optdepth},\\
    Q_N(\optdepth) &=& 0,\quad N\ge 2,\\ 
    \hat{Q}_1 &=& \frac{dQ_1}{d\optdepth} = e^{-\optdepth},\\
    \eta(\optdepth)=\sum_{N=1}^\infty \hat{Q}_N &=& \hat{Q}_1 = e^{-\optdepth} = Q_0(\optdepth),\quad m_\chi \lesssim 10^6\GeV. \label{eq:etalowm}
\end{eqnarray}
For $m_\chi\gtrsim 10^6\GeV$ (nucleon targets), $c_1=\hat{P}_1<1$, thus $\tilde{f}(s)\neq 0$. Note that, as $\hat{P}_1=c_1$ depends on $B$ and $\muFn$, these probabilities depend on the position directly, and not only through $\optdepth$, so this somehow invalidates our initial hypothesis. However, as long as $c_1$, i.e. $m^*$, does not vary significantly throughout the star, we can assume our  hypothesis is true locally, and continue with our approach. If this were not the case, the only possible approach would be to solve the Boltzmann transport equation.
Setting
\begin{equation}
    \tilde{f}(s) = \frac{1-\frac{1}{n^{*}}}{1+s}, 
\end{equation}
we obtain 
\begin{eqnarray}
    \tilde{Q}_N(s) 
    &=& \left(1-\frac{1}{n^{*}}\right)^{N-1}\left[\left(1-\frac{1}{n^*}\right)\frac{1}{(1+s)^{N+1}}+\frac{1}{n^{*}}\frac{1}{s(1+s)^{N}}\right],\\
    Q_N(\optdepth) &=& \left(1-\frac{1}{n^{*}}\right)^{N-1}\left[\left(1-\frac{1}{n^*}\right)\frac{\optdepth^{N}}{N!}e^{-\optdepth}+\frac{1}{n^{*}}\left(1-\frac{\Gamma(N,\optdepth)}{(N-1)!}\right)\right],\label{eq:qn}
\end{eqnarray}
where $\Gamma(N,\optdepth)$ is the incomplete Gamma function. 
We can distinguish two probabilities in Eq.~\ref{eq:qn}, the probability of DM to scatter $N$ times and be captured,
\begin{equation}
    Q_N^{cap}(\optdepth) = \left(1-\frac{1}{n^{*}}\right)^{N-1}\frac{1}{n^{*}}\left(1-\frac{\Gamma(N,\optdepth)}{(N-1)!}\right),
\end{equation}
and the probability to scatter $N$ times and not be captured
\begin{eqnarray}
    Q_N^{no \, cap}(\optdepth) &=& \left(1-\frac{1}{n^{*}}\right)^{N}\frac{\optdepth^{N}}{N!}e^{-\optdepth}.
\end{eqnarray}
In Fig.~\ref{fig:q10}, we show  $Q_N$, $Q_N^{cap}$  and $Q_N^{no \, cap}$ for constant cross section and  $N=20$. It is clear that there is a transition, where the probability of not being captured becomes relevant,  that occurs around $\optdepth\sim N$.

Note that the absorption probability after N scatterings is
\begin{eqnarray}
    Q_N(\infty) &=& Q_N^{cap}(\infty) = \frac{1}{n^{*}} \left(1-\frac{1}{n^{*}}\right)^{N-1},
\end{eqnarray}
and the cumulative probability is
\begin{eqnarray}
    \sum_{i=1}^N Q_i(\infty) &=& 1-\left(1-\frac{1}{n^*}\right)^{N} \sim 1- e^{-\frac{N}{n^*}},
\end{eqnarray}
which was chosen in Eq.~\ref{eq:pncum} as a fitting function for $\hat{P}_N$.

In order to calculate the probability to capture a DM particle after exactly $N$ scatterings, we need to plug inside the capture integral 
\begin{eqnarray}
    \hat{Q}_N &=& \frac{dQ_N^{cap}}{d\optdepth} = \frac{1}{n^{*}}\left(1-\frac{1}{n^{*}}\right)^{N-1}\frac{\optdepth^{N-1}}{(N-1)!}e^{-\optdepth} .
\end{eqnarray}
To calculate the total capture probability, we need to calculate explicitly the sum for all scatterings, namely
\begin{eqnarray}
    \eta(\optdepth) &=& \sum_{N=1}^\infty \hat{Q}_N(\optdepth) = \frac{1}{n^*}e^{-\frac{\optdepth}{n^*}},\quad m_\chi\gtrsim 10^6\GeV,\label{eq:etahighm}
\end{eqnarray}
and plug this factor inside the capture integral.
Note that for $n^*\rightarrow1$, the above expression tends to Eq.~\ref{eq:etalowm}. A problem arises, as $n^*$ is not fixed, but in fact depends on the position. To obtain a meaningful quantity,  we need to carefully address the radial dependence, which we do through 
\begin{eqnarray}
    \eta(\optdepth) &=&  \frac{1}{n^*(r)}e^{-\langle\frac{\optdepth}{n^*}\rangle},\quad m_\chi\gtrsim 10^6\GeV,
\end{eqnarray}
where $\langle\frac{\optdepth}{n^*}\rangle$ is calculated by plugging a factor $1/n^*(x)$ inside the integrals \ref{eq:taum} and \ref{eq:taup}. This means assuming that $N-1$ scatterings take place along the path over which is integrated, and the last scattering which causes the particle to be captured happens at the point $r$.

\subsection{Capture Rate in the large cross section limit}
\label{sec:largexslimit}

We now assume that the cross section is very large. In this limit,  $\optdepth^+(r,y)\gg\optdepth^{-}(r,y)$ and 
\begin{equation}
\eta \sim \int_0^1 dy e^{-\optdepth^-(r,y)}f_J(y), \quad m_\chi\gg 10^6\GeV.
\end{equation}
The capture rate integral has the form
\begin{eqnarray}
C=\int_0^{\Rstar} dr 4\pi r^2 n(r)\sigma f(r) \eta(r),
\end{eqnarray}
where $f(r)$ is some function for $r$, which can be obtained from Eqs.~\ref{eq:captureclsimplrel} and \ref{eq:omegampaulitext}, or from the approximated expression \ref{eq:csimplelargemtext}. 

Due to the factor $\eta$, the probability $Q_N$ is exponentially suppressed everywhere except at the surface of the star. Therefore, we expand the integral around $r=\Rstar$. Setting $r=\Rstar(1-\rho)$, and considering a thin layer of thickness $\epsilon \Rstar$, we obtain
\begin{eqnarray}
C&\sim&\int_{0}^\epsilon d\rho 4\pi \Rstar^3 n(\Rstar)\sigma f(\Rstar) \int_0^{1} dy f_J(y)\eta(\optdepth) \\
&\sim&4\pi \Rstar^3 n(\Rstar)\sigma f(\Rstar) \int_0^{1} f_J(y)dy \int_0^\epsilon d\rho \eta(\optdepth).
\end{eqnarray}
Close to the surface, we can expand $\optdepth^{-}$  in $r$, such that
\begin{eqnarray}
\optdepth &\sim& \frac{n(\Rstar)\sigma \Rstar \rho}{\sqrt{1-y^2}},
\end{eqnarray}
after which substituting $\rho$ with $\optdepth$ gives  
\begin{equation}
C=4\pi \Rstar^2 f(\Rstar) \int_0^{1} dy f_J(y)\sqrt{1-y^2} \int_0^{\optdepth(\epsilon)} d\optdepth \eta(\optdepth)
= \pi \Rstar^2 f(\Rstar) \int_0^{\optdepth(\epsilon)} d\optdepth \eta(\optdepth).\label{eq:cquasigeom}
\end{equation}
We then take the limit $\sigma\rightarrow\infty$, i.e.  $\optdepth(\epsilon)\rightarrow\infty$,
\begin{eqnarray}
C&\sim&\pi \Rstar^2 f(\Rstar) \int_0^\infty d\optdepth \eta(\optdepth) = \pi \Rstar^2 f(\Rstar) = C_\text{geom},
\end{eqnarray}
which is the geometric limit.

%%%%%%%%%%%%%%%%%%%%%%%%%%%%%%%%%%%%%%%%%%
%	BIBLIOGRAPHY
%%%%%%%%%%%%%%%%%%%%%%%%%%%%%%%%%%%%%%%%%%

\label{Bibliography}

\lhead{\emph{Bibliography}} % Change the page header to say "Bibliography"

\bibliography{Bibliography} % The references (bibliography) information are stored in the file named "Bibliography.bib"

\end{document}